\documentclass[prmat,reprint]{revtex4-1}

\usepackage{graphicx}  
\usepackage{dcolumn}   
\usepackage{bm}        
\usepackage{amssymb}   
\usepackage{amsmath}
\usepackage{comment}
\usepackage[utf8]{inputenc}
\usepackage{float}
\usepackage[dvipsnames]{xcolor}
\usepackage{graphicx}
\usepackage{comment}
\usepackage{bigints}

\newcommand{\pltime}{\tau_{\rm{pl}}}
\newcommand{\ps}{P^{\rm ss}}

\newcommand{\etau}{\delta \sigma_{\rm{u}}}

\newcommand{\Go}{G_0}

\newcommand{\tw}{t_{\rm w}}
\newcommand{\Gna}{G^{\rm na}}

\usepackage{pbox}
\usepackage{tabu}

\usepackage{mathtools}
\usepackage{makecell,tabularx}
\setcellgapes{3pt}

\newcolumntype{b}{X}
\newcolumntype{s}{>{\hsize=.5\hsize}X}

\newcommand{\bG}{\bar{G}^{*}}
\newcommand{\Gf}{G_f}
\newcommand{\Gammass}{\Gamma^{\rm{ss}}}
\newcommand{\deltatp}{\Delta \tilde{t}}

\begin{document}


\title{Mean field description of aging linear response in athermal amorphous solids} 



\author{Jack T. Parley}
 \email[Author to whom correspondence should be addressed: ]{jack.parley@uni-goettingen.de}
 \affiliation{Institut f{\"u}r Theoretische Physik, University of G{\"o}ttingen,
Friedrich-Hund-Platz 1, 37077 G{\"o}ttingen, Germany}

\author{Rituparno Mandal}
\affiliation{Institut f{\"u}r Theoretische Physik, University of G{\"o}ttingen,
Friedrich-Hund-Platz 1, 37077 G{\"o}ttingen, Germany}


\author{Peter Sollich}
\affiliation{Institut f{\"u}r Theoretische Physik, University of G{\"o}ttingen,
Friedrich-Hund-Platz 1, 37077 G{\"o}ttingen, Germany}
\affiliation{Department of Mathematics, King’s College London, London WC2R 2LS, UK}



\date{\today}

\begin{abstract}

We study the linear response to strain in a mean field elastoplastic model for athermal amorphous solids, incorporating the power-law mechanical noise spectrum arising from plastic events. In the ``jammed'' regime of the model, where the plastic activity exhibits a non-trivial slow relaxation referred to as \textit{aging}, we find that the stress relaxes incompletely to an age-dependent plateau, on a timescale which grows with material age. We determine the scaling behaviour of this aging linear response analytically, finding that key scaling exponents are universal and independent of the noise exponent $\mu$. For $\mu>1$, we find simple aging, where the stress relaxation timescale scales linearly with the age $t_{\rm w}$ of the material. At $\mu=1$, which corresponds to interactions mediated by the physical elastic propagator, we find instead a $t_{\rm w}^{1/2}$ scaling arising from the stretched exponential decay of the plastic activity. We compare these predictions with measurements of the linear response in computer simulations of a model jammed system of repulsive soft athermal particles, during its slow dissipative relaxation towards mechanical equilibrium, and find good agreement with the theory.
\end{abstract}


\maketitle 

\section{\label{sec:intro}Introduction}
Amorphous solids, including foams and emulsions used in everyday life, show rich and complex behaviour, and have long posed a challenge to theoretical progress due to their inherent disorder~\cite{nicolas_deformation_2018,bonn_yield_2017,berthier_theoretical_2011}. Many of these systems are effectively athermal because the constituent elements (be they droplets, bubbles or particles) are large enough for thermal fluctuations to be neglected. Progress in the understanding of the mechanical behaviour of such systems has been facilitated by elastoplastic models \cite{nicolas_deformation_2018}, which propose a mesoscopic approach. This is based on the substantial numerical and experimental evidence showing that local plastic (non-affine) rearrangements are the key to understanding deformation and flow in these systems~\cite{nicolas_deformation_2018,argon_plastic_1979,maloney_amorphous_2006,tanguy_plastic_2006,puosi_time-dependent_2014}. Elastoplastic models accordingly describe the dynamics of mesoscopic stress elements as consisting of periods of elastic loading interrupted by plastic relaxation events. This elastoplastic approach has been very successful in studying the yielding of amorphous solids under mechanical deformation ~\cite{lin_scaling_2014,lin_mean-field_2016,liu_driving_2016,fernandez_aguirre_critical_2018,ferrero_criticality_2019,ferrero_properties_2021,ferrero_yielding_2021,barlow_ductile_2020,parley_mean_2022}.

The {\em relaxation} dynamics of athermal amorphous solids, on the other hand, has received much less attention. Recent work~\cite{chacko_slow_2019,nishikawa_relaxation_2022,mandal_multiple_2020} has shown that model athermal suspensions of soft particles above jamming can display non-trivial slow dynamics, typically referred to as aging, as they perform gradient descent in the energy landscape~\footnote{Athermal gradient descent dynamics has also been studied recently below and close to jamming, both in particle simulations~\cite{nishikawa_relaxation_2021,olsson_relaxation_2022} and from the perspective of dynamical mean field theory~\cite{manacorda_gradient_2022}}.  This \textit{athermal} aging behaviour is to be contrasted with the aging of thermal colloidal glasses \cite{hunter_physics_2012,cloitre_rheological_2000} or spin glasses \cite{cugliandolo_evidence_1994}, which has been widely studied, using e.g.\ trap based models \cite{bouchaud_weak_1992} built around thermal activation, or record dynamics \cite{boettcher_aging_2018}. The importance of ``hotspots'' of non-affine relaxation, reminiscent of local plastic (Eshelby) events, during the athermal aging process~\cite{chacko_slow_2019} leads us instead to propose an elastoplastic approach to the problem.  

In a previous paper~\cite{parley_aging_2020} we introduced a mean field elastoplastic model and showed that it presents aging behaviour, characterised by a slow decay of the yield rate, i.e.\ the number of plastic events per unit time.
The model is mean field, treating stress propagation as a mechanical noise that is power-law distributed with exponent $\mu$, the physical elastic propagator corresponding to $\mu=1$. This extended the work of Lin and Wyart \cite{lin_mean-field_2016} in steady shear, where the success of the approach regarding the exponents associated with the yielding transition suggested that this is the correct mean field model in the sense that it applies in large dimensions. 

Here, we go beyond \cite{parley_aging_2020} and study the aging of the linear shear response of the model, which unlike the yield rate can be directly compared to stress measurements in particle-based simulations or experiments. We finally carry out such a comparison, taking as reference the aging soft athermal suspension mentioned above~\cite{chacko_slow_2019}, finding good agreement with the theory for $\mu=1$.

The paper is structured as follows. In Sec.~\ref{sec:mean_field}, 
we briefly recapitulate the mean field elastoplastic model introduced in~\cite{parley_aging_2020}. In Sec.~\ref{sec:background} we provide theoretical background on how the linear response, and in particular the viscoelastic moduli, are defined in the aging regime. We also set out how they can be calculated within our model. Next, in Sec.~\ref{sec:overview} we give an intuitive scaling argument that motivates our analytical results. In Sec.~\ref{sec:aging} we then derive these results in the aging regime, both in the time and in the frequency domain. Finally, in Sections~\ref{sec:NL} and~\ref{sec:MD} we specialise to the model with $\mu=1$, first checking our results within full non-linear simulations of the mean field model and then comparing the theory to stress measurements in an athermal particle system. We conclude with a discussion and outlook towards future research in Sec.~\ref{sec:discussion}.

\section{\label{sec:mean_field}Mean field elastoplastic model}

We recall here the most important features of the mean field elastoplastic model presented in \cite{parley_aging_2020}, referring the interested reader to the original paper. Following other elastoplastic descriptions \cite{nicolas_deformation_2018}, we consider the stress dynamics of mesoscopic blocks of the system as consisting of periods of elastic loading punctuated by plastic relaxation events, where the local stress is reset to $0$. We define a local yield threshold $\sigma_c$, so that the block located at site $i$ becomes plastic at a rate $\pltime
^{-1}$ if $|\sigma_i|>\sigma_c$, at which point all other blocks instantaneously receive a stress increment $\delta \sigma$ mediated by an elastic propagator $\mathcal{G}(\bm{r})$~\cite{picard_elastic_2004}. Neglecting spatial correlations, this stress propagation can then be captured as a mean-field mechanical noise, given by a distribution of stress increments $\rho(\delta \sigma)$. This distribution behaves for small arguments as $\rho \sim (A/N) |\delta\sigma|
^{-\mu-1}$, with $N$ the size of the system, i.e.\ the number of blocks, $\mu$ the noise exponent and $A$ the coupling parameter. For large $|\delta \sigma|$, it is cut off at a system size-independent upper cutoff $\etau=(2A/\mu)^{1/\mu}$ that corresponds physically to the stress increment caused by yielding in a directly neighbouring block.

\begin{figure}\center{\includegraphics[width=0.35\textwidth]{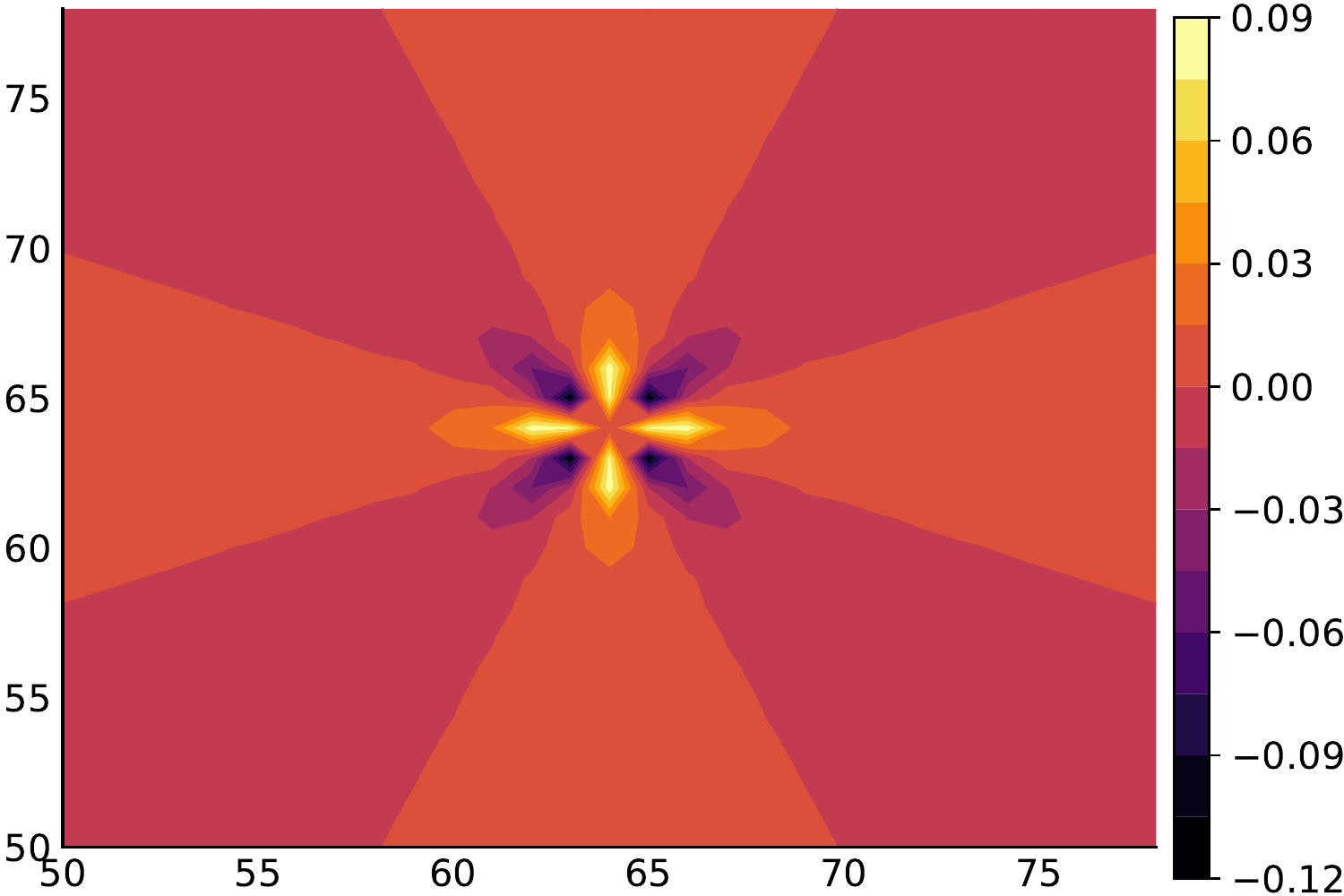}}
\caption{\label{fig:propa}Stress propagation caused by a localised plastic relaxation event in $2d$. Colour map shows the stress propagator elements after a unit stress drop at the center of a $128 \times 128$ square lattice (see \cite{parley_aging_2020} for details; for the purpose of the colour map we have set the central stress propagator element to zero). Data courtesy of Suzanne M. Fielding.}
\end{figure}

The model contains two key parameters, $\mu$ and $A$. The noise exponent $\mu$ is given by $\mu=d/\beta$. Here, $d$ is the spatial dimension, while $\beta$ is the decay exponent of the propagator $\mathcal{G}\sim r^{-\beta}$ with $r=|\bold{r}|$. The stress propagation from a localised plastic event is known~\cite{picard_elastic_2004} to be long-range (with $\beta=d$), and to have a spatially alternating sign (with e.g. a quadrupolar form in $2d$, see Fig.~\ref{fig:propa}). If as discussed above one considers stress propagation from isolated plastic events, this implies $\mu=1$. From a more coarse-grained perspective, it has been argued that mechanical noise accumulated within some fixed time interval should be considered as arising from collections of avalanches \cite{fernandez_aguirre_critical_2018,ferrero_criticality_2019,ferrero_properties_2021,ferrero_yielding_2021}, which leads to a mean field model with $1<\mu<2$. We will therefore develop our analysis for generic exponent values $\mu$ in the range $1\leq \mu\leq 2$.

The second model parameter, i.e.\ the coupling constant $A$, can also be derived \cite{parley_aging_2020} from two different perspectives. In a lattice model with one block per site $A$ is fixed by the form of the Eshelby propagator for the given lattice geometry (e.g.~$A\simeq 0.32$ for a square $2$D lattice). If instead one views the constituent blocks of the system as weak zones at randomly distributed sites, $A$ depends on the strength of the elastic interactions and on the density of such sites. We will therefore also treat it as a tunable parameter.

The master equation describing the mean field elastoplastic dynamics described above can be shown to be~\cite{parley_aging_2020} 
\begin{eqnarray}\label{ME}
     \partial_t P(\sigma,t)&=&-G_0\dot{\gamma}\partial_{\sigma}P(\sigma,t)
     \nonumber\\
     &&{}+A \Gamma(t)\int_{\sigma-\etau}^{\sigma+\etau}\mathrm{d}\sigma' \ \frac{P(\sigma',t)-P(\sigma,t)}{|\sigma-\sigma'|^{\mu+1}}
    \nonumber\\ &&{}-\frac{\theta(|\sigma|-\sigma_c)}{\pltime}P(\sigma,t)+\Gamma(t)\delta(\sigma)
 \end{eqnarray}
where the yield rate is defined as
\begin{equation}
    \Gamma(t)=\frac{1}{\pltime}\int_{-\infty}^{\infty}\theta(|\sigma|-\sigma_c)P(\sigma,t)\mathrm{d}\sigma
\end{equation}
The first term on the right hand side of (\ref{ME}) describes elastic loading of the blocks by external shear strain with shear rate $\dot\gamma$, with $G_0$ the shear modulus; the second one captures the redistribution of stress caused by yield events, and the third and fourth terms represent the local yield events for $|\sigma|> \sigma_c$ that cause the stress to be reset to zero. As also shown in \cite{parley_aging_2020}, the master equation (\ref{ME}) for general $\mu $ becomes that of the well known H\'{e}braud-Lequeux (HL) model with coupling constant $\alpha$ \footnote{In taking the limit $\mu\rightarrow 2$, one scales $A$ to zero as $A\sim 2-\mu$ so that the second moment of the jump distribution $\alpha_{\rm eff}=A/(2-\mu)\left(2A/\mu\right)^{2/\mu-1}$ goes to a finite limiting value $\alpha_{\rm{HL}}$ corresponding to the coupling parameter of the HL model.} for $\mu\rightarrow 2$
\begin{eqnarray}\label{hl_equation}
\frac{\partial P(\sigma,t) }{\partial t} &=&-G_0 \dot{\gamma}\frac{\partial P}{\partial\sigma}
\label{HL}\\
&&{}+\alpha \Gamma(t)\frac{\partial^2 P}{\partial \sigma^2}-\frac{\theta(|\sigma|-\sigma_c)}{\pltime}P+\Gamma(t)\delta(\sigma)
\nonumber
\end{eqnarray}

\begin{figure}\center{\includegraphics[width=0.35\textwidth]{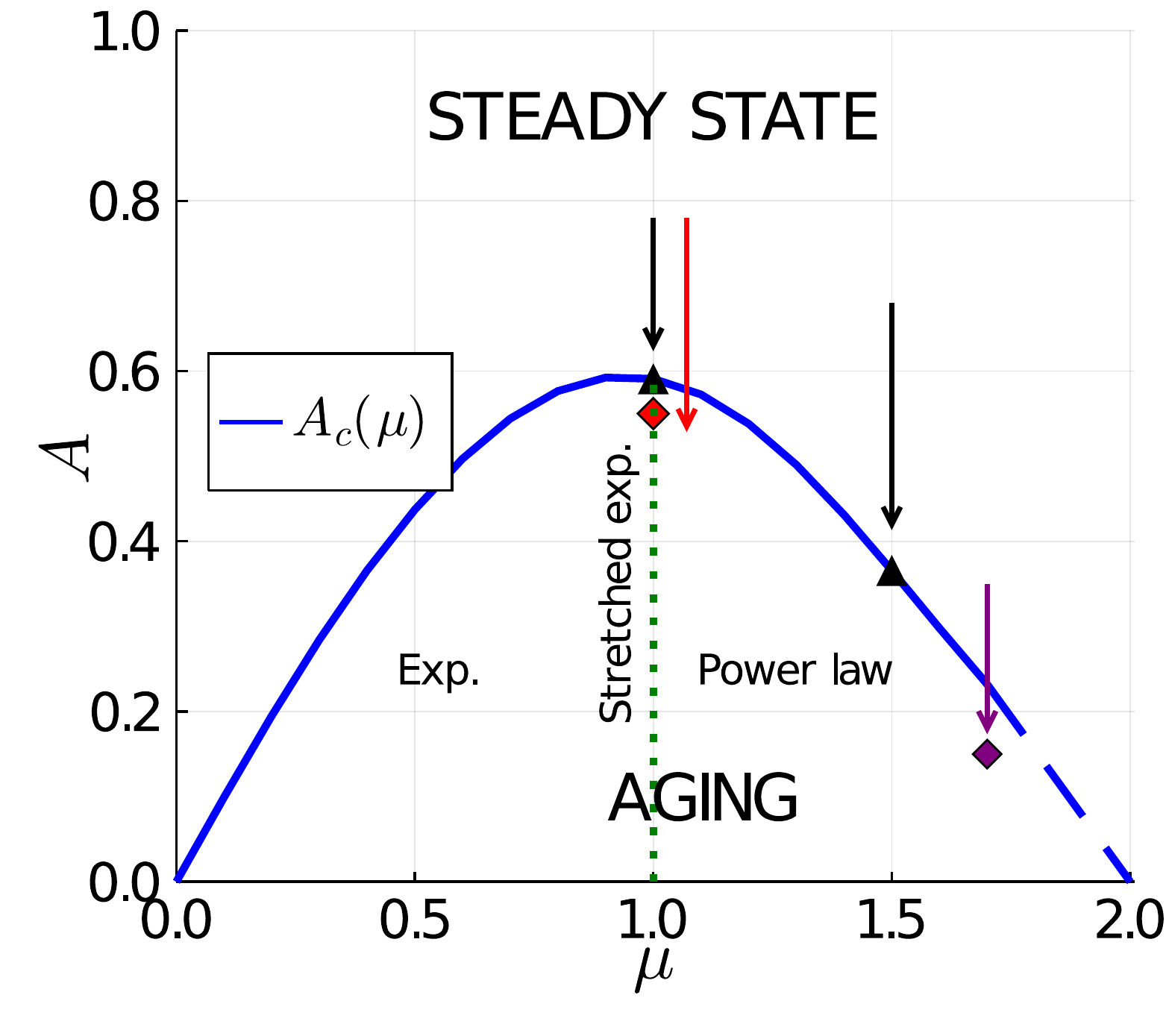}}
\caption{\label{fig:LR_points} Phase diagram of the model in the $(\mu,A)$ plane \cite{parley_aging_2020}, defined by the curve $A_c(\mu)$ (blue). The two black triangles 
($\mu=1.0, A=A_c(1.0)$ and $\mu=1.5,     A=A_c(1.5)$) and arrows indicate the numerical parameter values for which we study aging at criticality in App.~\ref{app:criticality}. The purple ($\mu=1.7, A=0.15$) and red ($\mu=1.0, A=0.55$) diamonds and arrows show the cases studied for aging in the glass phase  in Sec.~\ref{sec:aging}. 
}
\end{figure}

We summarise briefly the phase diagram of the model in the $(\mu,A)$ plane, studied in detail in~\cite{parley_aging_2020}. There, the critical coupling curve $A_c(\mu)$ (reproduced in Fig.~\ref{fig:LR_points}) separating the two phases of the model was found numerically, presenting a bell-shaped form with a peak at $\mu \simeq 1$. For $A>A_c(\mu)$, the system is in a ``liquid'' phase behaving as a Newtonian fluid $\Sigma=\eta\dot{\gamma}$ under applied shear; here and throughout the macroscopic stress is taken as the average $\Sigma(t)=\int \mathrm{d}\sigma \ \sigma P(\sigma,t)$. Without shear, the system is able to sustain a steady state with finite yield rate $\Gammass>0$, behaving essentially as a Maxwell fluid with a finite relaxation time. The latter diverges as $A\rightarrow A_c^{+}$, with anomalous non-Maxwellian behaviour arising as this critical point is approached (see below). The existence of such a steady state within the model has been argued to be unphysical \cite{agoritsas_relevance_2015}, given that external driving should be necessary to maintain the dissipative plastic events. On the other hand, elastoplasticity has been shown to play an important role also in unsheared systems, particularly for long-range dynamic facilitation in supercooled liquids below the mode-coupling temperature~\cite{chacko_elastoplasticity_2021}. The unsheared steady state regime may therefore be relevant in such a context, although one would presumably need to generalize the model discussed here to explicitly include the thermal activation of plastic events (along the lines of \cite{popovic_thermally_2021}).


We will in any case focus mainly on the aging regime below. In this glassy phase for $A<A_c(\mu)$, there is no steady state with $\Gamma >0$ in the absence of shear, and the yield rate decays as the system approaches an initial condition dependent frozen-in stress distribution $Q_0(\sigma)\equiv P_0(\sigma,t\rightarrow \infty)$ (see e.g.\ Fig.~2 in \cite{parley_aging_2020}). This  distribution was shown to exhibit \cite{lin_mean-field_2016,parley_aging_2020} pseudogap scaling near the yield threshold, $Q_0(\sigma)\sim (\sigma_c-|\sigma|)^{\mu/2}$. 
This behaviour is found also in the steady state stress distribution on the liquid side in the limit $\Gamma \rightarrow 0$, and is in agreement with the results of MD simulations~\cite{shang_elastic_2020}.

In \cite{parley_aging_2020}, we studied the slow decay of $\Gamma(t)$ by evolving the unperturbed dynamics starting from an initial distribution with enough unstable sites. This was argued to represent the dynamics of the system after an initial preparation, such as stirring, shear melting or a sudden change in density \cite{parley_aging_2020}. If the system is athermal, the ensuing dissipative dynamics is driven by rearrangements that can only be triggered by events taking place elsewhere in the system, as described here. The yield rate was found to decay as a power law $\Gamma(t)\sim t^{-\mu/(\mu-1)}$ for $1<\mu<2$, a stretched exponential $\Gamma(t)\sim e^{-B\sqrt{t}}$ for $\mu=1$ and an exponential for $\mu<1$, reflecting the relative importance of far-field and near-field events as the range of the stress propagator is varied \cite{parley_aging_2020}. The different regimes are sketched in Fig.~\ref{fig:LR_points}, where we indicate also the different parameter values for which we will study the linear shear response numerically in this paper. We include among these two parameter values pertaining to the case of \textit{critical aging}, i.e.\ relaxation at criticality $A=A_c(\mu)$, where the yield rate decays as $\Gamma(t)\sim t^{-1}$ for all $\mu$~\cite{parley_aging_2020}. 

\section{\label{sec:background}Theoretical background}

We consider in this section the linear shear rheology of an amorphous system relaxing after preparation at time $t=0$. We assume that a small step strain $\gamma(t)=\gamma_0 \theta(t-t_{\rm w})$ (with $\gamma_0 \ll 1$) is applied at a certain switch-on time, which we denote as the waiting time $t_{\rm w}$. The corresponding shear stress is given by the linear  constitutive equation
\begin{equation}
    \sigma(t)=\int_{-\infty}^{t} G(t,t')\dot{\gamma}(t')\mathrm
{d}t'
\end{equation}
where $G(t,t')$ is the so-called stress relaxation function. From this response to a step strain one may then derive the linear response to more complex perturbations such as oscillatory strain, as described below.

Within our mean field elastoplastic model, the shear perturbation manifests itself via its effect on the dynamics of the stress distribution $P(\sigma,t)$.
In the generic aging case, both the unperturbed distribution $P_0(\sigma,t)$ and the unperturbed yield rate $\Gamma_0(t)$ will depend on time. We then expand the perturbed solution $P(\sigma,t)$ of the master equation (\ref{ME}) for $t>t_{\rm w}$ as
\begin{equation}
    P(\sigma,t)=P_0(\sigma,t)+\gamma_0 \, \delta P(\sigma,t)+\mathcal{O}(\gamma_0^2)
    \label{P_perturb}
\end{equation}
Likewise, for the yield rate we may write
\begin{equation}
    \Gamma(t)=\Gamma_0(t)+\gamma_0 \,\delta \Gamma (t)+\mathcal{O}(\gamma_0^2)
\end{equation}
To simplify the analysis we now assume 
as in \cite{sollich_aging_2017,parley_aging_2020} that the system preparation leads to a symmetric initial stress distribution $P_0(\sigma,0)$. The unperturbed dynamics preserves this symmetry, so that $P_0(\sigma,t)=P_0(-\sigma,t) \ \forall t$. With this assumption, one may show as in \cite{sollich_aging_2017} that the first order correction to the yield rate $\delta \Gamma (t) $ vanishes. This simply follows from the invariance of the time evolution of the master equation (\ref{ME}) under joint sign reversal of $\sigma$ and $\gamma_0$, which implies that $\delta P(\sigma,t)$ must be an odd function of $\sigma$, so that
\begin{equation}
    \delta \Gamma (t)=\frac{1}{\pltime}\int_{-\infty}^{\infty}\mathrm{d}\sigma\ \theta(|\sigma|-\sigma_c)\delta P(\sigma,t)=0
\end{equation}
If we now insert the perturbed form (\ref{P_perturb}) of $P(\sigma,t)$ into the master equation (\ref{ME}), we find at $\mathcal{O}(\gamma_0)$ and for $t>t_{\rm w}$ the following equation for the perturbation: 
\begin{eqnarray}\label{deltap}
     \partial_t \delta P(\sigma,t)&=&A \Gamma(t)\int_{\sigma-\etau}^{\sigma+\etau}\mathrm{d}\sigma' \ \frac{\delta P(\sigma',t)-\delta P(\sigma,t)}{|\sigma-\sigma'|^{\mu+1}} \nonumber
    \\&&-
    \frac{\theta(|\sigma|-\sigma_c)}{\pltime}\delta P(\sigma,t)
\end{eqnarray}
The initial condition for this is found by integrating (\ref{ME}) in a small time interval around $t=t_{\rm w}$, giving
\begin{equation}\label{ic}
    \delta P(\sigma,t_{\rm w})=-\partial_{\sigma}P_0(\sigma,t_{\rm w})
\end{equation}
Since we identify the macroscopic stress with the average over the local distribution, once we have found $\delta P(\sigma,t)$ the linear stress relaxation function can be computed as 
\begin{equation}\label{G}
    G(t,t_{\rm w})=\int_{-\infty}^{\infty}\mathrm{d}\sigma \ \sigma \delta P(\sigma,t)=2\int_0^{\infty}\mathrm{d}\sigma \ \sigma\delta P(\sigma,t)
\end{equation}
where the second equality follows from the anti-symmetry of $\delta P$. Using the initial condition (\ref{ic}) and bearing in mind  that $P_0(\sigma,\tw)$ is normalised we have the initial value $G(t_{\rm w},t_{\rm w})=G_0$.

The steady state and aging stress relaxation are distinct in their dependence on the waiting time $t_{\rm w}$. If the unperturbed system is already prepared in a steady state, $P_0(\sigma,t)=\ps(\sigma)$ and $\Gamma_0(t)=\Gamma^{\rm ss}$ are independent of time and we find as expected a time translation invariant (TTI) stress relaxation function $G(t,t_{\rm w})=G(t-t_{\rm w})\equiv G(\Delta t)$. In the aging regime, on the other hand, this invariance is lost and $G(t,t_{\rm w})$ in general depends on both time arguments.

A similar distinction may be made in the frequency response, for which we follow the generic discussion in \cite{fielding_ageing_2000}. For TTI systems, we may write the response to an oscillatory strain $\gamma(t)=\Re[\gamma_0 e^{i (\omega t+\phi)}] $ as $\sigma (t)=\Re[G^{*}(\omega)\gamma_0 e^{i (\omega t+\phi)}]$, 
where the viscoelastic spectrum $G^{*}(\omega)=G'(\omega)+iG''(\omega)$ is proportional to the Fourier transform of $G(\Delta t)$. In aging systems \cite{fielding_ageing_2000}, the viscoelastic spectrum generically depends on three arguments: the oscillatory frequency $\omega$, the time $t$ when the stress is measured and the waiting time $t_{\rm w}$. One finds
\begin{multline}\label{aging_Gstar}
    G^*(\omega,t,t_{\rm w})\\
    =G(t,t_{\rm w})e^{-i \omega(t-t_{\rm w})}+i \omega \int_{t_{\rm w}}^{t}\mathrm{d}t' \ G(t,t')e^{-i \omega (t-t')}
\end{multline}
In the limit where $\omega (t-t_{\rm w}) \gg 1$ (many oscillations before the stress measurement) and $\omega t_{\rm w} \gg 1$ (large waiting time), equation (\ref{aging_Gstar}) may approach the \textit{forward spectrum} $\Gf^{*}(\omega,t)$. This is 
calculated by assuming the strain is applied from the measurement time $t$ {\em into the future}:
\begin{equation}\label{FS}
    G_{f}^{*}(\omega,t)=i \omega \int_{t}^{\infty}{\rm d}t' \ G(t',t)e^{-i \omega (t-t')}
\end{equation}
We will show, both numerically and analytically (in App.~\ref{app:FS}), that this limiting behaviour holds in our elastoplastic model. Note that generally we also require the condition $\omega \ll 1/{\pltime}$ to stay within the range of applicability of the model, which does not include e.g.\ dissipative effects from solvent viscosity that would become relevant at higher frequencies.

Finally, we propose an alternative approach for numerically calculating the aging frequency response $G^*(\omega,t,t_{\rm w})$, which helps to reduce oscillations that appear when using directly the original expression (\ref{aging_Gstar}). This approach is inspired by experimental work \cite{purnomo_glass_2008,purnomo_linear_2006,purnomo_rheological_2007} and is closer to how the frequency response is measured in reality, where one needs to measure the relative phase and amplitude across several periods. We take the stress signal $\sigma(t)$ and correlate it with the strain signal $\gamma(t)$ over a time window of $m$ periods around an observation time $t$ . We denote this averaged response by $\bG(\omega,t,t_{\rm w})$ 
\begin{equation}\label{purno}
    \bG(\omega,t,t_{\rm w})=\frac{\omega}{m \pi \gamma_0}\int_{t-\frac{m \pi}{\omega}}^{t+\frac{m \pi}{\omega}}\mathrm{d}t' \ \sigma (t') e^{-i(\omega t' + \phi)} 
\end{equation}
where as usual $\bG$ can be separated into $\bG=\bar{G}'+i \bar{G}''$. If we then express $\sigma(t)$ in terms of the \textit{unaveraged} moduli $G^{*}=G'+i G''$, the above expression becomes:
\begin{widetext}
\begin{eqnarray}\label{averaged_Gstar}
\bG(\omega,t,t_{\rm w})&=&\frac{\omega}{m\pi}\Bigg(\int_{t-\frac{m\pi}{\omega}}^{t+\frac{m\pi}{\omega}}\mathrm{d}t' \ \left(\cos(\omega t'+\phi)^2 G'(\omega,t',t_{\rm w})-\frac{1}{2}\sin(2(\omega t'+\phi))G''(\omega,t',t_{\rm w})\right) \nonumber\\&&{}+
i \int_{t-\frac{m\pi}{\omega}}^{t+\frac{m\pi}{\omega}}\mathrm{d}t' \ \left(\sin(\omega t'+\phi)^2 G''(\omega,t',t_{\rm w})-\frac{1}{2}\sin(2(\omega t'+\phi))G'(\omega,t',t_{\rm w})\right)\Bigg)
\end{eqnarray}
\end{widetext}
The oscillations in $G^{*}(\omega,t,t_{\rm w})$ are of frequency $\omega$ (see also Fig.~\ref{fig:osci_FS} in App.~\ref{app:FS}). They are thus orthogonal to the constant and $2 \omega$ kernels in the averaging formula above, and therefore no longer present in the resulting averaged moduli.

To calculate $\bG$ in practice, we express it directly in terms of the age-dependent relaxation function $G(t,t')$. In order to simplify this expression we make a particular choice for the phase of the strain signal $\gamma(t)=\Re[\gamma_0 e^{i (\omega t+\phi)}]$, fixing $\phi=-\omega t_{\rm w}-\pi/2$.
This ensures that $\gamma(t)=\gamma_0\sin\left(\omega(t-t_{\rm w})\right)$ and hence that the applied strain starts continuously from zero, leading to the simplified result~\footnote{We note for clarity that this special choice of phase is made solely to simplify the expression (\ref{Gav_time}), and does not in itself contribute to reducing the oscillations in $G
^{*}(\omega,t,t_{\rm w})$. The reduction of oscillations is accomplished by the averaging, and is independent of the choice of phase $\phi$.}
\begin{widetext}
\begin{equation}\label{Gav_time}
    \bG(\omega,t,t_{\rm w}) 
    =\frac{\omega}{m\pi}\int_{t-\frac{m\pi}{\omega}}^{t+\frac{m\pi}{\omega}}\mathrm{d}t' \  \big[\sin{\left(\omega(t'-t_{\rm w})\right)}
    +i\cos{\left(\omega(t'-t_{\rm w})\right)}\big]
    \int_{t_{\rm w}}^{t'}\mathrm{d}t'' \  G(t',t'') \   \omega \cos{\left(\omega(t''-t_{\rm w})\right)}
\end{equation}
\end{widetext}
which we will use for the numerical results shown in Sec.~\ref{sec:aging}. This form can also be obtained directly from (\ref{purno}) with the appropriate choice of the phase angle.



\section{\label{sec:overview}Overview of analytical results}
Before we turn to analyse the aging linear response in detail, we give a brief overview of the analytical results, highlighting universal features that are independent of the noise exponent $\mu$. Here and in the following, we set $\sigma_c=1$ and $\pltime=1$, providing the stress and time units. In addition, without loss of generality we set $\Go=1$, so that $G(t_{\rm w},t_{\rm w})=1$. This is not a choice of stress units (the unit of stress being set by the yield threshold); rather it represents a numerical constant that can simply be absorbed into the applied strain. The \textit{amount of stress} that has been relaxed up to time $t$, due to plastic events, can then be written as
\begin{eqnarray}\label{1_G_initial}
    G(t_{\rm w},t_{\rm w})-G(t,t_{\rm w})&=&1-G(t,t_{\rm w})\\&=&\int_{-\infty}^{\infty}\sigma \left(\delta P(\sigma,t_{\rm w})-\delta P(\sigma,t)\right)\mathrm{d}\sigma \nonumber
\end{eqnarray}
where in the second line we have used (\ref{G}) and the normalisation of $\sigma \,\delta P(\sigma,t_{\rm w})$ stemming from (\ref{ic}). We will denote the total (asymptotic) amount of stress the system is able to relax as
\begin{equation}\label{1_G}
    1-G_\infty(t_{\rm w})\equiv 1-G(t \rightarrow \infty,t_{\rm w})
\end{equation}
For a system which is able to relax fully, this quantity is thus unity.

The intuition behind our analytical results is given mainly by the following argument. Both in steady state and in aging, after the step strain is applied the relaxation is at first purely confined to two small symmetric regions around the boundaries $\sigma=\pm 1$. The two symmetric boundary layers make an equal contribution to the ensuing stress relaxation, so for the following discussion we focus on the positive boundary layer around $\sigma=1$, corresponding to $1-\sigma \ll 1$. In this region, blocks are close enough to instability so that their stress can diffuse across the boundary set by the yield threshold in the short time regime, and a significant decay in $\delta P(\sigma,t)$ takes place. More precisely, up to a time $t$ we expect the diffusion due to mechanical noise to result in a stress scale
\begin{equation}\label{Delta_sigma}
    \Delta \sigma\sim \left(\int_{t_{\rm w}}
^{t}\Gamma(t')\mathrm{d}t'\right)^{H}
\end{equation}
given by the Hurst exponent $H=1/\mu$, and the corresponding form of the yield rate $\Gamma(t)$. From (\ref{1_G_initial}), this means that (assuming $\delta P(\sigma,t)$ has decayed enough, see also App.~\ref{app:delta_P}) the amount of stress relaxed up to time $t$ is essentially given by the integral of the initial condition $\delta P(\sigma,t_{\rm w})$ over the range of stress $\Delta \sigma$ below the yield threshold (note that $\sigma \approx 1$ in this range). 

\begin{figure}\center{\includegraphics[width=0.35\textwidth]{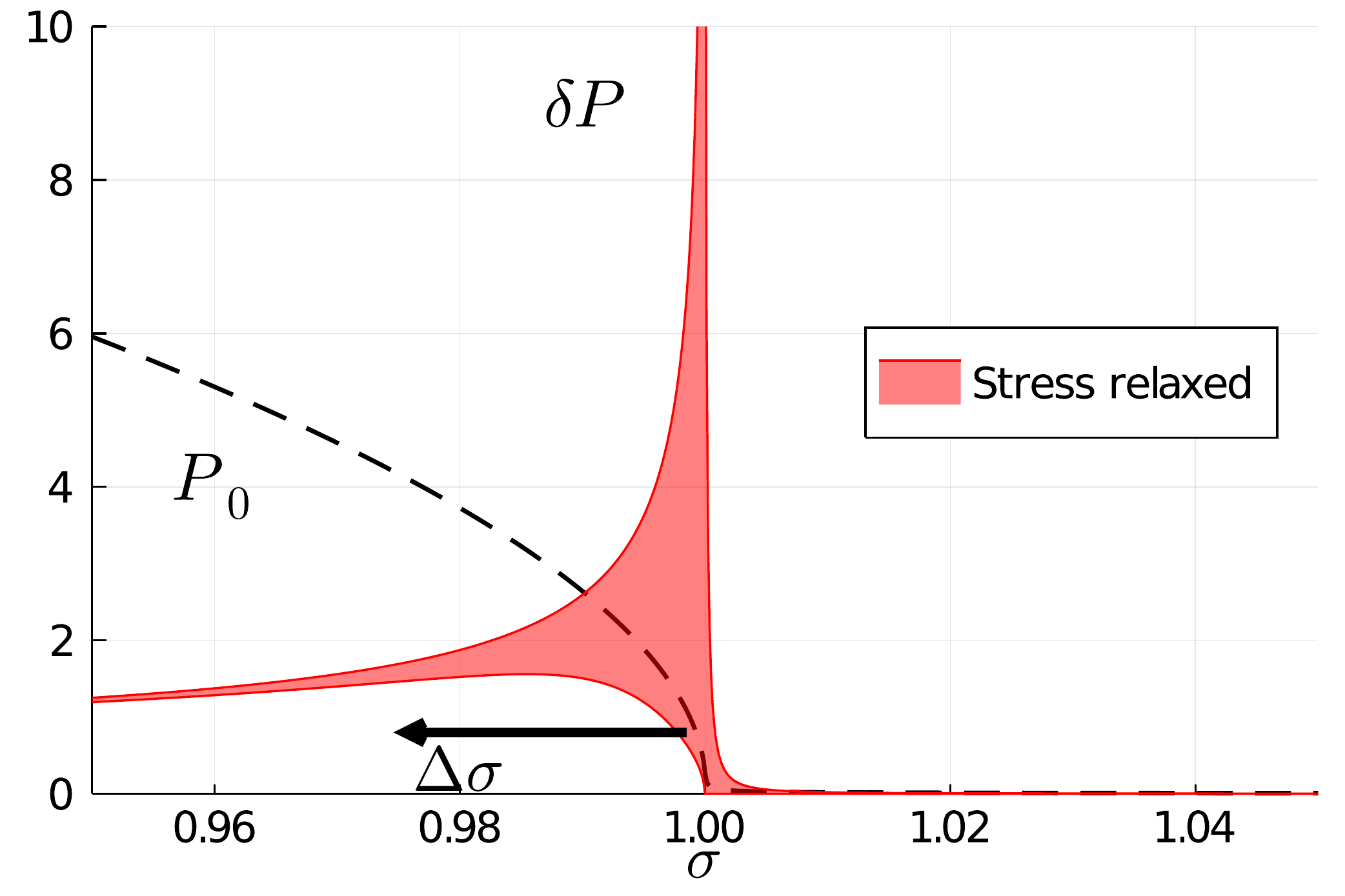}}
\caption{\label{fig:sketch}Sketch displaying the singular behaviours $P_0(\sigma,t_{\rm w})\sim (1-\sigma)^{\mu/2}$ and $\delta P(\sigma,t_{\rm w})\sim (1-\sigma)^{\mu/2-1}$ for $1-\sigma \ll 1$ (positive boundary layer), in this case for $\mu=1$. $\delta P$ decays significantly on the scale $\Delta \sigma$, so that the main contribution to the stress relaxation is the shaded area (the stress is actually the integral of $\sigma \delta P$, but $\sigma \approx 1$ in the relevant region). The negative boundary layer (at $\sigma=-1$, not shown) makes an equal contribution, with both terms in the integral ($\sigma$ and $\delta P$) changing sign. $P_0$ has been amplified by a factor of $50$ for visibility (dashed line).}
\end{figure}

We recall that this initial condition is given by the derivative of the unperturbed distribution  (\ref{ic}). Now, both the unperturbed steady state close to the arrest transition, and the unperturbed aging distribution at long times ($\Gamma(t)\ll 1$), display a pseudogap behaviour $P_0(\sigma,t_{\rm w})\sim (1-\sigma)^{\mu/2}$ for $1-\sigma \ll 1$ (see Sec.~\ref{sec:mean_field}). This means that the initial condition for the stress distribution perturbation has the scaling $\delta P(\sigma,t_{\rm w})\sim (1-\sigma)^{\mu/2-1}$ (see Fig.~\ref{fig:sketch}). To find the amount of stress relaxation, we need to integrate this over the scale $\Delta \sigma$, so that
\begin{equation}\label{integral}
    1-G(t,t_{\rm w})\sim \Delta \sigma^{\frac{\mu}{2}}\sim {\left(\int_{t_{\rm w}}^{t}\Gamma(t')\mathrm{d}t'\right)}^{\frac{1}{2}} \quad \forall \mu
\end{equation}
Remarkably, then, the exponent $1/2$ relating the amount of stress relaxation to the number of yield events is {\em universal} across all values of the exponent $\mu$. 

The detailed analytical results in the time domain -- derived below -- are displayed in Table \ref{T1} and can be related to the intuitive arguments above as follows. In the aging regime, the integral on the right hand side of (\ref{integral}) converges to a finite value. The relaxation is therefore confined to a range of stresses near the yield threshold 
and does not extend to the remainder or ``bulk'' of the stress distribution at long times. Thus the system is not able to relax the stress caused by the initial shear strain completely; instead the stress decays to a finite plateau.

For the steady state near the arrest transition, where $\Gamma(t)=\Gamma^{\rm ss} \ll 1$, Eq.~(\ref{integral}) implies an anomalous relaxation $1-G(\Delta t)\sim {\Delta t}^{1/2}$ at short times. This eventually gives way to an exponential relaxation characteristic of a Maxwell fluid (see App.~\ref{app:SS} for details). In the case of critical aging, treated in App. \ref{app:criticality}, the relaxation does extend to the bulk at long times but is given by a power-law decay instead of an exponential. Turning to the frequency domain, results for which are displayed in Table~\ref{T2}, the ubiquity of the exponent $1/2$ is evident in the behaviour of the loss modulus; as explained below, this simply mirrors the short time behaviour in the time domain. 

We saw above that the exponent $H\mu/2=1/2$ characterising the relaxation of stresses near the yield threshold is universal, i.e.\ independent of the exponent $\mu$ characterising the noise distribution. Interestingly, this universality can be traced back to a link between exponents of self-affine processes first proposed in~\cite{zoia_asymptotic_2009}. The exponent $\mu/2$ (denoted $\phi$ in~\cite{zoia_asymptotic_2009}) characterises the behaviour near an absorbing boundary, the yield threshold. This is related to the persistence exponent $\theta$, which describes the algebraic decay $\sim t
^{-\theta}$ of the probability of no return to an initial value, through $\theta=H\mu/2$. The persistence exponent $\theta$, in turn, can be shown via the Sparre-Andersen theorem \cite{andersen_fluctuations_1954,bray_persistence_2013} to take the universal value $\theta=1/2$ for any random walk with a symmetric jump distribution. This corresponds to the $1/2$ exponent we will find throughout the present analysis, albeit without the interpretation in terms of persistence.


\section{\label{sec:aging}Aging regime}
In the regime $A<A_c(\mu)$, where the system ages, one expects the decaying plastic activity to lead also to an aging linear response, given that there are fewer and fewer rearrangements available to relax the stress caused by the applied step strain. In the following we treat separately the cases $1<\mu<2$ and $\mu=1$, where \cite{parley_aging_2020} the yield rate decays respectively as a power law and as a stretched exponential (see also Fig.~\ref{fig:LR_points}). In both cases we will find that because the integral of $\Gamma(t)$, which represents the total number of plastic events that will occur in the system, remains finite then the stress relaxation function decays incompletely from unity to a plateau. On the other hand, the scaling with age of both the plateau and the typical time taken to reach it, which are the main focus of interest of our study, will depend on the exponent $\mu$.

\subsection{$1<\mu<2$}
\subsubsection{Intuitive argument in time domain}
\begin{figure}\center{\includegraphics[width=0.42\textwidth]{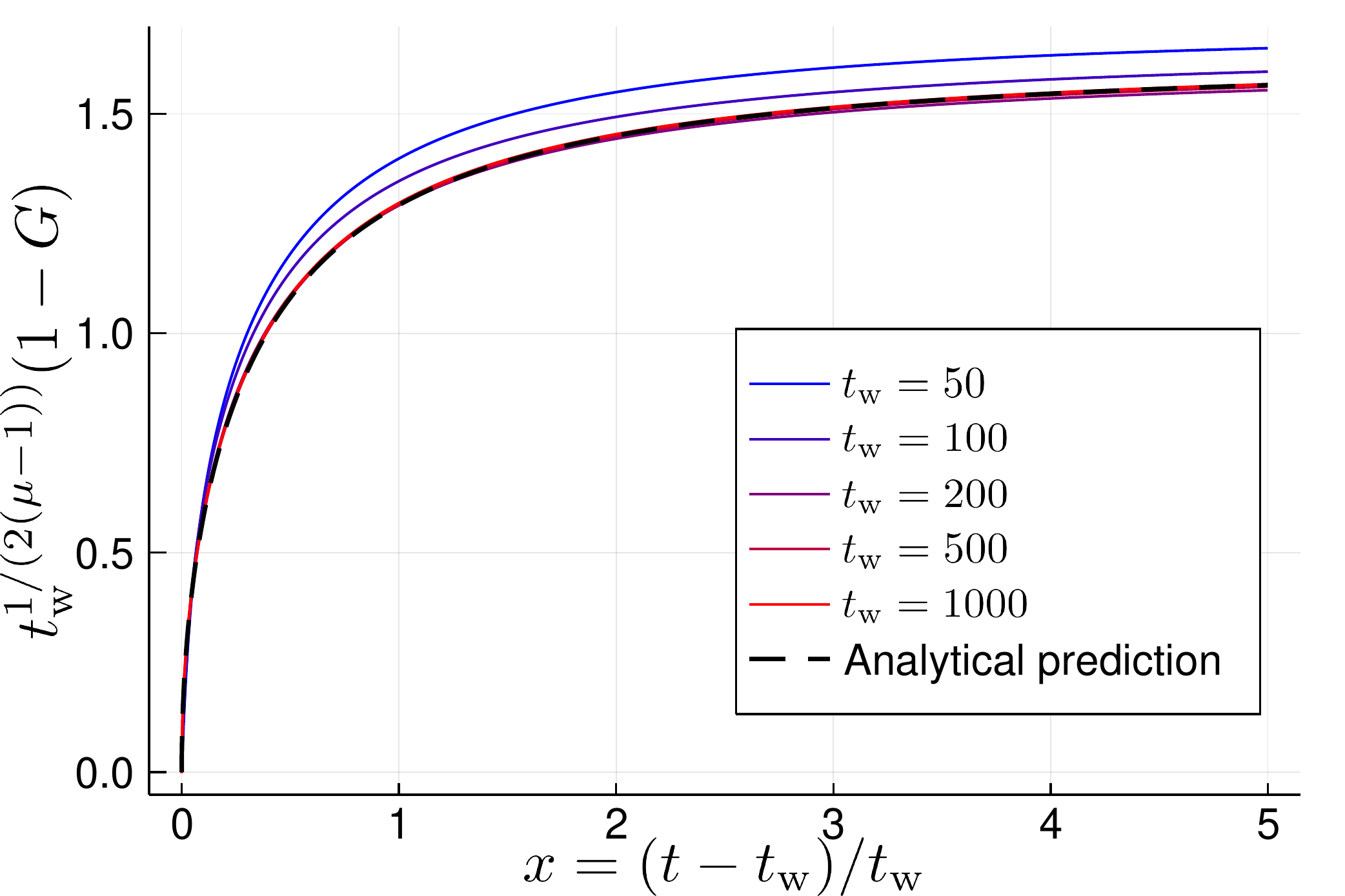}}
\caption{\label{fig:G_17}Stress relaxation obtained from numerically solving the linearised equation (\ref{deltap}) for $\mu=1.7$ and $A=0.15$, starting from initial conditions extracted at different $t_{\rm w}$ from the unperturbed aging dynamics. The curves collapse following (\ref{Ginf_12}) and (\ref{G_mu_12}) for $t_{\rm w}=200$ and above.}
\end{figure}

In the regime $1<\mu<2$, it was shown~\cite{parley_aging_2020} that at long times the yield rate ages as a power law with exponent $\Gamma(t)\sim t^{-\mu/(\mu-1)}$. We now explore the consequences of this using the same intuitive argument as in Sec.~\ref{sec:overview}, referring the reader to App.~\ref{app:delta_P} for a more detailed analysis of the full stress distribution perturbation $\delta P(\sigma,t)$. As already noted in Sec.~\ref{sec:overview}, the whole relaxation is now confined to the initial regime around the boundary layers $|\sigma|\approx 1$. Taking into account that $\delta P(\sigma,t_{\rm w})\sim(1-\sigma)
^{\mu/2-1}$ for large $t_{\rm w}$ (where $P_0(\sigma,\tw)$ is already close to $Q_0(\sigma)$), we have as before 
that $1-G(t,\tw)\sim \Delta \sigma^{\mu/2}$ with $\Delta \sigma={\left(\int_{t_{\rm w}}^{t}\Gamma (t')\mathrm{d}t'\right)}^{1/\mu}$. For waiting times large enough for $\Gamma(t)$ to have entered the asymptotic regime we therefore have that
\begin{eqnarray}\label{G_mu_12}
    1-G(t,\tw)&\sim& {\left(\int_{t_{\rm w}}^{t}\Gamma (t')\mathrm{d}t'\right)}^{\frac{1}{2}} \nonumber \\
    &\approx& c\, t_{\rm w}^{-\frac{1}{2(\mu-1)}}\sqrt{1-(1+x)^{-\frac{1}{\mu-1}}}
\end{eqnarray}
with $c$ an initial condition-dependent constant. The dependence on the measurement time $t$ can be expressed entirely via the rescaled time difference $x=(t-t_{\rm w})/t_{\rm w}$, implying {\em simple aging} where relaxation timescales grow linearly with the age $\tw$. We also see from~(\ref{G_mu_12}) that the amount of stress relaxation $1-G$ saturates to a plateau, which we denote as 
\begin{equation}\label{Ginf_12}
    1-G_{\infty}(t_{\rm w})=ct_{\rm w}^{-\frac{1}{2(\mu-1)}}
\end{equation}
To check these scaling predictions we compare them to direct numerical solutions of the time evolution~(\ref{deltap}), for the case $\mu=1.7$, $A=0.15$. We extract initial conditions $\delta P(\sigma,t_{\rm w})$ in Eq.~(\ref{ic}) from numerics for the unperturbed system~\footnote{Here and in what follows we use, as in \cite{parley_aging_2020}, the steady state with $\Gamma=0.134$ as initial distribution for the unperturbed aging dynamics.}, at different waiting times $\tw$. For the shorter waiting times up to $t_{\rm w}=200$  we include pre-asymptotic effects by using the full form of $\Gamma(t)$ measured in the unperturbed dynamics before it enters the asymptotic power law (at around $t\simeq 400$),
while for longer waiting times we use directly a fit of the asymptotic behaviour of $\Gamma(t)$~\footnote{We note that, as discussed in~\cite{parley_aging_2020}, in the unperturbed numerics the power law asymptote of $\Gamma(t)$ is eventually cut off exponentially by the fact that the required discretization of the $\sigma$-axis can no longer resolve the boundary layer.}. 
Plotting the resulting stress relaxation $1-G$ vs $t-t_{\rm w}$, while rescaling the time axis by $t_{\rm w}$ and the stress relaxation axis by the appropriate power of $t_{\rm w}$ from (\ref{Ginf_12}), we find that the rescaled curves practically collapse onto each other and show very good agreement with the asymptotic expression (\ref{G_mu_12}) for $t_{\rm w}=200$ and above (see Fig.~\ref{fig:G_17}). The curves below $t_{\rm w}=200$ converge monotonically towards the asymptotic form, with the deviations from the latter arising from the pre-asymptotic behaviour of $\Gamma(t)$, plus potentially stress relaxation extending beyond the boundary layers $|\sigma|\approx 1$, which is not accounted for in our analytical arguments.

\subsubsection{Frequency domain}

As discussed in Sec. \ref{sec:overview}, in aging systems one may in general introduce an age-dependent frequency response $G^*(\omega,t,t_{\rm w})$ (\ref{aging_Gstar}), with $t$ the time of measurement. We show in App.~\ref{app:FS} that for our model $G^*(\omega,t,t_{\rm w})$ 
does approach the forward spectrum $G_{f}^{*}(\omega,t)$ (\ref{FS}) in the limits $\omega (t-t_{\rm w})\gg 1$, $\omega t \gg 1$ (with $\omega \ll 1$)
discussed above in Sec.~\ref{sec:background}. The forward spectrum in turn is found to take the asymptotic form
\begin{equation}\label{FS_17}
    \frac{G_{f}^{*}(\omega,t)}{1-G_{\infty}(t)}\sim 1-(1-i)c\left(\frac{1}{\mu-1}\right)^{\frac{1}{2}}\sqrt{\frac{\pi}{8}}w^{-\frac{1}{2}}
\end{equation}
where we have defined a rescaled frequency $w\equiv \omega t$, and $c$ is the same initial condition dependent constant as in (\ref{G_mu_12}).

The two main features of the aging moduli~(\ref{FS_17}) directly reflect the behaviour~(\ref{G_mu_12}) in the time domain. Firstly, we see that we need to rescale the magnitude of the moduli by $1-G_{\infty}(t)$, which corresponds to the finite total amount of relaxation the system can undergo, and decays in time as the power law given in Eq.~(\ref{Ginf_12}). On the other hand, we find that once the decaying total relaxation is taken into account, the frequency response becomes a function of $w\equiv\omega t$ only. This rescaling reflects the simple aging scaling of the typical relaxation time we found in the time domain. 

\begin{figure}\center{\includegraphics[width=0.42\textwidth]{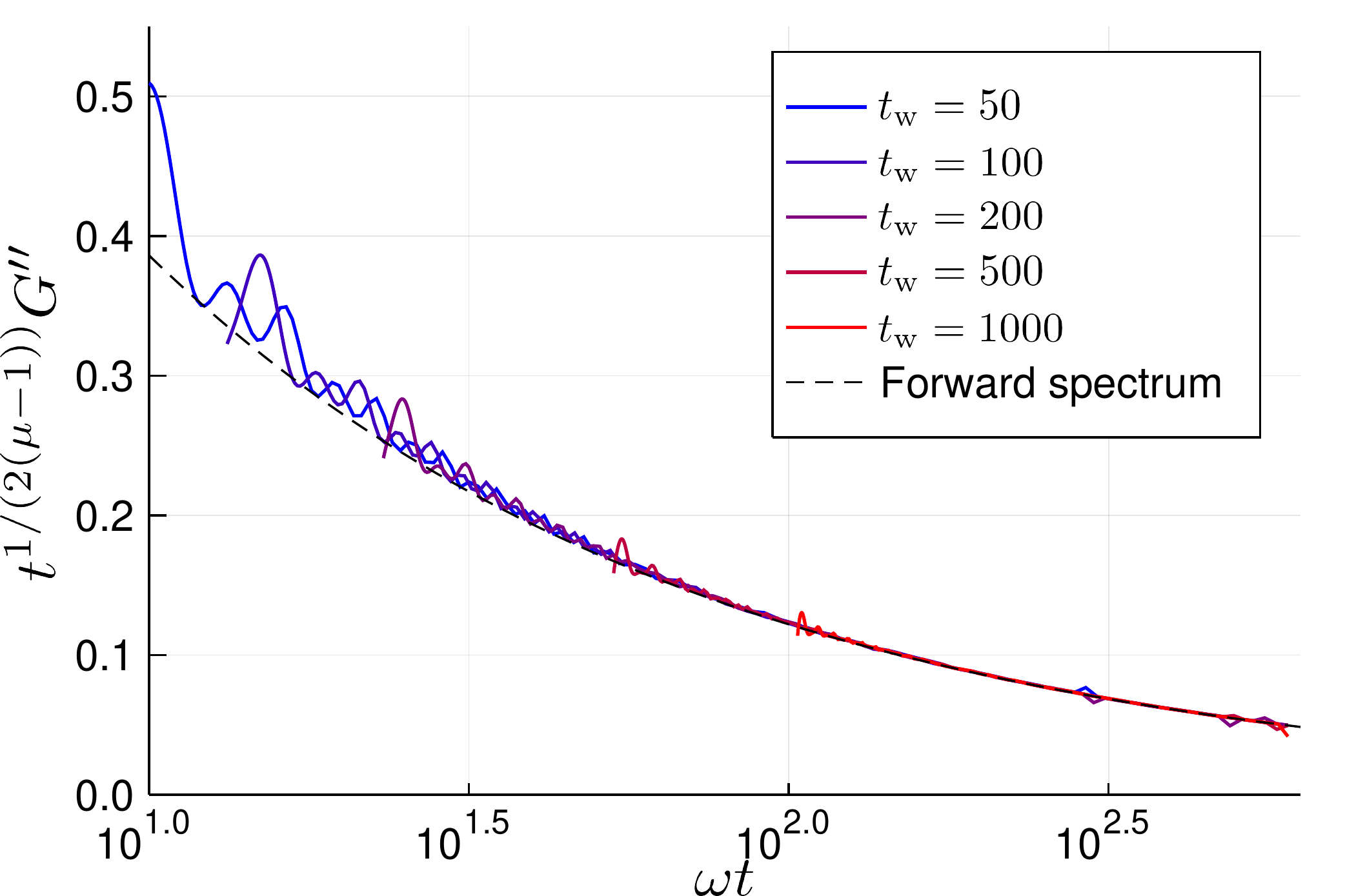}}
\caption{\label{fig:frequency_17}Loss modulus calculated using the averaged form (\ref{Gav_time}) from the stress relaxation at different waiting times for $\mu=1.7$, $A=0.15$. We see a good agreement with the forward spectrum (\ref{FS_17}) for large $t_{\rm w}$ and after enough oscillations.}
\end{figure}

As discussed in Sec. \ref{sec:background}, for the purpose of numerically computing the aging frequency response we use the averaged form $\bG(\omega,t,t_{\rm w})$ given in~(\ref{Gav_time}). Focussing on the same case $\mu=1.7$, $A=0.15$ we fix the frequency to $\omega=0.1$ and calculate the integrals in (\ref{Gav_time}) numerically, inserting directly the asymptotic form in the time domain (\ref{G_mu_12}) for a range of different waiting times (see Fig.~\ref{fig:frequency_17}). We choose $m=1$ (the results are very similar for $m=2$,$4$), implying that we are averaging over one period around each observation time $t$, which we choose in the range from $t_{\rm w}+\pi/\omega$ to $t_{\rm max}=6000$. In Fig.~\ref{fig:frequency_17} we show the resulting loss modulus, which indeed approaches the asymptotic behaviour~(\ref{FS_17}) for large enough $t_{\rm w}$ and after enough oscillations. 

\subsection{$\mu=1$}

\subsubsection{Stress relaxation function}

In the marginal case $\mu=1$, it was found \cite{parley_aging_2020} that for a system relaxing in the glassy phase the yield rate decays at long times as a stretched exponential $\Gamma(t)\sim e^{-B \sqrt{t}}$, with a constant $B$ that depends on the initial condition. Following again the scaling argument~(\ref{integral}) for the relaxation within the boundary layer, we have in this case that 
\begin{eqnarray}\label{G_1}
    &&\frac{1-G(t,\tw)}{1-G_{\infty}(t_{\rm w})} \equiv H(x,t_{\rm w})
     \nonumber \\{}&=&{\left(1-e^{-B(\sqrt{t_{\rm w}+x\sqrt{t_{\rm w}}}-\sqrt{t_{\rm w}})}\frac{1+B\sqrt{t_{\rm w}+x\sqrt{t_{\rm w}}}}{1+B\sqrt{t_{\rm w}}}\right)}^{1/2} \nonumber \\
    {}&\simeq& \sqrt{1-e^{-Bx/2}} \quad {\rm for} \quad t_{\rm w}\gg 1
\end{eqnarray}
where the rescaled time difference is now $x \equiv (t-t_{\rm w})/\sqrt{t_{\rm w}}$, and the value $1-G_{\infty}(t_{\rm w})$ at which the amount of stress relaxation saturates is
\begin{equation}\label{Ginf_mu_1}
    1-G_{\infty}(t_{\rm w})=ce^{-\frac{B}{2}\sqrt{t_{\rm w}}}{\left(B\sqrt{t_{\rm w}}+1\right)}^{1/2}
\end{equation}
with $c$ again an initial condition dependent constant.

\begin{figure}\center{\includegraphics[width=0.42\textwidth]{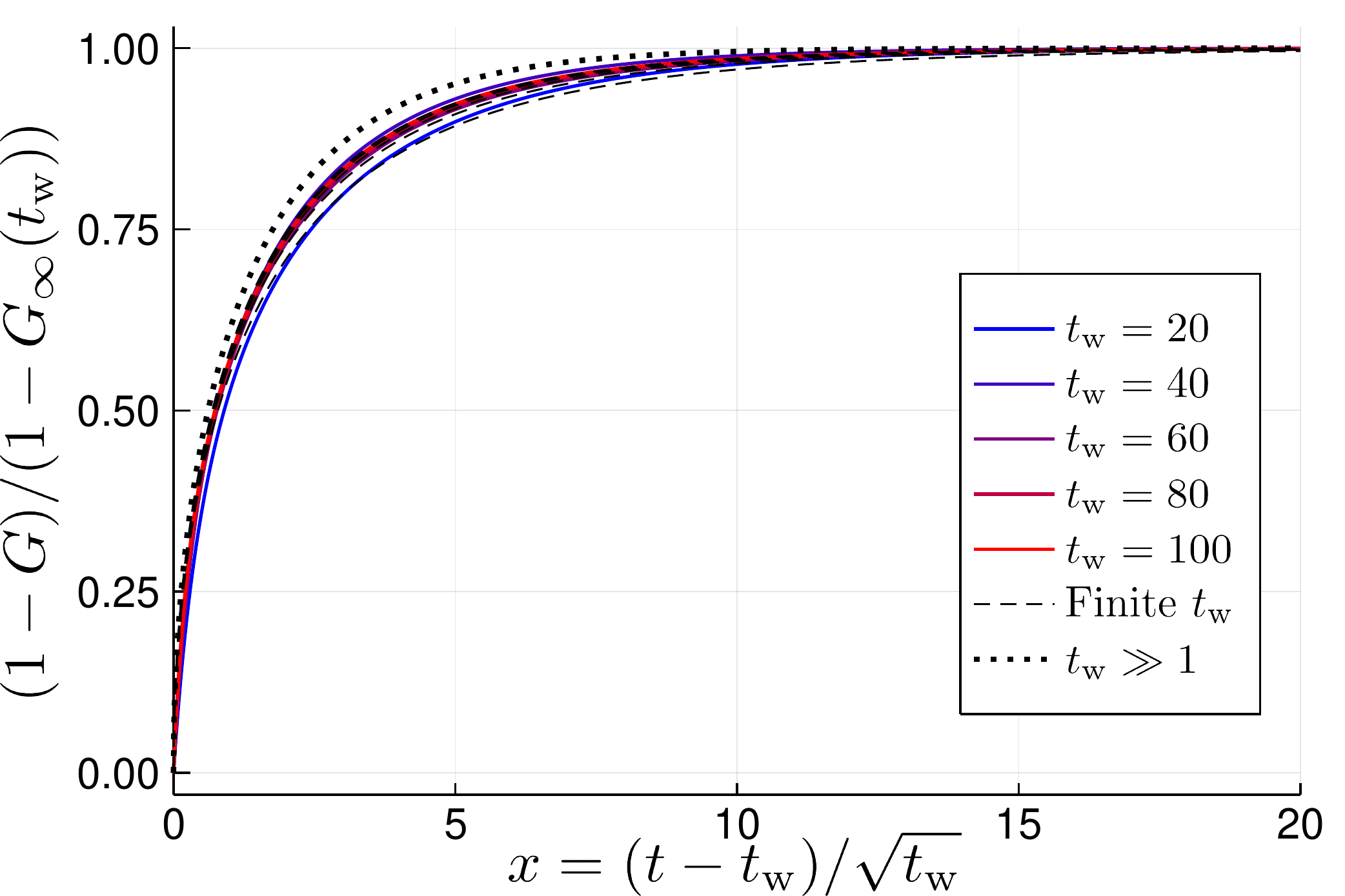}}
\caption{\label{fig:G_mu_1}Stress relaxation obtained from numerically solving the linearised equation (\ref{deltap}) for $\mu=1$ and $A=0.55$, starting from initial conditions extracted at different $t_{\rm w}$ from the unperturbed aging dynamics. With the appropriate rescalings, the curves are indistinguishable from the finite-$t_{\rm w}$ prediction $H(x,t_{\rm w})$ (\ref{G_1}), which approaches the asymptotic expression~(\ref{lin_1}) for $t_{\rm w}\rightarrow \infty$ (dotted line).}
\end{figure}

The case $\mu=1$, therefore, no longer follows simple aging, and we find instead a square root scaling $x=(t-t_{\rm w})/\sqrt{t_{\rm w}}$ of the relaxation times with age. This scaling, as well as the large $t_{\rm w}$ expression for $1-G$ in the last line of Eq.~(\ref{G_1}), may be found alternatively  by linearising the stretched exponential decay of $\Gamma(t)$ around $t_{\rm w}$ in the expression for the stress relaxation, i.e.\
\begin{eqnarray}\label{lin_1}
    1-G &\sim& {\left(\int_{t_{\rm w}}^{t} \mathrm{d} t' \ e^{-B\sqrt{t'}}\right)}^{\frac{1}{2}}\simeq {\left(\int_{t_{\rm w}}^{t} \mathrm{d}t' \ e^{-B\left(\sqrt{t_{\rm w}}+\frac{t'-t_{\rm w}}{2\sqrt{t_{\rm w}}}\right)}\right)}^{\frac{1}{2}} \nonumber\\
    {}&\simeq& (1-G_{\infty}(t_{\rm w}))\sqrt{1-e^{-Bx/2}}
\end{eqnarray}
In Fig.~\ref{fig:G_mu_1} we compare again with numerical results from Eq.~(\ref{deltap}), for the case $\mu=1$, $A=0.55$. The value of $B$ is fitted from the unperturbed dynamics, which in this case enters the stretched exponential regime already for $t\gtrsim 20$~\footnote{As done above for $\mu=1.7$, we extrapolate the asymptote of $\Gamma(t)$ to later times than we had access to in the unperturbed numerics, due to the same discretisation limit described there (the boundary layer becoming even harder to resolve for $\mu=1$).} so that there are no pre-asymptotic corrections from $\Gamma(t)$, and we study a range of waiting times from $t_{\rm w}=20$ to $100$. We find essentially perfect agreement with the finite-$t_{\rm w}$ form in (\ref{G_1}), which approaches the asymptotic expression for $t_{\rm w}\rightarrow \infty$~(\ref{lin_1}) as $t_{\rm w}$ increases. This approach can be shown from (\ref{G_1}) and (\ref{lin_1}) to be monotonic, with the leading order correction decaying as $\sim t_{\rm w}^{-1/2}$. 

\subsubsection{Frequency domain}

To investigate the aging frequency response, we proceed as in the case $1<\mu<2$. The aging moduli again approach the forward spectrum, which is now given by (see App.~\ref{app:FS}),
\begin{equation}\label{FS_1}
    \frac{G_{f}^{*}(\omega,t)}{1-G_{\infty}(t)}\sim 1-(1-i)c\sqrt{\frac{B}{2}}\sqrt{\frac{\pi}{8}}w^{-\frac{1}{2}}
\end{equation}
with a rescaled frequency $w=\omega t^{1/2}$.

Again, as for $\mu>1$, the aging frequency-dependent moduli directly reflect the behaviour~(\ref{G_1}) in the time domain. It is important to note that although (\ref{FS_1}) and (\ref{FS_17}) look similar, the rescaled frequency $w$ is different in the two cases. The shared $w^{-1/2}$ behaviour is a genuine commonality, on the other hand, stemming as it does from the universality discussed in Sec.~\ref{sec:overview}.

\begin{figure}\center{\includegraphics[width=0.42\textwidth]{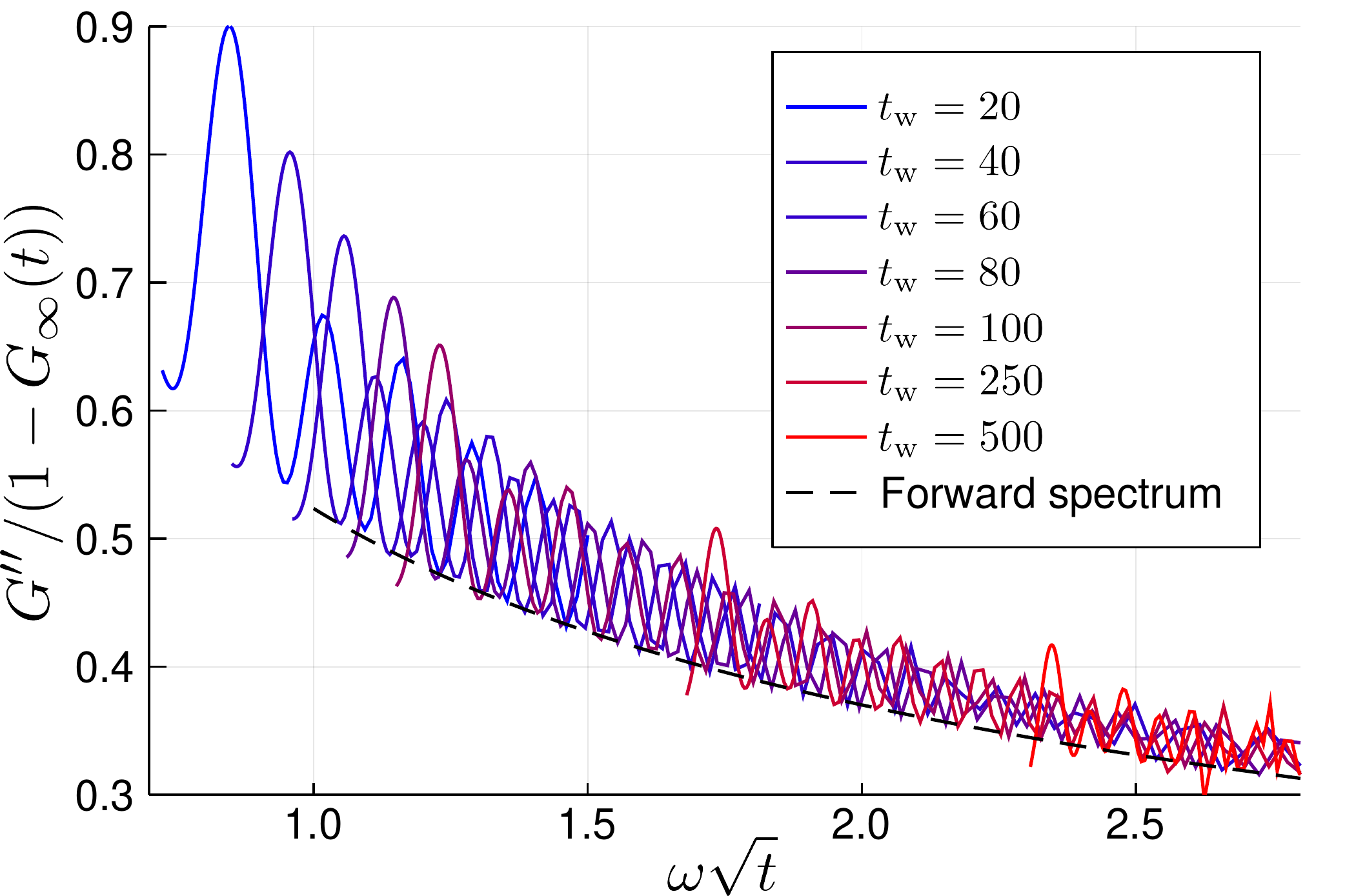}}
\caption{\label{fig:frequency_mu_1}Loss modulus calculated using the averaged form (\ref{Gav_time}) from the stress relaxation at different waiting times for $\mu=1$, $A=0.55$. The forward spectrum (\ref{FS_1}) is approached for large $t_{\rm w}$ and after enough oscillations.}
\end{figure}

Finally, we numerically compute the aging frequency response, using again the averaged form~(\ref{Gav_time}), for the case $\mu=1$, $A=0.55$ considered above. We choose $m=1$, so that we average over one period around the observation time. In contrast to the case $1<\mu<2$, where results were independent of $m$ (for $m=2$, $4$), here the averaging is sensitive to $m$ due to the rapidly decaying magnitude $1-G_{\infty}(t)$, which leads to a bias in the results for larger $m$. For $m=1$, we see in Fig.~\ref{fig:frequency_mu_1} that the loss modulus does indeed approach the asymptotic form~(\ref{FS_1}) after enough oscillations.

\section{(Weakly) Non-linear behaviour ($\mu=1$)}\label{sec:NL}

We next study numerically the non-linear response to step strain of the model. This will allow us to check that the linear theory developed so far does indeed hold for $\gamma_0 \ll 1$, and will also shed light on the extent of this linear regime. Furthermore, the predictions we will obtain for the non-linear effects will in themselves be interesting for the comparison to the MD data in Sec.~\ref{sec:MD}.  

The non-linear, strain and age-dependent response function to a step strain $\gamma_0\, \theta(t-t_{\rm w})$ is written as
\begin{equation}
    \sigma(t)=\gamma_0 G(t,t_{\rm w};\gamma_0)
\end{equation}
which defines the nonlinear stress relaxation function $G(t,t_{\rm w};\gamma_0)$.
In order for our discussion to be relevant also to the MD simulations presented in Sec.~\ref{sec:MD} we focus on $\mu=1$, with a slightly higher value of the coupling ($A=0.58$) than the one shown in Fig.~\ref{fig:G_mu_1}. This provides us with a wider time range (up to around $t=400$) in which to study aging properties before the yield rate becomes too small to resolve numerically. 

We now consider a range of waiting times within this asymptotic regime, and study the non-linear response to a range of step strains. To do this, we now evolve the \textit{full} master equation (\ref{ME}) after application of a step strain. In our discrete numerical setup, this amounts to shifting the initial distribution $P_0(\sigma,t_{\rm w})$ by a number of grid points $\gamma_0/\Delta \sigma$, 
where $\Delta \sigma$ is the stress discretisation.
The smallest step amplitude we can reliably explore is then some small multiple of $\Delta\sigma$, in our case  $\gamma_0=5\times 10^{-4}$ (corresponding to $4\,\Delta\sigma$). Importantly, in the ensuing dynamics $\Gamma(t)$ is \textit{perturbed} by the strain, in contrast to the linear theory where $\Gamma(t)=\Gamma_0(t)$. 

\begin{figure}\center{\includegraphics[width=0.42\textwidth]{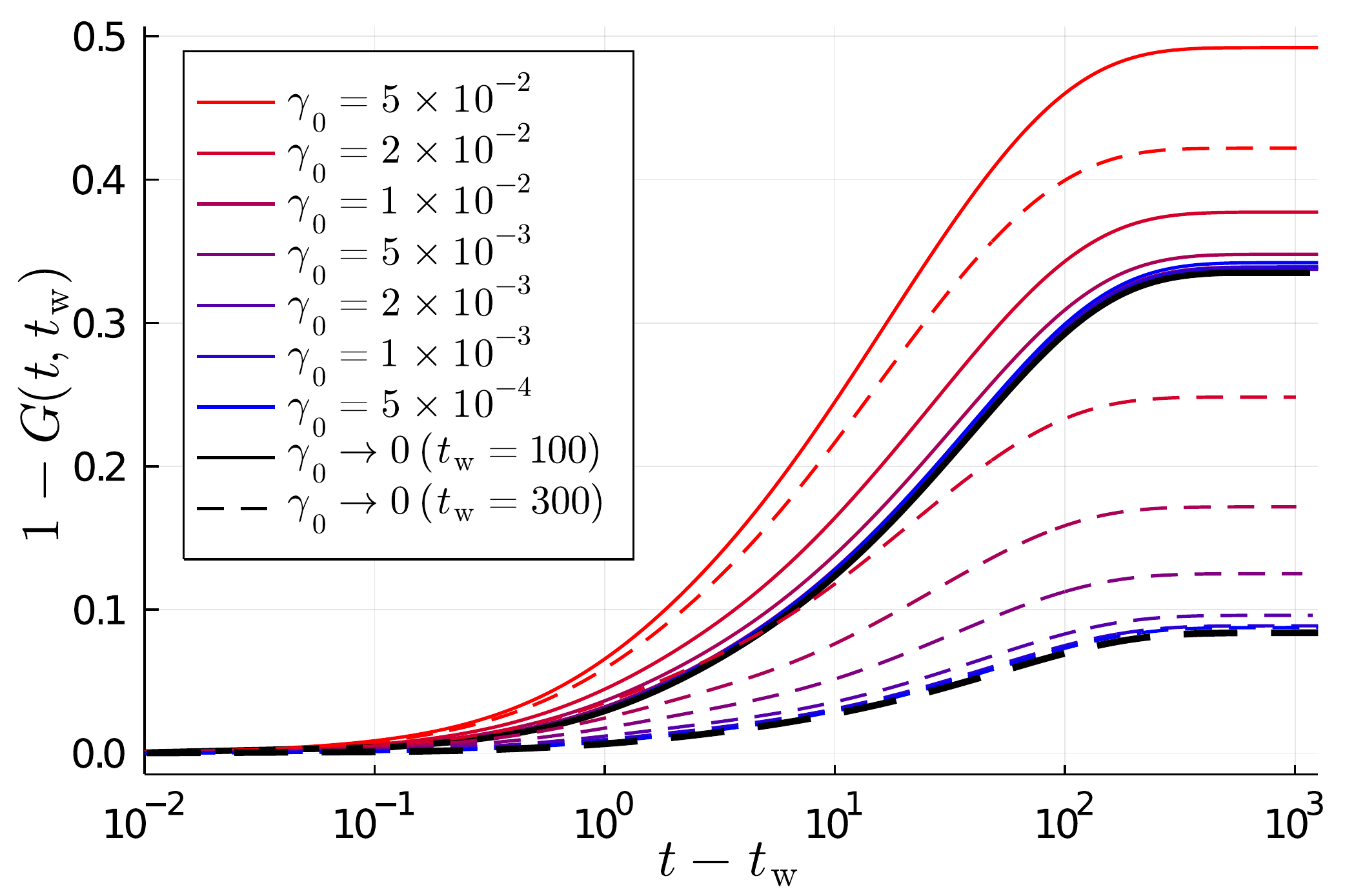}}
\caption{\label{fig:tw_100_300}Stress relaxation following a step strain applied at $t_{\rm w}=100$ (full lines) and $t_{\rm w}=300$ (dashed lines), obtained from evolving the full master equation (\ref{ME}), for step strains $\gamma_0$ ranging from $5 \times 10^{-2}$ (red) down to $5 \times 10^{-4}$ (blue). This is compared with the relaxation obtained from the linearised equation~(\ref{deltap}). For $t_{\rm w}=100$ one finds agreement for $\gamma_0\lessapprox 10^{-2}$; for $t_{\rm w}=300$, on the other hand, agreement with the linear theory holds only for $\gamma_0\lessapprox 2\times 10^{-3}$. 
}
\end{figure}



In Fig.~\ref{fig:tw_100_300} we show the non-linear response function for a range of strains $\gamma_0 \in (5\times 10^{-4},5 \times 10^{-2})$, for two waiting times $t_{\rm w}=100$ and $t_{\rm w}=300$. On the same plot, we display also the prediction from the linear theory for each $t_{\rm w}$, evaluated by solving Eq.~(\ref{deltap}) using as input the unperturbed $\Gamma_0(t)$. One notices first that for both waiting times, the smallest step strain amplitudes do indeed give a response function that matches the prediction of the linear theory. However, we see clearly that for the later waiting time more of the strain step values deviate from linear response. In other words, the extent of the linear regime shrinks considerably at later waiting times. To study this more in detail, we take
the measured asymptotic relaxations for each $t_{\rm w}$ and interpolate them to obtain $1-G_\infty(t_{\rm w};\gamma_0)$ as a function of $\gamma_0$ (see Fig.~\ref{fig:Ginf_spline} in App.~\ref{app:NL}). From here we identify the linear regime as extending up to $\gamma_{\rm max}(t_{\rm w})$, which we define by setting a threshold (10 $\%$) on the relative deviation of the amount of stress relaxation with respect to the linear value; fixing a threshold for the relative deviations of the plateaus $G_{\infty}(t_{\rm w},\gamma_0)$ themselves leads to similar results. A naive expectation for the scaling of $\gamma_{\rm max}(t_{\rm w})$ would be to consider the initial perturbation to the yield rate caused by the step strain, which (see below)
is of order $\sim \gamma_0^{1+\mu/2}=\gamma_0^{3/2}$. For the linear regime one then expects the condition $\gamma_0^{3/2}\ll \Gamma(t_{\rm w})$ and hence the scaling $\gamma_{\rm max}(t_{\rm w})\sim {\left(\Gamma(t_{\rm w})\right)}^{2/3}$. In Fig.~\ref{fig:gmax} in App.~\ref{app:NL} we show that the measured $\gamma_{\rm max}(t_{\rm w})$ agrees well with this prediction.

We now proceed to study the non-linear effects on the total amount of relaxation at long times, and on the temporal evolution of the rescaled relaxation function, which we recall saturates at this final value. In Sec.~\ref{sec:aging} we derived analytical expressions for the linear response limit of both of these quantities, given in~(\ref{Ginf_mu_1}) and~(\ref{G_1}), respectively. 

\begin{figure}\center{\includegraphics[width=0.42\textwidth]{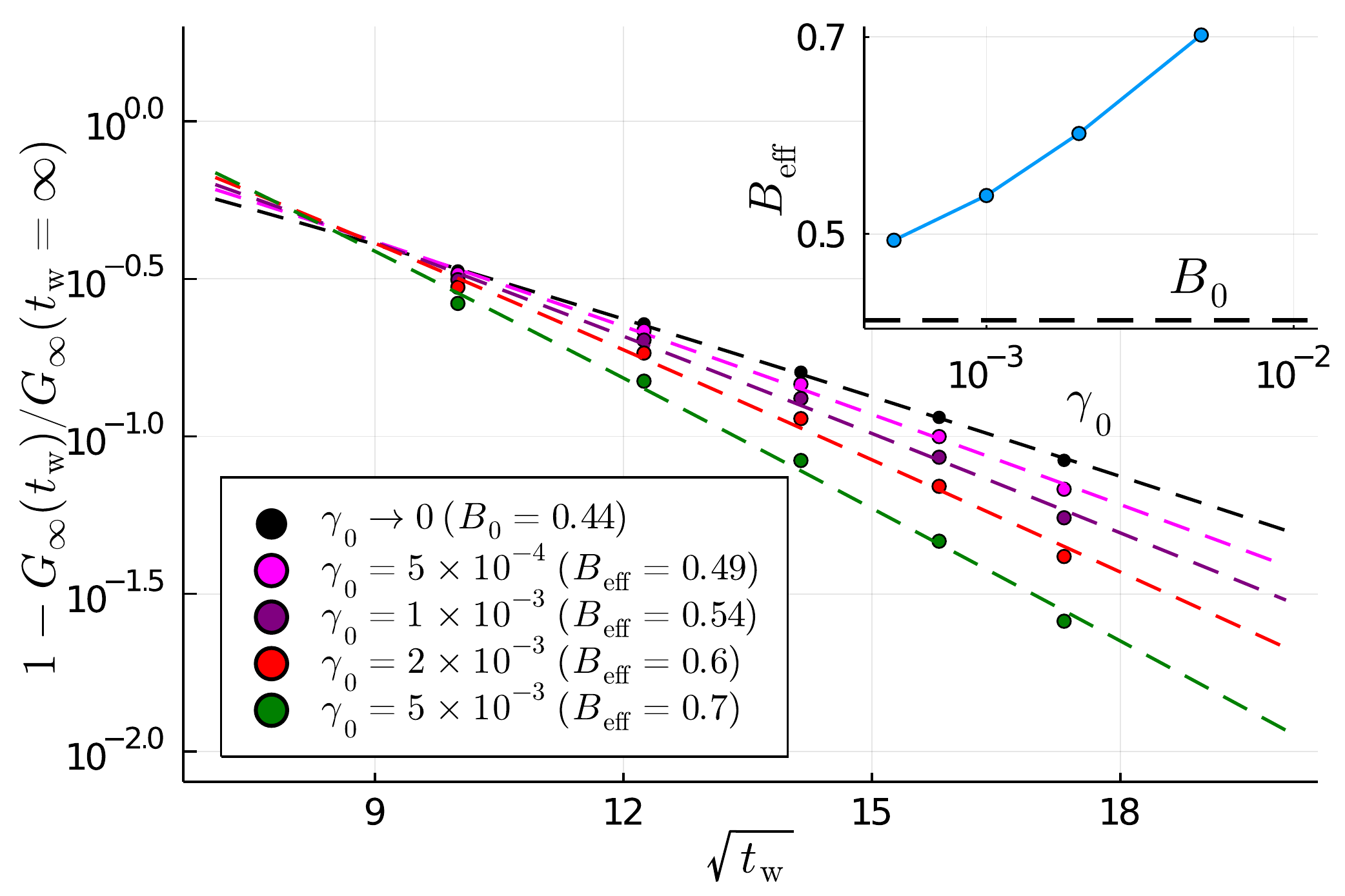}}
\caption{\label{fig:Ginfs_MF}Plateau modulus values $G_{\infty}(t_{\rm w};\gamma_0)$ extracted from the non-linear step strain numerics at different waiting times, rescaled by the $t_{\rm w} \rightarrow \infty$ modulus for each $\gamma_0$. Dashed lines show the analytical expression~(\ref{Ginf_mu_1}), with a fitted value $B_{\rm eff}(\gamma_0)$ that grows for larger step strain. Inset shows fitted values of $B_{\rm eff}$ versus $\gamma_0$.}
\end{figure}

Starting with the plateau at which the relaxation saturates, we first point out a qualitative difference in the non-linear case. For finite $\gamma_0$, there is now a non-zero stress relaxation even for $t_{\rm w}\rightarrow\infty$, where the stress distribution is frozen and all blocks are stable, so that $G_{\infty}(t_{\rm w} \rightarrow \infty;\gamma_0)<1$. To account for this, in Fig.~\ref{fig:Ginfs_MF} we rescale the plateau values by the $t_{\rm w} \rightarrow \infty$ plateau, so that we plot $1-G_{\infty}(t_{\rm w};\gamma_0)/G_{\infty}(t_{\rm w} \rightarrow \infty;\gamma_0)$, which by construction does decay to zero with increasing $\tw$ for all $\gamma_0$. As expected, the values from the linear regime agree well with the prediction~(\ref{Ginf_mu_1}), using for $B$ the value $B_0$ extracted from the unperturbed numerics $\Gamma_0\sim e^{-B_0\sqrt{t}}$. Surprisingly, we see that also the data for non-linear $\gamma_0$ (shown are four values up to $\gamma_0=5 \times 10^{-3}$) are well described by the expression~(\ref{Ginf_mu_1}), but with a higher ``effective" value of $B$ that we denote $B_{\rm eff}$. $B_{\rm eff}$ increases with $\gamma_0$, implying that the final plateau value for $t_{\rm w}\rightarrow\infty$ is approached already at shorter waiting times for larger step strains. 

\begin{table*}[!ht]
    \begin{tabularx}{\linewidth}{ssbb}
   & Scaling \newline variable & $1-G_{\infty}(t_{\rm w})$ & Stress response  
   \\ \hline
     Aging \newline $1<\mu\leq2$ & $x=\frac{t-t_{\rm w}}{t_{\rm w}}$ & $ct_{\rm w}^{-\frac{1}{2(\mu-1)}}$ & $\frac{1-G(x)}{1-G_{\infty}(t_{\rm w})}=\sqrt{1-(1+x)^{-\frac{1}{\mu-1}}}$ \\
     \hline
     Aging \newline $\mu=1$ & $x=\frac{t-t_{\rm w}}{\sqrt{t_{\rm w}}}$ & $c e^{-B\sqrt{t_{\rm w}}/2}$ $\sqrt{B\sqrt{t_{\rm w}}+1}$ & $\frac{1-G(x)}{1-G_{\infty}(t_{\rm w})}\sim \sqrt{1-e^{-Bx/2}}$
          \\ \hline
     Fluid \newline Near AT, $\forall \mu$  & $\Delta t =t-t_{\rm w}$ & $1$ & Short time: $1-G(\Delta t) \sim {\Delta t}^{1/2}$ \newline Long time: $G(\Delta t)\sim e^{-\Delta t/\tau}$ \\ \hline
     Critical aging $\forall \mu$ & $x=\frac{t-t_{\rm w}}{t_{\rm w}}$ & $1$ & Short time: $1-G(x)\sim \sqrt{\ln(1+x)}$ \newline Long time: $G(x)\sim x^{-1/\mu}$ \\ 
\hline
\end{tabularx}
\caption{\label{T1}Summary of analytical results in the time domain. The total amount of relaxation (second column) is defined by (\ref{1_G}); $c$ is an initial condition-dependent constant. The fluid state approaching the arrest transition (AT) (for $A\gtrsim A_c$), and the critical aging case (for $A=A_c$), are treated in Appendices \ref{app:SS} and \ref{app:criticality} respectively, where the short and long time regimes are properly defined. Note that the stress response at short times follows in all cases $1-G\sim x^{1/2}$ in the corresponding scaling variable, reflecting the universal $1/2$ exponent discussed in Sec.~\ref{sec:overview}. 
}
    \end{table*}
    
\begin{table*}
    \begin{tabularx}{\linewidth}{ssb}
   & Scaling \newline variable & Loss modulus  \\
         \hline
     Aging \newline $1<\mu \leq 2$ & $w=\omega t$ & $\frac{G''(w)}{1-G_{\infty}(t)}\sim \left(\frac{1}{\mu-1}\right)^{1/2} \sqrt{\frac{\pi}{8}} w^{-1/2}$ \\ \hline
     Aging \newline $\mu=1$ & $w=\omega \sqrt{t}$ & $\frac{G''(w)}{1-G_{\infty}(t)}\sim \left(\frac{B}{2}\right)^{1/2} \sqrt{\frac{\pi}{8}} w^{-1/2}$ \\
\hline
\end{tabularx}
\caption{\label{T2}Analytical results for the aging frequency response. The asymptotic expressions hold for $\omega (t-t_{\rm w})\gg 1$, $\omega t \gg 1$ (with $\omega \ll 1$), as detailed in the text. Although the scaling variables are different, we note the common exponent $-1/2$, which is simply a consequence of the universal behaviour in the short time regime.}
    \end{table*}

The final plateau value and the corresponding stress relaxation $1-G_{\infty}(t_{\rm w} \rightarrow\infty;\gamma_0)$ are purely non-linear features, because in the linear theory $\Gamma_0(t_{\rm w}\rightarrow\infty)=0$ and no more relaxation takes place. We can construct a lower bound on $1-G_{\infty}(t_{\rm w} \rightarrow\infty;\gamma_0)$ in the following way. Neglecting the effect of stress redistribution, which can trigger additional yield events, we can consider the proportion of blocks that are made unstable by the initial step strain $\gamma_0$. These lie in the stress interval $\sigma \in (1-\gamma_0,1)$. The distribution $P_0(\sigma,t_{\rm w})$ behaves as $P_0\sim q_0 (1-\sigma)^{\mu/2}$ for $\sigma \lesssim 1$, 
giving to leading order in $\gamma_0$ a stress relaxation
\begin{equation}\label{l_bound}
    1-G_{\infty}(t_{\rm w} \rightarrow\infty;\gamma_0)\gtrsim q_0 \,\frac{1}{1+\mu/2}\, \gamma_0^{\mu/2}
\end{equation}
The same argument also shows that the perturbation to the yield rate is $\sim \gamma_0^{1+\mu/2}$, as given above. Our data do indeed lie above this lower bound, and approach it as $\gamma_0 \rightarrow0$ (see Fig.~\ref{fig:bound_modulus} in App.~\ref{app:NL}).

\begin{figure}\center{\includegraphics[width=0.42\textwidth]{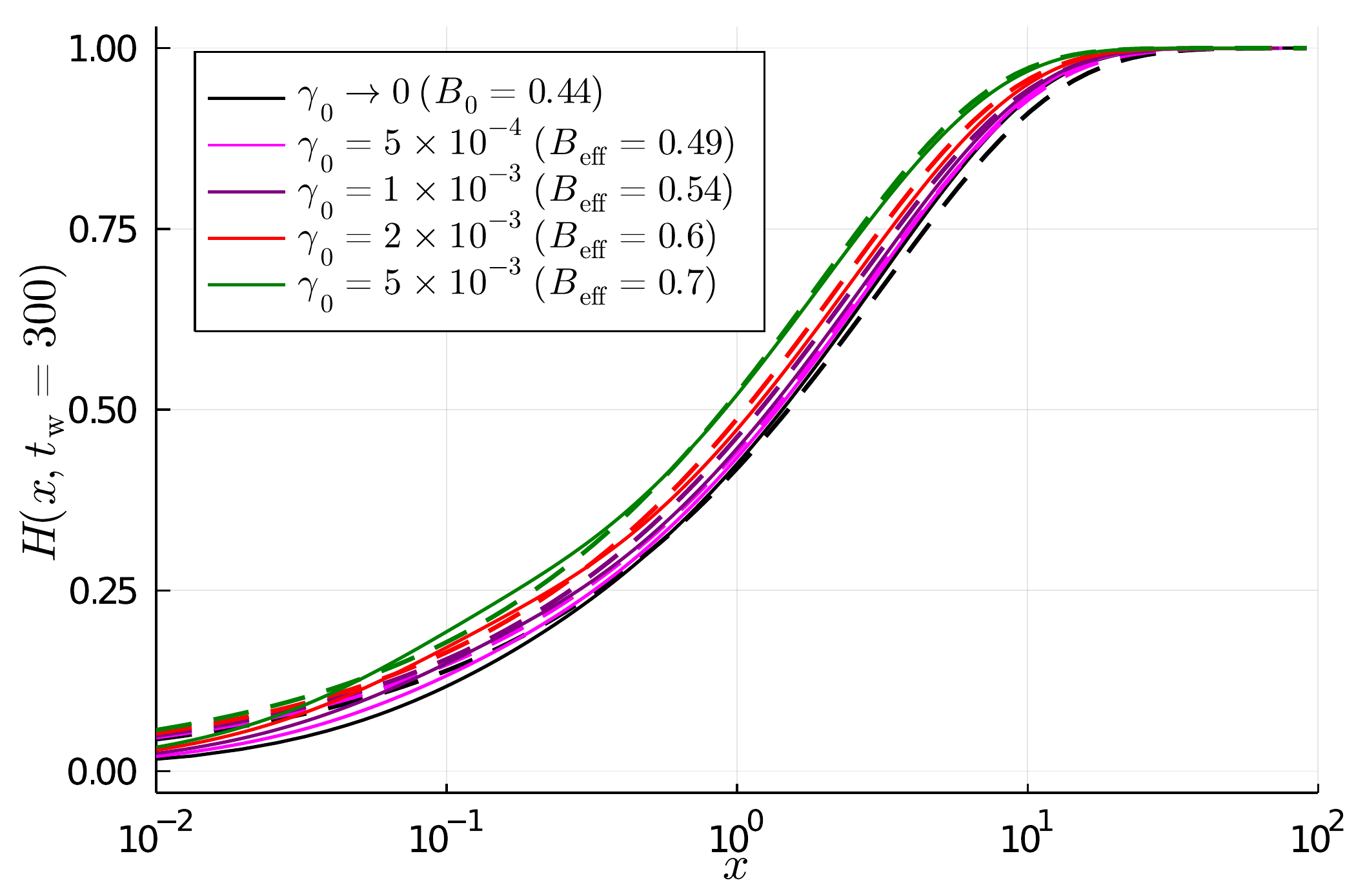}}
\caption{\label{fig:H(x)}Stress relaxation at $t_{\rm w}=300$ for different $\gamma_0$, rescaled in each case by the final amount of relaxation. The time axis is rescaled to $x=(t-t_{\rm w})/\sqrt{t_{\rm w}}$. Dashed lines show the finite-$t_{\rm w}$ expression for the stress relaxation (\ref{G_1}), using the values $B_{\rm eff}(\gamma_0)$ extracted from Fig.~\ref{fig:Ginfs_MF}.}
\end{figure}

Finally, we turn to the temporal evolution of the relaxation function. We show this in Fig.~\ref{fig:H(x)} for $t_{\rm w}=300$ and the same four values of $\gamma_0$ as above, along with the linear response. In each case we rescale the stress by the final plateau value, and the time as $x=(t-t_{\rm w})/\sqrt{t_{\rm w}}$. In this representation, the linear response indeed follows the expression (\ref{G_1}) for $H(x,t_{\rm w})$ derived in Sec.~\ref{sec:aging}, with the same value of $B_0$. Interestingly, even the non-linear relaxations can be fitted very well by the same expression (\ref{G_1}), but (as for the plateau decay) with a higher $B_{\rm eff}$ value, which again increases for larger $\gamma_0$ so that larger step strains accelerate the dynamics. We note that, unlike in the linear theory, in the non-linear case the $B_{\rm eff}(\gamma_0)$ values inferred from the plateau decays do not necessarily have to describe also the full dynamics. For later waiting times (as is the case shown in Fig.~\ref{fig:H(x)}), however, we find that the same $B_{\rm eff}(\gamma_0)$ values fitted from the plateau decays do in fact provide a good fit for the full time evolution at each step strain $\gamma_0$.


Summarising, we have found that the extent of the linear regime shrinks considerably at later waiting times. However, we have also found that even in the (weakly) non-linear case, both the plateau decay and the stress dynamics are still well described by the linear theory through (\ref{Ginf_mu_1}) and (\ref{G_1}), but with effective constants
$B_{\rm eff}(\gamma_0)>B_0$. We therefore see that the application of non-linear step strains effectively leads to faster dynamics. The same effect will be observed in the MD simulations discussed in the following section.

\section{Comparison with MD simulations}\label{sec:MD}
We now compare our mean field prediction to molecular dynamics simulations of a model athermal solid. For this we consider a bidisperse assembly of soft harmonic spheres at high packing fraction $\phi=1$ (well above jamming), immersed in an effective solvent. This model has been used widely in the literature \cite{durian_bubble-scale_1997}, and is considered an appropriate description of, for example, dense emulsions, foams or microgels suspensions comprising droplets, bubbles or particles of typical radius $R\gtrsim 1\mu m$, in the athermal regime \cite{chacko_slow_2019}. Neglecting inertia and explicit hydrodynamic interactions, the \textit{unperturbed} dynamics of the system, starting from an initial condition with significant overlap between the spheres, is simply a gradient descent in the energy landscape. This dissipative dynamics was studied in \cite{chacko_slow_2019} (see also \cite{nishikawa_relaxation_2022}), where it was shown to present a slow (power--law) decay of the energy and velocity, which was referred to as \textit{athermal aging}. Here we study the linear response of the system to a step strain  at different waiting times $t_{\rm w}$ during this aging process; further simulation details may be found in App.~\ref{app:MD}.    

An important difference in the particle system is that even mechanically stable (frozen) system configurations, which are reached for $t_{\rm w}\rightarrow\infty$ (in our simulations, this limit is reached at $t_{\rm w} \approx5\times10^{5}$) show a finite stress relaxation; see Fig.~\ref{fig:raw} in App.~\ref{app:MD}. In fact, for any $t_{\rm w}$ there is always a
non-affine relaxation, simply due to the particles recovering a state of force balance after the application of the step strain. This \textit{reversible} non-affine motion can be expressed analytically in terms of the Hessian of the current energy minimum following~\cite{maloney_amorphous_2006}, given that at small strain it does not involve any plastic yielding. However, for this same reason it is not accounted for within our elastoplastic description (see more in the discussion). To be able to compare with our theory, we therefore need to factor out this non-affine relaxation and focus only on the relaxation due to plastic events.

\begin{figure}\center{\includegraphics[width=0.42\textwidth]{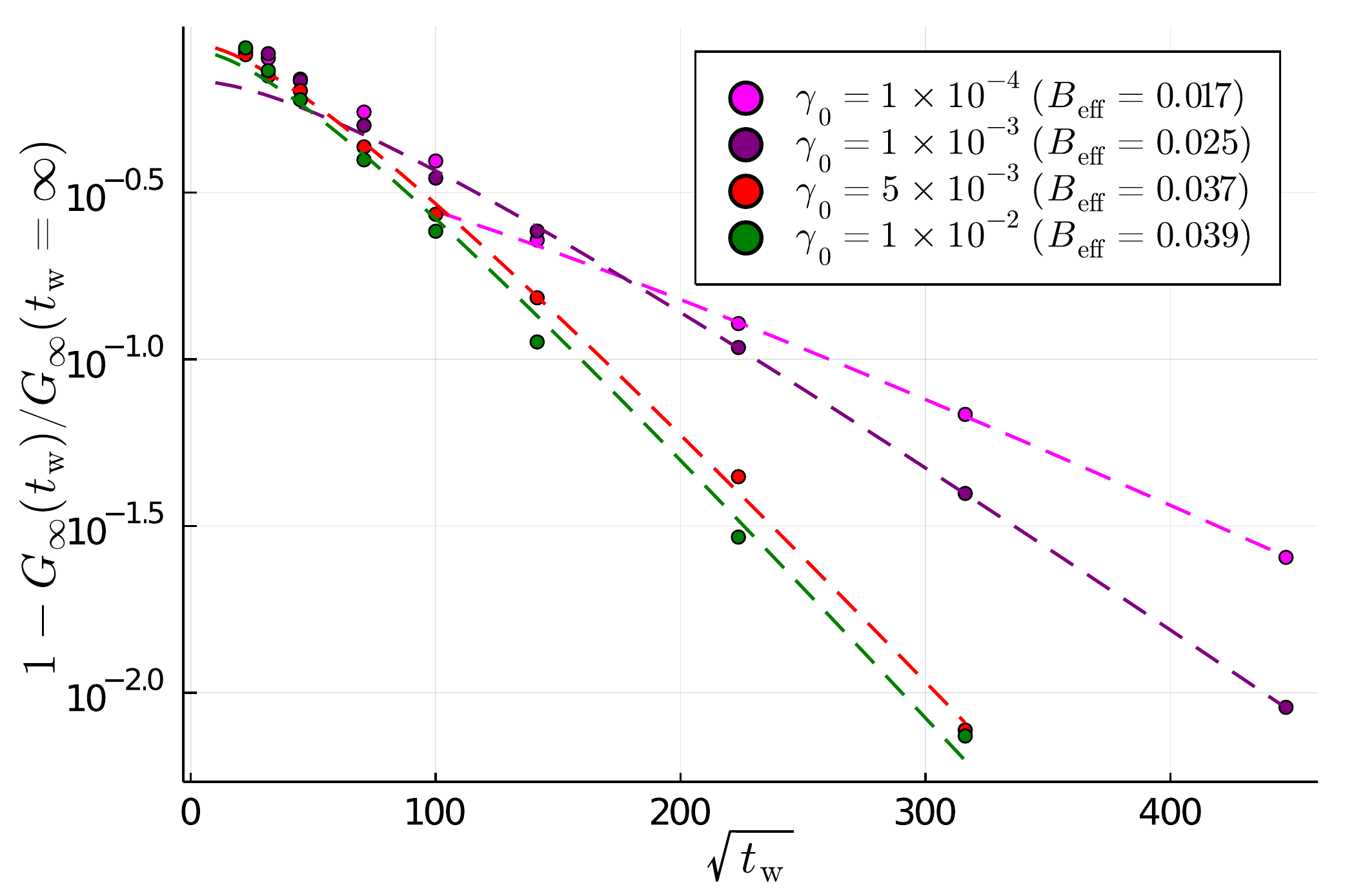}}
\caption{\label{fig:Ginfs_Ritu}Plateau modulus values $G_{\infty}(t_{\rm w};\gamma_0)$ extracted from the MD simulation, rescaled by the $t_{\rm w}\rightarrow \infty$ modulus for each $\gamma_0$. Dashed lines show the analytical expression (\ref{Ginf_mu_1}), with a fitted value $B_{\rm eff}(\gamma_0)$ that grows with increasing step strain.}
\end{figure}

In the case of the plateau values, which we consider first, this is taken care of automatically by proceeding as in the evaluation of the theory (Fig.~\ref{fig:Ginfs_MF}): we rescale by the $t_{\rm w}=\infty$ relaxation, considering again $1-G_{\infty}(t_{\rm w},\gamma_0)/G_{\infty}(t_{\rm w}=\infty,\gamma_0)$ for various values of the step strain (see Fig.~\ref{fig:Ginfs_Ritu}). For all $\gamma_0$ we fit the analytical form (\ref{Ginf_mu_1}), extracting an effective value of $B$ in each case. We see that, on the one hand, the data agree well with the (modulated) stretched exponential form~(\ref{Ginf_mu_1}) in all cases; on the other hand, we find the same trend as in mean field, with the effective $B$ increasing with the strain $\gamma_0$.

We next turn to the full temporal dynamics of the stress relaxation function. Here, we need firstly to account for the $t_{\rm w}=\infty$ relaxation, which we assume is purely due to the non-affine part. We denote this as $\Gna (\Delta t)$, formally defined as $\lim_{\tw\to\infty} G(\tw+\Delta t,\tw)$. We then consider the ratio between the full stress relaxation function and the non-affine relaxation purely due to the recovery of force balance:
\begin{eqnarray}\label{Gpl}
    G^{\rm pl}(t,t_{\rm w})&\equiv& \frac{G(t,t_{\rm w})}{\Gna(t-t_{\rm w})}, \\  \mathrm{with} \quad \Gna (\Delta t)&\equiv& \lim_{\tw\to\infty} G(\tw+\Delta t,\tw) \nonumber
\end{eqnarray}
so that for an infinitely aged system, $G^{\rm pl}=1$ and the response is purely elastic as in our mean field model (for small applied strain).

\begin{figure}\center{\includegraphics[width=0.42\textwidth]{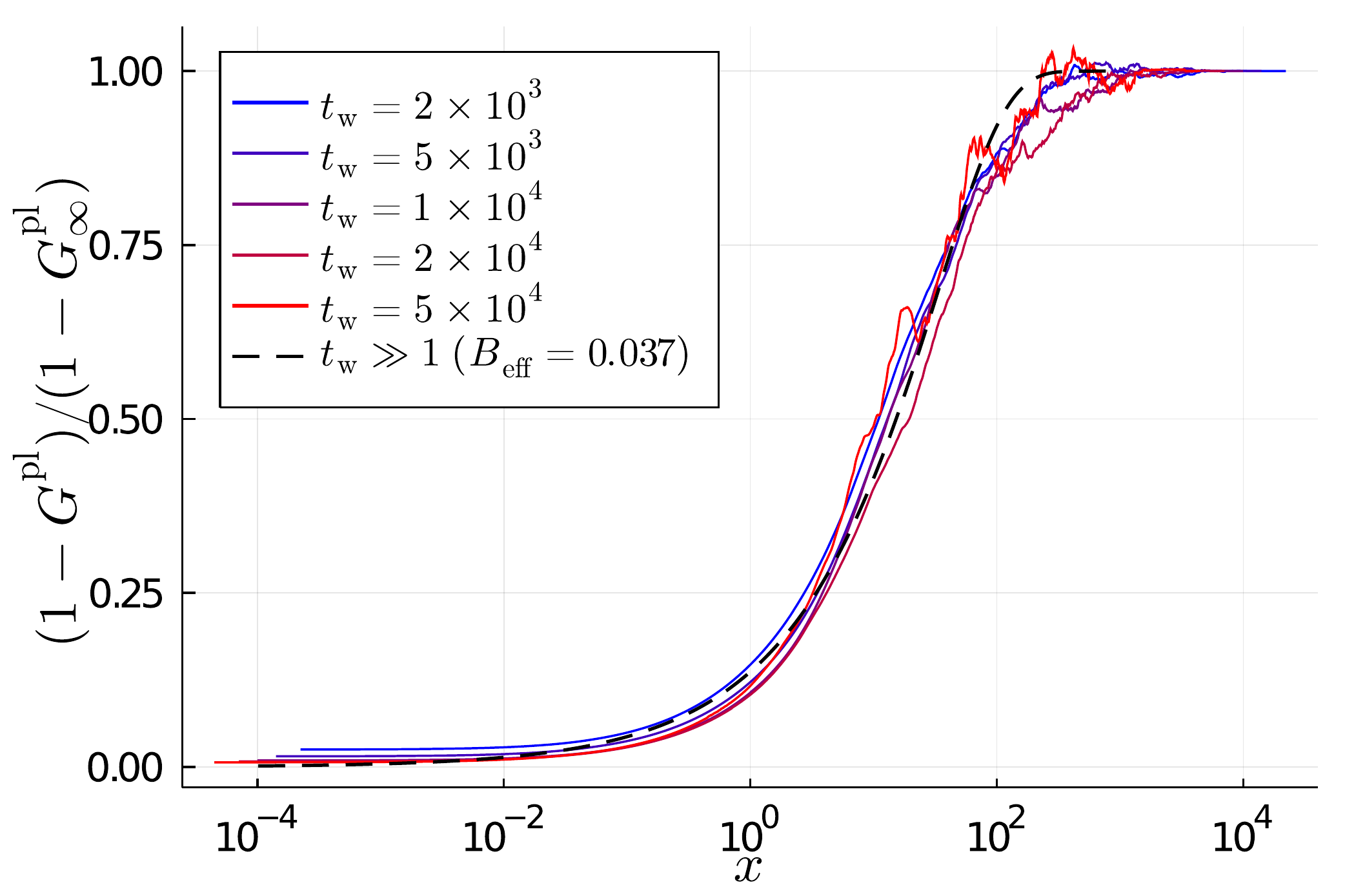}}
\caption{\label{fig:collapse}Plastic stress relaxation in the MD simulation with $\gamma_0=5 \times 10^{-3}$ for different waiting times, obtained by applying (\ref{Gpl}). The time axis is rescaled to $x=(t-t_{\rm w})/\sqrt{t_{\rm w}}$. Dashed line shows the $t_{\rm w}\gg 1$ expression for the stress relaxation (\ref{G_1}), evaluated using $B=B_{\rm eff}(\gamma_0)$ extracted from the plateau values (Fig.~\ref{fig:Ginfs_Ritu}).}
\end{figure}

We show the result for $\gamma_0=5 \times 10^{-3}$ in Fig.~\ref{fig:collapse}. For the plot we rescale $1-G^{\rm pl}(t,t_{\rm w})$ by the asymptotic plastic relaxation $1-G_{\infty}^{\rm pl}(t_{\rm w})$ corresponding to each $t_{\rm w}$, in order to compare with the rescaled form (\ref{G_1}) of the theoretical prediction, which we recall varies from $0$ to $1$. We find a very good collapse of the curves by rescaling the time axis as $(t-t_{\rm w})/\sqrt{t_{\rm w}}$. More importantly, the asymptotic form of (\ref{G_1}) for large $t_{\rm w}$ fits excellently the data, \textit{with} the corresponding value of $B$ fitted from the plateau decay (see Fig.~\ref{fig:Ginfs_Ritu}).

Overall, figures \ref{fig:Ginfs_Ritu} and \ref{fig:collapse} point to a good agreement with the theory for $\mu=1$. We show here only the temporal data for $\gamma_0=5 \times 10^{-3}$, obtained by averaging over $N_{\rm rep}=128$ repetitions. For the smaller step strains, at large waiting times, even with $N_{\rm rep}=1280$ the numerical signal is not clear enough to study the full stress relaxation up to our largest $t_{\rm w}$. For $\gamma_0=10^{-3}$ we nonetheless find a similar collapse to Fig.~\ref{fig:collapse}, with the corresponding value of $B_{\rm eff}$ from Fig.~\ref{fig:Ginfs_Ritu}, at least up to $t_{\rm w}=2\times 10^{4}$. This supports the expectation that the results in Fig.~\ref{fig:collapse} should also be representative of the behaviour for smaller step strain values, the only difference being the slightly slower dynamics (smaller $B$).



\section{Discussion and outlook}\label{sec:discussion}
In this paper we have studied the aging linear shear response within the framework of a mean field elastoplastic model of amorphous solids, introduced previously in~\cite{parley_aging_2020}. The main feature of this model was the incorporation of mechanical noise due to stress propagation, which was argued to be power-law distributed with exponent $\mu$. Here, we have found analytically the long-time form of the aging step response $G(t,t_{\rm w})$ for the different values of $\mu$, along with the aging frequency response; these are summarised in Tables \ref{T1} and \ref{T2}. The theoretical predictions for $\mu=1$, which is the exponent describing the physical elastic propagator, were then compared against data from MD simulations of a model athermal system in its aging regime, finding good correspondence with the theory. In the following discussion, we first discuss separately the theoretical results in the context of other aging phenomena, before commenting further on the comparison to the MD simulation and to possible experiments.

From a purely theoretical perspective, it is interesting to compare the athermal aging response found here with ``classical'' aging phenomena, studied particularly in spin glasses~\cite{cugliandolo_evidence_1994}. As in~\cite{fielding_ageing_2000}, we refer to the step strain response in our model as \textit{aging} due to the fact that the stress relaxation takes place on timescales that grow with the age $t_{\rm w}$ of the system. An important difference, however, is that our results cannot be fitted to the general form advocated by Cugliandolo and Kurchan~\cite{cugliandolo_out--equilibrium_1994}, where $G(t,t_{\rm w})=G[h(t)/h(t_{\rm w})]$, $h(t)$ being an effective clock. This is due to several key assumptions in~\cite{cugliandolo_out--equilibrium_1994} that are violated here. For starters, our model \textit{does not} have weak long-term memory, nor is the response function related to any correlation function. Weak long-term memory refers to the property that if a perturbation (in this case, a step strain) is applied for a short time and then turned off, the system is able to forget this perturbation asymptotically. This is not the case here, due to the incomplete relaxation which leads to frozen-in stress. This is all in contrast with the soft glassy rheology model~\cite{sollich_rheology_1997,sollich_rheological_1998,fielding_ageing_2000}, where the yielding through effective activation always leads to full relaxation at long times (thus ensuring weak long-term memory), and the aging response can be cast into the Cugliandolo-Kurchan form~\cite{cugliandolo_out--equilibrium_1994}.

Turning to the comparison with the model athermal suspension considered in Sec.~\ref{sec:MD}, it would firstly be interesting to extend our elastoplastic description in order to account for the non-affine relaxation, which we recall we removed from the data for our comparison. Presumably, what would need to be added to our current picture is the heterogeneity of elastic moduli in the material, which would imply the system falls out of force balance after application of a step strain.

In order to connect further the mesoscopic model to the model particle system, an obvious route would be to study in detail the statistics of plastic events within the MD simulations. An important detail we left aside in Sec.~\ref{sec:MD} concerns the evolution of the system properties during the aging process: as studied in~\cite{chacko_slow_2019}, for later times the root mean square velocity decreases, and the active ``hotspots'' where non-affine relaxation occurs grow in size. One may then also expect the parameters of the corresponding elastoplastic model not to be constant. In fact, by considering the squared ratio of the constants $B$ measured in MD and mean field, it is in principle possible to infer the value of $\pltime$ in MD time units. Given that the coupling $A$ is also unknown, we may take a range of $B$ values measured in the mean field model ($B=0.4$ to $1.7$, as $A$ is decreased), which along with the simulation value $B\approx0.037$ in Fig. \ref{fig:collapse} would yield $\pltime \ \in \ (123,\ 2000)$ in MD time units. It would be interesting to measure the plastic timescale in the MD simulation and check whether it lies in the above-mentioned range, and stays roughly constant at least for the range of waiting times in Fig.~\ref{fig:collapse}. 

Another avenue for exploring the mesoscopic assumptions of the elastoplastic model would be to employ a frozen-matrix method~\cite{noauthor_notitle_nodate} as in \cite{puosi_probing_2015,ruscher_residual_2020}, to obtain direct information on the full local stress distributions. Although the results in Sec.~\ref{sec:MD}, in particular the good fits of the plateau and stress dynamics with the same value of $B$ shown in Figs.~\ref{fig:Ginfs_Ritu} and~\ref{fig:collapse}, already provide good support for the boundary layer dynamics described here, probing the distributions themselves would of course provide stronger evidence, and would shed more light on further questions such as the value of the coupling $A$.

As regards experiments, it would certainly be interesting to compare the theory with measurements on aging suspensions. Carbopol microgels \cite{agarwal_signatures_2019,lidon_power-law_2017}, for instance, which are considered to be prototypical of athermal dynamics, could be a good candidate. Linear viscoelastic moduli in these systems would be interesting to measure, as was done in \cite{purnomo_glass_2008,purnomo_linear_2006,purnomo_rheological_2007} for a class of thermosensitive suspensions,
whose behaviour could be captured by the predictions of the soft glassy rheology model. 

In future work on the modelling side, one aspect that could be studied is the behaviour for $\mu<1$. We expect this to be physically less relevant, and not to present genuine aging, but the mathematical analysis could generate interesting insights into how the scalings presented in Sec.~\ref{sec:overview}, in particular Eq.~(\ref{Delta_sigma}), break down for $\mu<1$.
An obvious direction for extending the model would be to study the effect of disorder on the aging described here and in~\cite{parley_aging_2020}. This could be done by introducing a distribution $\rho(\sigma_c)$ of yield barriers as in~\cite{agoritsas_relevance_2015,parley_mean_2022}; in this way there would be aging not only in stress, but also as a result of mesoscopic regions transitioning to deeper energy minima with higher yield barriers.
    
\begin{acknowledgements}
We thank Suzanne M. Fielding for providing the illustrative data in Fig.~\ref{fig:propa}. This project has received funding from the European Union’s Horizon 2020 research and innovation programme under Marie Skłodowska-Curie grant agreement No 893128.
\end{acknowledgements}

\appendix

\section{\label{app:SS}Steady state linear response approaching the arrest transition}

\begin{figure}\center{\includegraphics[width=0.45\textwidth]{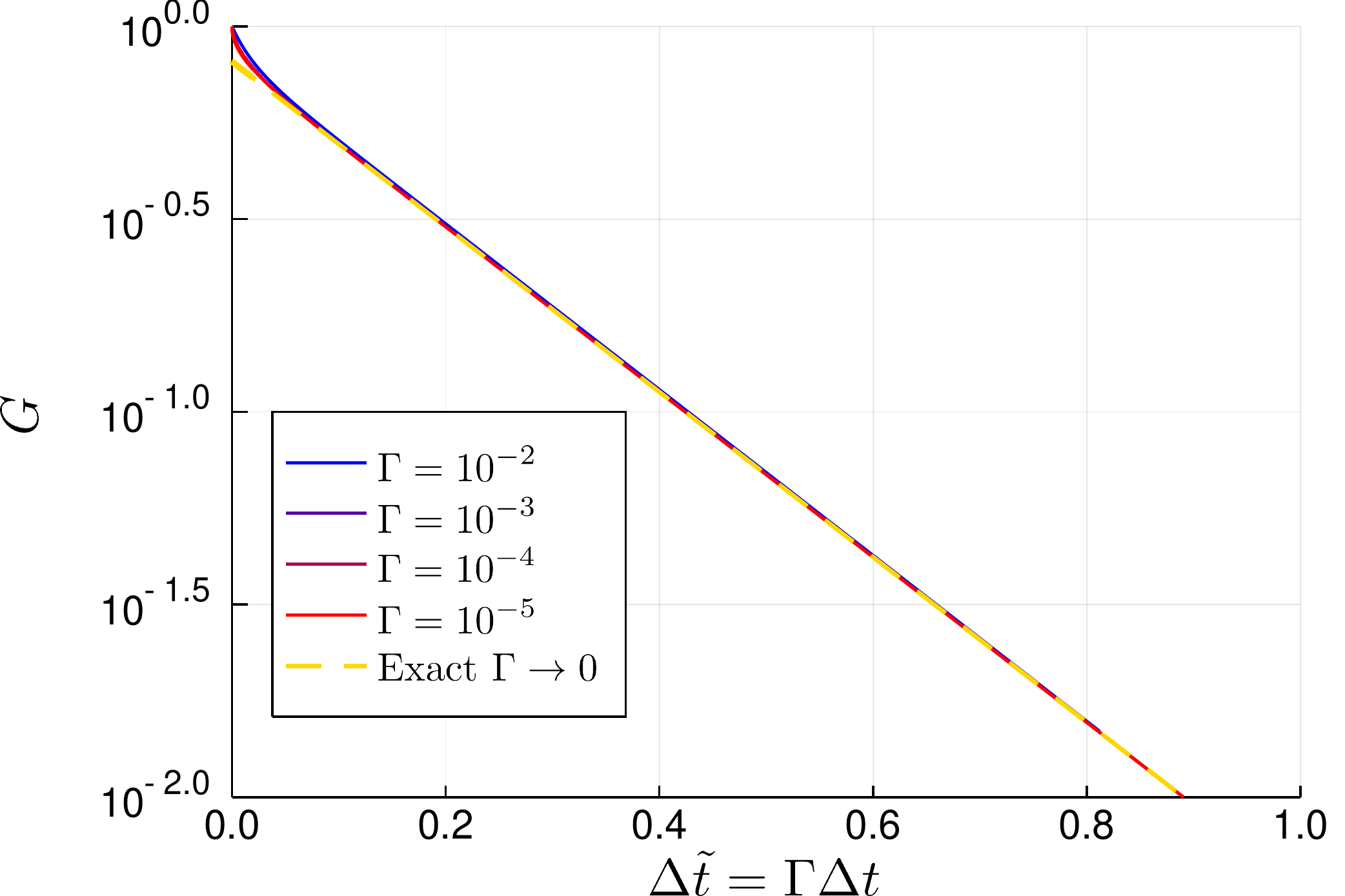}}
\caption{\label{fig:SS_longtime}Stress relaxation in the long time regime in the HL model, for the steady state approaching the arrest transition ($\Gammass \ll 1$). For $\Delta \tilde{t}\gg 2/\pi^2$, we find an exponential relaxation, purely dominated by the first term in the summation (\ref{HL_Gt}).}
\end{figure}

\begin{figure}\center{\includegraphics[width=0.45\textwidth]{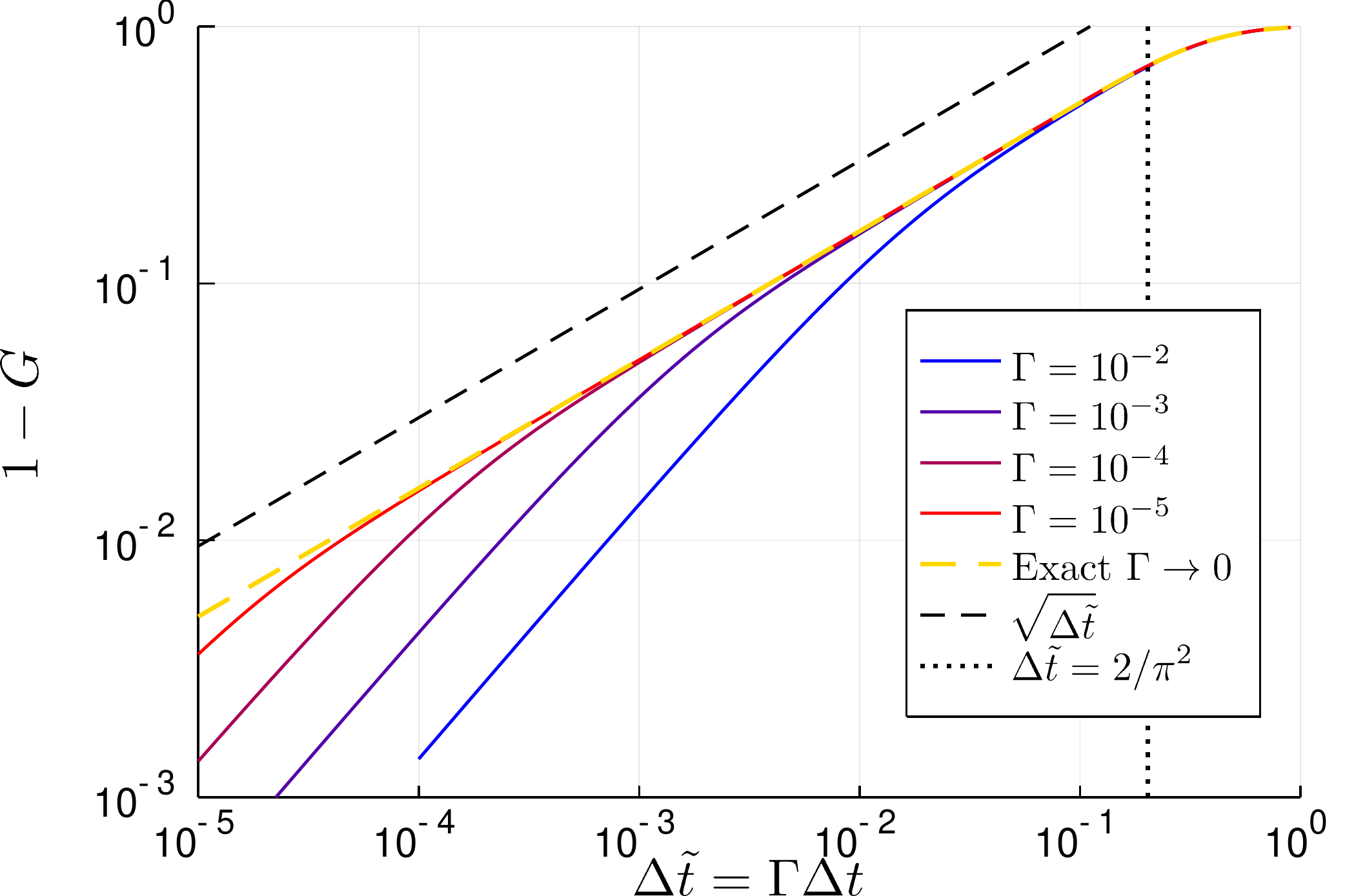}}
\caption{\label{fig:SS_shorttime}Same data as in Fig.~ \ref{fig:SS_longtime}, but plotted in the short time regime. For $\Gamma\ll \Delta \tilde{t}\ll \tilde{\tau}$, we see the development of a power-law regime $1-G\sim (\Delta \tilde{t})^{1/2}$, as predicted from the analytical form~(\ref{HL_Gt}) for $\Gamma \rightarrow 0$.}
\end{figure}

We first consider here the linear response in the steady state regime, where as explained in Sec.~\ref{sec:background} we expect TTI to hold. We discuss first the HL case ($\mu=2$), where insight may be gained through analytical arguments. Although some expressions for the steady state linear frequency response are provided in the original paper \cite{hebraud_mode-coupling_1998}, we focus here on the critical behaviour approaching the arrest transition. As $\alpha_c$ is approached from above, the diffusive dynamics of the local stress becomes more and more sluggish with the yield rate disappearing quadratically as $\Gammass\sim (\alpha-\alpha_c)^2$ \cite{agoritsas_relevance_2015,parley_aging_2020}, meaning there are fewer plastic rearrangements to fluidise the system. In the limit where $\Gammass \rightarrow 0$, one may replace the yielding term in equation~(\ref{deltap}) by absorbing boundary conditions at $\sigma=\pm 1$. One can then map the problem to that of a diffusing particle in a box (see also \cite{parley_aging_2020}), which can be solved by the technique of separation of variables. Given the antisymmetry of $\delta P(\sigma,t)$ described in Sec.~\ref{sec:background}, the solution is given by the asymmetric eigenmodes. Rescaling the time difference (we recall $\Delta t =t-t_{\rm w}$) by the yield rate as $\deltatp=\Delta t\,\Gamma$, we find
\begin{equation}\label{HL_Gt}
    G(\deltatp)= \frac{8}{\pi^2}\sum_{m, \rm{odd}}\frac{1}{m^2}e^{-\alpha m^2 \pi^2 \deltatp}
\end{equation}
This 
stress relaxation function separates into two different relaxation regimes. This is shown in Figs.~\ref{fig:SS_longtime} and~\ref{fig:SS_shorttime} where, along with the exact limiting form~(\ref{HL_Gt}), we plot the results of numerically integrating equation (\ref{deltap}) for values of $\Gammass$ between $10^{-2}$ and $10^{-5}$, starting from the steady state and using a pseudospectral method (for details see App.~E.1 of~\cite{parley_aging_2020}). 
At long times the relaxation is dominated by the slowest asymmetric eigenmode with absorbing boundary conditions, whose eigenvalue we write as $\lambda_1
^{(2)}$ for $\mu=2$. For 
$\Delta \tilde{t}\gg \tilde{\tau}$ with $\tilde{\tau}=1/(\alpha \lambda_1^{(2)})\approx 2/\pi^2$ one then finds an exponential relaxation (Fig.~\ref{fig:SS_longtime}). On the other hand, in the short time regime $\Gamma \ll \Delta \tilde{t}\ll\tilde{\tau}$, 
we find that $1-G(\deltatp)\sim (\deltatp)^{1/2}$ (Fig.~\ref{fig:SS_shorttime}), reflecting the singular behaviour of the summation~(\ref{HL_Gt}).

\begin{figure}\center{\includegraphics[width=0.45\textwidth]{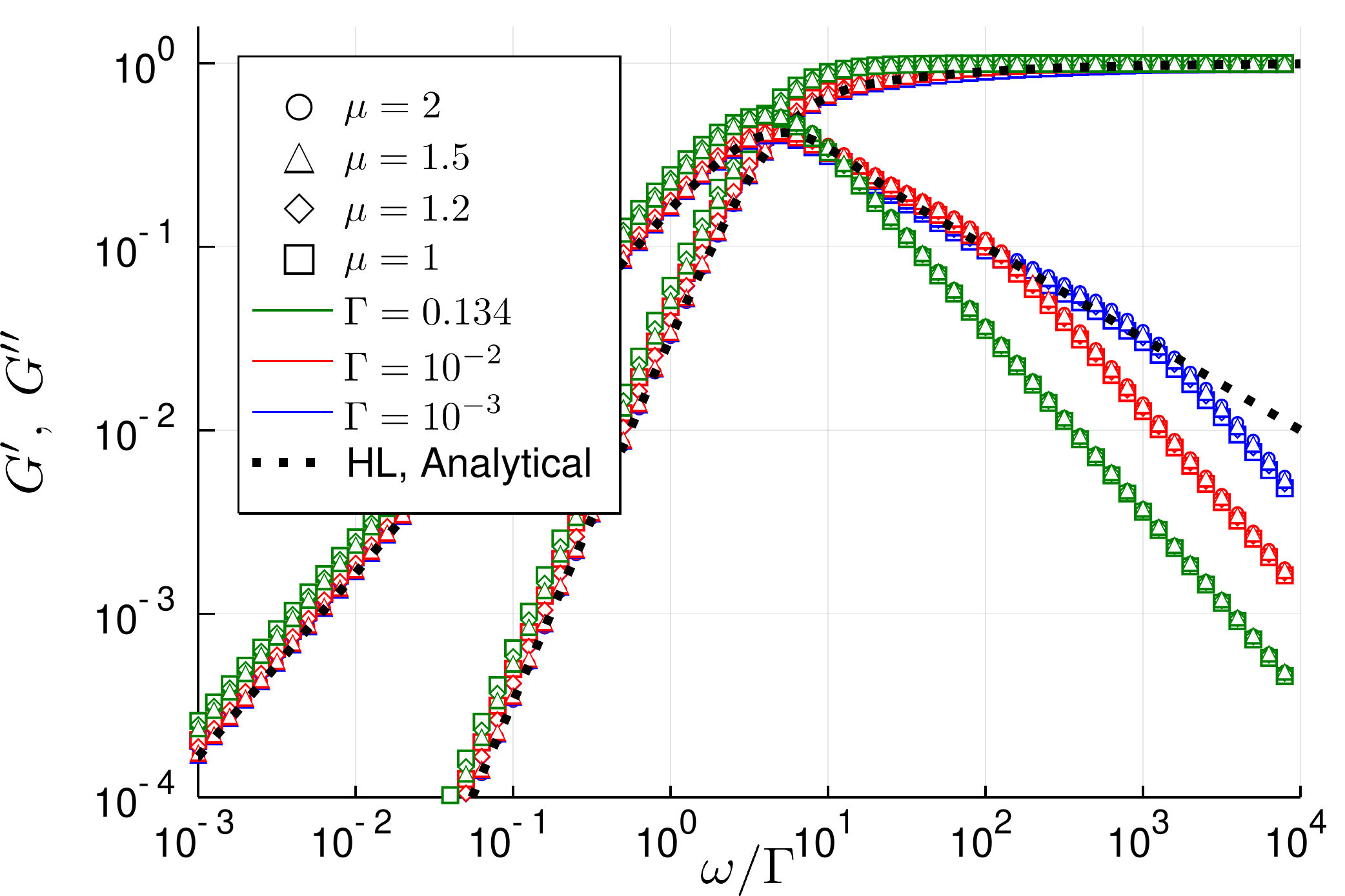}}
\caption{\label{fig:frequency_SS}Viscoelastic moduli in steady state for different values of $\mu$ and $\Gammass$, obtained via spectral decomposition of the corresponding operator. The moduli are collapsed by rescaling the frequency as $\tilde{\omega}=\omega/\Gamma$. Dotted lines show the analytical predictions for $G'$ and $G''$ as $\Gammass\rightarrow0$ in the HL model (\ref{HL_Gstar}).}
\end{figure}

Looking next at the viscoelastic behaviour for $\mu=2$ in the frequency domain, we know from Eq.~(\ref{HL_Gt}) that with a rescaled frequency $\tilde{\omega}=\omega/\Gamma$, the viscoelastic moduli in the limit $\Gamma \rightarrow 0$ are given by
\begin{eqnarray}\label{HL_Gstar}
    G^*(\tilde{\omega})&=&G'(\tilde{\omega})+iG''(\tilde{\omega}) \\ &&{}=8\sum_{m,\rm{odd}}\frac{1}{\alpha^2m^4\pi^4+\tilde{\omega}^2}\left(\frac{\tilde{\omega}^2}{\pi^2 m^2}+i\alpha \tilde{\omega}\right)\nonumber
\end{eqnarray}
Reflecting the behaviour in the time domain, this results in a loss modulus $G''(\tilde{\omega})$ peaked at $\tilde{\omega}\sim \tilde{\tau}^{-1}$, with a non-Maxwellian behaviour $G''(\tilde{\omega})\sim \tilde{\omega}^{-1/2}$ (as mentioned in \cite{hebraud_mode-coupling_1998}) for $\tilde{\tau}^{-1}<\tilde{\omega}<1/\Gamma$ (see dotted line in Fig.~\ref{fig:frequency_SS}). The same power law in this range of frequencies also appears in the elastic modulus as $1-G'(\tilde{\omega})\sim \tilde{\omega}^{-1/2}$.

We now turn to study other values of the noise exponent $0<\mu<2$. For convenience we do this in the frequency domain, where, instead of solving each time the PDE~(\ref{deltap}), we can compute the viscoelastic spectrum directly by diagonalising a discretised form \cite{buldyrev_average_2001} of the operator on the right of~(\ref{deltap}); for details see App.~E.2 in \cite{parley_aging_2020}. The results are shown in Figure \ref{fig:frequency_SS}, where we consider values of $\Gammass=0.134$, $10^{-2}$ and $10^{-3}$ and consider a range of different $\mu$.

The surprising and a priori unexpected result in Fig. \ref{fig:frequency_SS} is that the moduli show the same form also for $\mu<2$, with the same power law $G''(\tilde{\omega})\sim \tilde{\omega}^{-1/2}$ for the loss modulus. With hindsight this simply mirrors the behaviour in the short time regime, which as argued in Sec.~\ref{sec:overview} turns out to have the universal form $1-G(\Delta t)\sim {\Delta t}^{1/2}$ for all $\mu$.

\section{Critical aging}\label{app:criticality}

\begin{figure}\center{\includegraphics[width=0.42\textwidth]{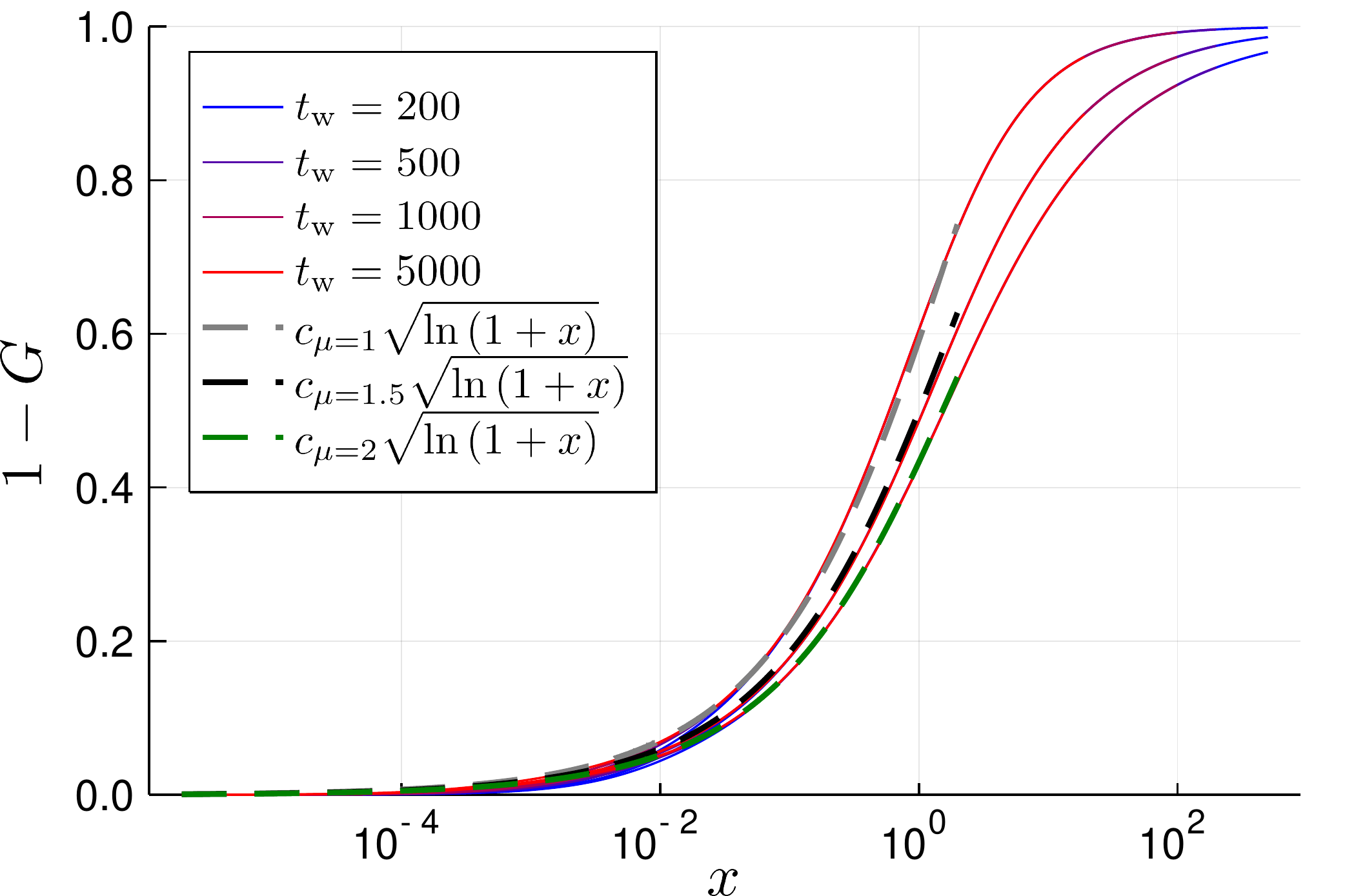}}
\caption{\label{fig:critical_st}Stress relaxation in the short time regime for the critical aging case $A=A_c(\mu)$ (for $\mu=2$, $\alpha=\alpha_c$), found by numerically solving (\ref{deltap}), with initial conditions obtained from the unperturbed dynamics. Curves for different $t_{\rm w}$ collapse essentially on top of each other when plotted against the rescaled time difference $x=(t-t_{\rm w})/t_{\rm w}$, following (\ref{log_critical}); $c_{\mu}$ is a $\mu$-dependent prefactor.}
\end{figure}

We consider for completeness the special case of a relaxation at precisely the critical value of the coupling $A=A_c$ (or $\alpha=\alpha_c$ in the HL model), which we refer to as critical aging. As discussed briefly in \cite{parley_aging_2020}, one finds from the analysis for $1\leq\mu\leq 2$ that the yield rate decays as $\Gamma(t)\sim 1/t$, irrespective of the value of $\mu$. For the short time regime (Fig.~\ref{fig:critical_st}), following the same arguments as in Sec.~\ref{sec:overview}, this implies an initial relaxation -- arising from stress diffusion near the yield threshold -- growing as $1-G(t,t_{\rm w})\sim\left(\int_{t_{\rm w}}
^{t}\Gamma(t')\mathrm{d}t'\right)^{1/2}$. This can be written in terms of the scaling variable $x=(t-t_{\rm w})/t_{\rm w}$, so that one has simple aging and 
\begin{equation}\label{log_critical}
    1-G\sim \sqrt{\ln(1+x)} \quad  \rm{for} \quad  x \ll 1 
\end{equation}
For the yield rate at criticality, one expects that in fact the prefactor of the asymptotic behaviour $\Gamma(t)\sim d_1(\mu)/t$ will be initial condition independent for a given $\mu$, given that the total number of yield events (given by the integral of $\Gamma(t)$) diverges and so all memory of the initial condition is lost. In fact, as we will show now for the HL model, this prefactor is related to the lowest asymmetric eigenvalue $\lambda_1^{(\mu)}$ of the $\mu$-dependent propagator with absorbing boundary conditions at $|\sigma|=1$, by the relation $d_1(\mu)=1/(\mu \lambda_1^{(\mu)})$. The boundary conditions are non-local for $\mu<2$, i.e.\ must be imposed for all $|\sigma|>1$~\cite{zoia_fractional_2007}; the eigenvalue $\lambda_1^{(\mu)}$ is defined in Eq.~(\ref{egv_eq}) below.

In the HL case $\mu=2$, we can show this link on the basis of the scaling analysis in \cite{sollich_aging_2017}. For the case of a relaxation at $\alpha=\alpha_c$, the exponent parameters in \cite{sollich_aging_2017} take the values $l=1$ and $s=2$. The frozen-in distribution, on the other hand, acquires a simple form composed of two line segments, $Q_0(\sigma)=1-|\sigma|$. The leading order corrections in the interior ($|\sigma|<1$) and in the exterior ($|\sigma|>1$; where the right and left exterior tails are symmetric, we write only the right one, i.e.\ $\sigma>1$) are given by
\begin{eqnarray}
    P(\sigma,t)&=&Q_0(\sigma)+t^{-\frac{1}{2}}Q_1(\sigma) \quad |\sigma|<1  \\
    P(\sigma,t)&=&t^{-\frac{1}{2}}R_1(z) \quad \quad \quad \quad \quad \sigma >1 
\end{eqnarray} 
with $z=t^{1/2}(\sigma-1)$. Continuity of the distribution and its derivative imply the boundary conditions
\begin{eqnarray}
    Q_1(1)&=&R_1(0) \label{bc_1}\\
    \partial_{\sigma}Q_0(1)&=&\partial_{z}R_1(0)=-1 \label{bc_2}
\end{eqnarray}
We consider now the master equation (\ref{hl_equation}) in the exterior, with $\alpha=\alpha_c=1/2$ and $\Gamma(t)=d_1/t$. Applying also the boundary condition (\ref{bc_2}), we have that $R_1(z)=\sqrt{d_1}e^{-z/\sqrt{d_1}}$. From the master equation in the interior, we find that 
\begin{equation}\label{Q1}
    \partial_{\sigma}^2 Q_1(\sigma)+\lambda Q_1(\sigma)=0
\end{equation}
where $\lambda \equiv 1/(2 d_1)$. The boundary condition (\ref{bc_1}) implies that $Q_1(-1)=Q_1(1)=\sqrt{d_1}$. Furthermore, normalisation of $P(\sigma,t)$ requires that $\int Q_1 \mathrm{d}\sigma=0$, so that integration of (\ref{Q1}) yields $\partial_{\sigma}Q_1(-1)=\partial_{\sigma}Q_1(1)$. Altogether, Eq.~(\ref{Q1}) and the boundary conditions imply that $\lambda=m^2\pi^2$, $m \in \mathbb{N}$. Given that we are dealing with the first correction, we expect $m=1$ so that $\lambda=\lambda_1
^{(2)}$, and therefore $d_1=1/(2\lambda_1^{(2)})$. For $1\leq\mu<2$ the leading order correction scales as $P=Q_0+t^{-1/\mu}Q_1$, so that one expects $d_1=1/(\mu \lambda_1^{(\mu)})$ from the same analysis. We confirm this only numerically as a full derivation would be difficult due to the presence of non-local boundary conditions.

We now show how the prefactor $d_1(\mu)$ leads to the long-time difference ($x\gg 1$) scaling of the stress relaxation function $G(x)\sim x^{-1/\mu}$. The perturbation $\delta P(\sigma,t)$ follows the dynamics (\ref{deltap}), which in the interior reads
\begin{equation}\label{deltaP_Ac}
    \partial_t \delta P(\sigma,t)=\frac{d_1(\mu)}{t}\int_{\sigma-\etau}^{\sigma+\etau}\frac{\delta P(\sigma',t)-\delta P(\sigma,t)}{|\sigma-\sigma'|^{1+\mu}}\mathrm{d}\sigma'
\end{equation}
At long times we expect (as in the case $A>A_c$ in App.~\ref{app:SS}) the stress profile to be dominated by the slowest asymmetric eigenmode $\psi_1(\sigma)$, so that $\delta P(\sigma,t \gg 1)\approx f(t)\psi_1(\sigma)$ 
where $\psi_1(\sigma)$ satisfies
\begin{equation}\label{egv_eq}
    \int_{\sigma-\etau}^{\sigma+\etau} \frac{\psi_1(\sigma')-\psi_1(\sigma)}{|\sigma-\sigma'|^{1+\mu}}\mathrm{d}\sigma'=-\lambda_1^{(\mu)}\psi_1(\sigma)
\end{equation}
with $\psi_1 (\sigma)=0 \ \forall \ |\sigma|>1$, i.e.\ absorbing boundary conditions. Inserting the ansatz into (\ref{deltaP_Ac}), we find using $d_1(\mu)\lambda_1^{(\mu)}=1/\mu$
\begin{equation}
    \frac{\partial \ln f}{\partial \ln t}=-\frac{1}{\mu}
\end{equation}
so that $\delta P(\sigma,t)\simeq \psi_1(\sigma) t^{-1/\mu}$ at long times. Considering (from the short time regime) that we have simple aging, this implies $G(x)\sim x^{-1/\mu}$ in the long time regime as claimed.

\begin{figure}\center{\includegraphics[width=0.42\textwidth]{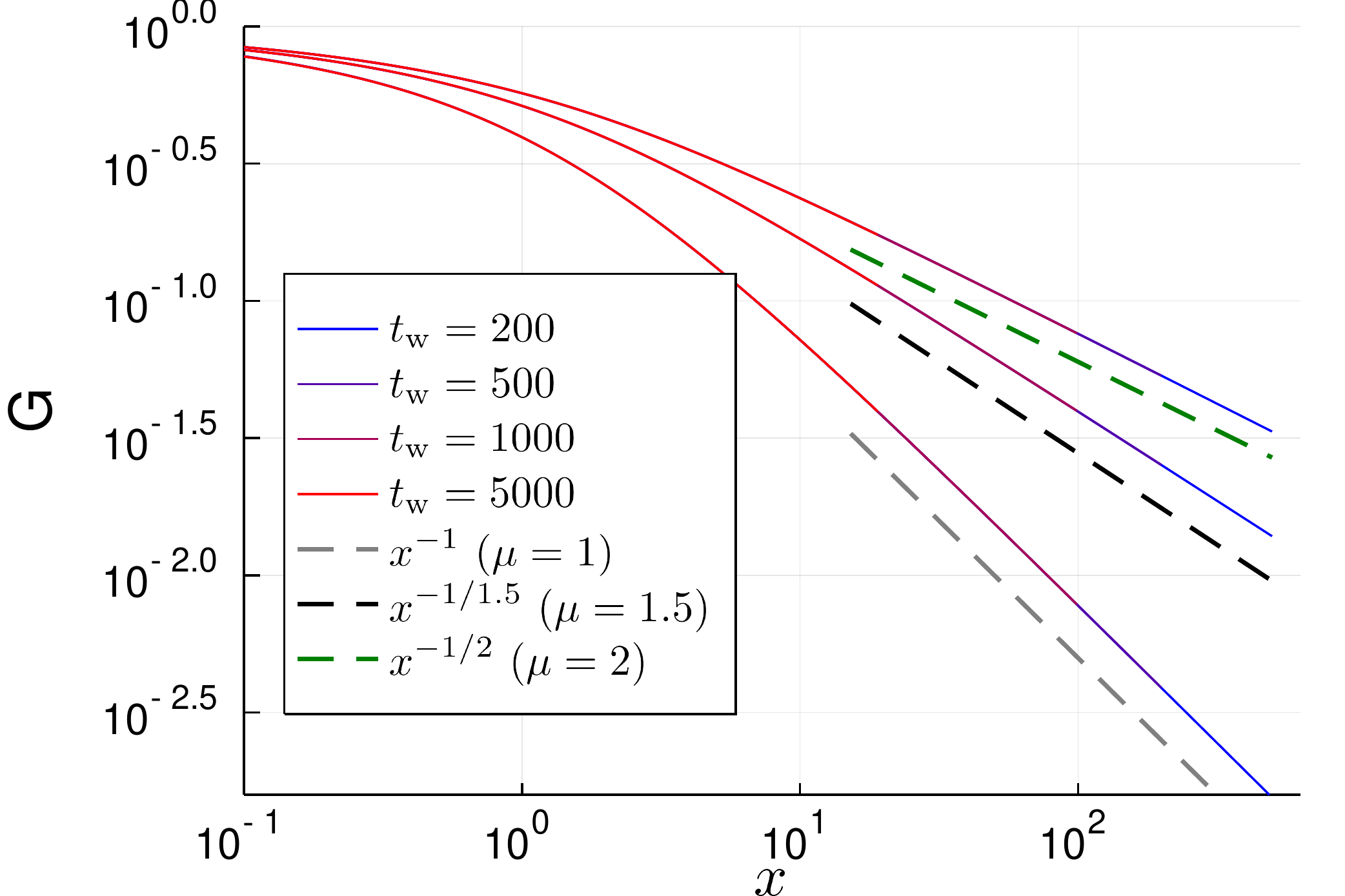}}
\caption{\label{fig:critical_lt}Same as Fig. \ref{fig:critical_st}, but in the long time regime. For $x\gg 1$, the curves follow the predicted power-law relaxation with $\mu$-dependent exponent $G(x)\sim x^{-1/\mu}$.}
\end{figure}

In Figs.~\ref{fig:critical_st} and \ref{fig:critical_lt} we show stress relaxation functions obtained for a range of $t_{\rm w}$, for a system relaxing at $A_c(\mu)$ from an initial distribution with enough unstable blocks at $t=0$. We then evolve Eq.~(\ref{deltap}) to find the aging stress relaxation function.  For $x \gg 1$, we see from Fig.~\ref{fig:critical_lt} that indeed $G(x)\sim x^{-1/\mu}$. Interestingly, a power-law stress relaxation was also found at the jamming transition point in the particle simulations of \cite{saitoh_stress_2020}, with a critical behaviour $G(t)\sim t^{-1/2}$ (which would be recovered for $\mu=2$). It is important, however, to note that in \cite{saitoh_stress_2020} the step response is studied starting from initial conditions that have already fully relaxed to mechanical equilibrium (via an energy minimization algorithm), whereas here we are considering the stress response during the physical relaxation process towards this inherent state.

\section{Scaling of $\delta P(\sigma,t)$}\label{app:delta_P}

\begin{figure}\center{\includegraphics[width=0.42\textwidth]{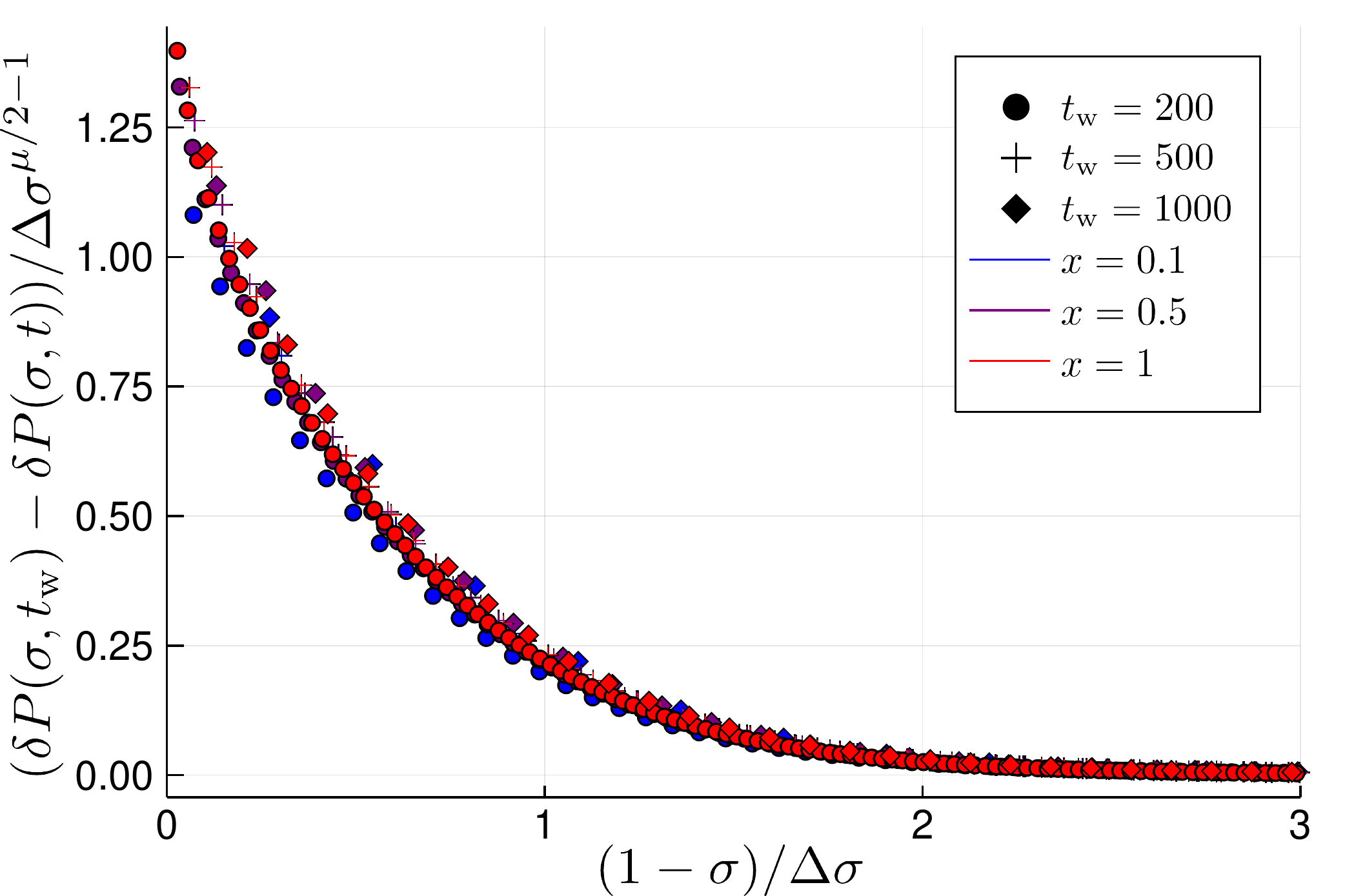}}
\caption{\label{fig:scaling_interior}Decay of the linear perturbation $\delta P(\sigma,t)$ in the interior boundary layer. Data is for the same case ($\mu=1.7$, $A=0.15$) as shown in Sec.~\ref{sec:aging}. Symbols show different $t_{\rm w}$ (shapes) at different $x=(t-t_{\rm w})/t_{\rm w}$ (colors), which all collapse as detailed in the text.}
\end{figure}

We give more details here regarding the scaling of the stress distribution perturbation $\delta P(\sigma,t)$ that leads to the result~(\ref{G_mu_12}) for the stress relaxation function. First of all, as in Sec.~\ref{sec:overview} we may write Eq.~(\ref{G}) as
\begin{equation}\label{1_G_2}
    1-G(t,t_{\rm w})=\int_{-\infty}^{\infty}\sigma \left(\delta P(\sigma,t_{\rm w})-\delta P(\sigma,t)\right)\mathrm{d}\sigma
\end{equation}
In the aging regime, $\delta P(\sigma,t)$ is practically frozen in the interior \footnote{There is, potentially, a contribution from relaxation around the origin $\sigma=0$, but we have checked numerically that this gives a sub-leading contribution.} $|\sigma|<1$ away from $\sigma=1$, 
while the relaxation in the exterior tails $|\sigma|>1$ will be shown below to be sub-leading. The leading contribution to the integral~(\ref{1_G_2}) will come from two symmetric interior boundary layers, on the left and the right. Focusing on the positive one at $\sigma=1$, we introduce as a division between interior and boundary layer a fixed stress interval $\epsilon$ such that $\Delta \sigma (t_{\rm w},x)\ll \epsilon \ll 1$, $\forall t_{\rm w},x$, where $\Delta \sigma (t_{\rm w},x)$ is the width of the interior boundary layer at time $t=t_{\rm w}(1+x)$ for a perturbation applied at $t_{\rm w}$. The leading contribution to~(\ref{1_G_2}) will then be given by
\begin{equation}
    1-G\simeq 2 \int_{1-\epsilon}^{1} \sigma \left(\delta P(\sigma,t_{\rm w})-\delta P(\sigma,t)\right)\mathrm{d}\sigma
\end{equation}
One expects the difference $\delta P(\sigma,t_{\rm w})-\delta P(\sigma,t)$ to become a scaling function of the width $\Delta \sigma$ within the interior boundary layer. In addition, given that $\delta P(\sigma,t)$ drops significantly within this layer, one expects the height of the function itself to scale as ${\Delta \sigma}^{\mu/2-1}$, which is inherited from the height of the initial distribution at $\sigma \sim 1-\Delta \sigma$; recall that the initial condition of the perturbation scales as $\delta P(\sigma,t_{\rm w})\sim (1-\sigma)^{\mu/2-1}$ near the boundary. We then have that
\begin{multline}
    1-G \simeq 2(\Delta \sigma)^{\mu/2-1} \int_{1-\epsilon}^{1}\sigma f\left(\frac{1-\sigma}{\Delta \sigma}\right)\mathrm{d}\sigma\\
    =(\Delta \sigma)^{\mu/2}2\int_{0}^{\frac{\epsilon}{\Delta \sigma}}f(z)\mathrm{d}z-(\Delta\sigma)^{\mu/2+1}2
    \int_{0}^{\frac{\epsilon}{\Delta \sigma}}zf(z)\mathrm{d}z
\end{multline}
where we performed the change of variable $z=(1-\sigma)/\Delta \sigma)$. As $  \Delta \sigma \ll \epsilon$, the $\epsilon$-dependence in the integrals disappears and we are left with $1-G \simeq (\Delta \sigma)^{\mu/2}$ to leading order, confirming the result~(\ref{G_mu_12}) given above. In Fig.~\ref{fig:scaling_interior}, we check for various values of $t_{\rm w}$ and $x$ (with $\mu=1.7$, $A=0.15$ as in Fig. \ref{fig:G_17}) the above scaling of $\delta P(\sigma,t)$ in the interior boundary layer, finding a very good collapse. 

\begin{figure}\center{\includegraphics[width=0.42\textwidth]{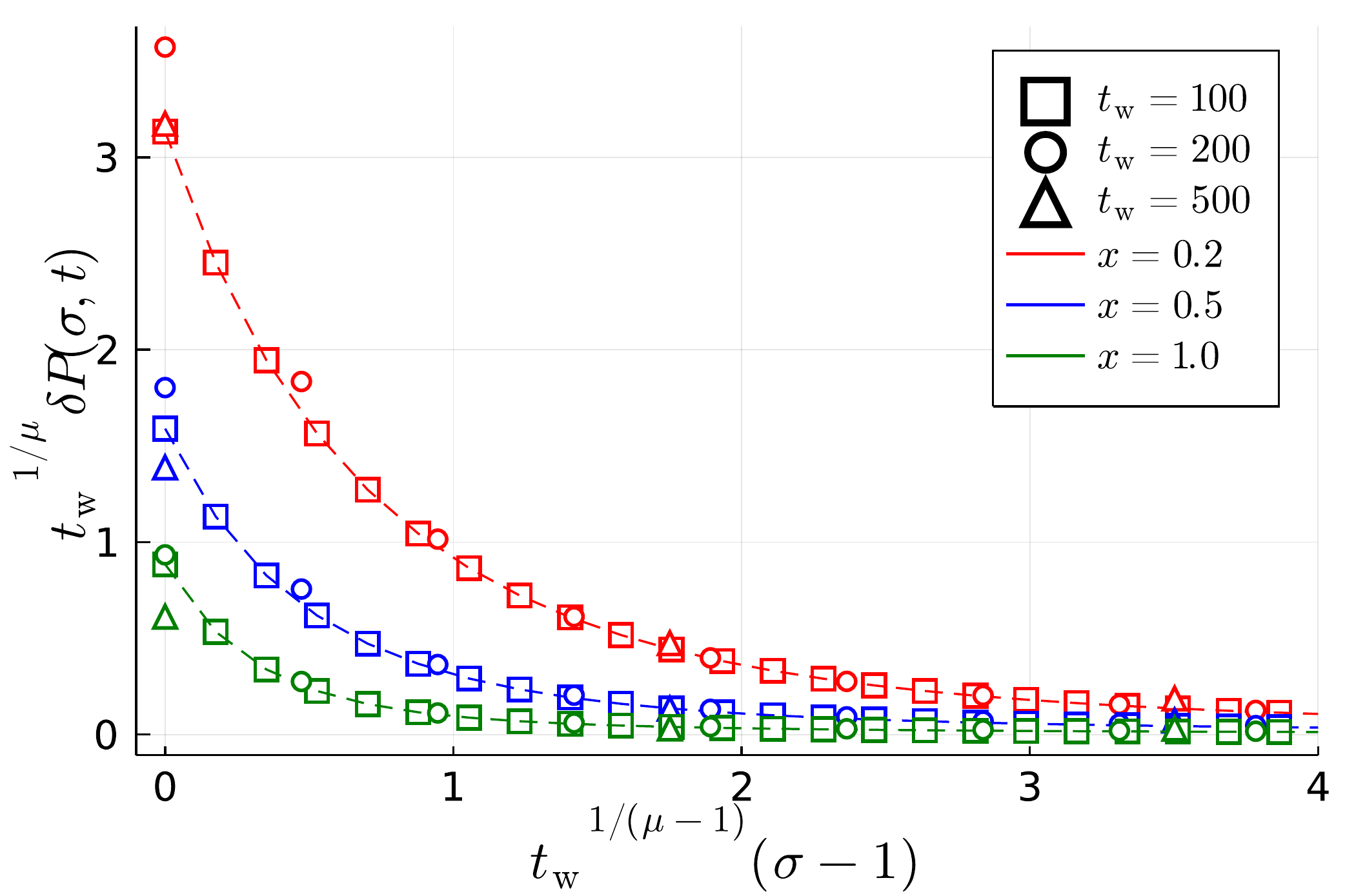}}
\caption{\label{fig:exterior_tail}Decay of the external tail of $\delta P(\sigma,t)$, again for $\mu=1.7$, $A=0.15$. Symbols show different $t_{\rm w}$ (shapes) at different $x=(t-t_{\rm w})/t_{\rm w}$ (colors), which all collapse as detailed in the text.}
\end{figure}

Finally, we show that the exterior tail contribution to the integral~(\ref{1_G_2}) 
is indeed sub-leading. We find that, as in the HL model~\cite{sollich_aging_2017}, at a fixed value of $x$ the exterior tails of $\delta P(\sigma,t)$ may be collapsed by rescaling their width and the height by appropriate powers of $t_{\rm w}$. For the $\sigma$-axis, we know already that the distribution $\delta P(\sigma,t_{\rm w})$ inherits the scaling of the boundary layer in the unperturbed dynamics. There it was shown \cite{parley_aging_2020} that $\Gamma \sim t
^{-\mu/(\mu-1)}$, while the boundary layer width scaled as $\Gamma^{1/\mu}$, so that we expect the exterior tail to have a width that evolves with $x$ on a scale $\mathcal{O}(t_{\rm w}^{-1/(\mu-1)})$. Turning now to the scaling of the height of the tail $\delta P(\sigma,t)$, we find numerically that the boundary value $\delta P(1,t)$, decays as a power law $(t-t_{\rm w})^{-1/\mu}$ beyond time differences of order unity $t-t_{\rm w}>\mathcal{O}(1)$, i.e.\ $x>\mathcal{O}(t_{\rm w}
^{-1})$, leading to a height evolving with $x$ on a scale $t_{\rm w}^{-1/\mu}$. This is confirmed numerically in Fig. \ref{fig:exterior_tail} for various values of $t_{\rm w}$ and $x$, again running the dynamics with $\mu=1.7$, $A=0.15$. Overall, these scalings imply that the contribution from the exterior tail is indeed sub-leading, given that it is of order $t_{\rm w}^{-1/(\mu-1)-1/\mu}$, which is small compared to ${\Delta \sigma}^{\mu/2}=\mathcal{O}(t_{\rm w}^{-1/(2(\mu-1))})$. This holds also as we approach the marginal case $\mu\rightarrow 1$, as the (negative) exponent of the leading contribution is smaller by a factor of 2.

\section{Forward spectrum}\label{app:FS}

In this appendix we provide details on the derivation of the asymptotic forms~(\ref{FS_17}) and~(\ref{FS_1}) of the forward spectrum defined in Eq.~(\ref{FS}), and discuss how the full $t_{\rm w}$-dependent aging spectrum~(\ref{aging_Gstar}) approaches this limit.

We consider first the case $1<\mu<2$, and assume $t_{\rm w}$ is large enough for expression (\ref{G_mu_12}) to hold, that is we take
\begin{equation}
    G(t,t_{\rm w})=1-c t_{\rm w}^{-\frac{1}{2(\mu-1)}}\sqrt{1-(1+x)^{-\frac{1}{\mu-1}}}
\end{equation}
with $x=(t-t_{\rm w})/t_{\rm w}$. We now insert this expression into (\ref{aging_Gstar}). Following \cite{sollich_aging_2017}, we introduce the new variables $w\equiv \omega t$ and $w'\equiv\omega (t-t_{\rm w})$. After some algebra, (\ref{aging_Gstar}) can be rewritten as 
\begin{multline}\label{re_Gstar}
    G^*(\omega,t,t_{\rm w})=1-\frac{c}{t^{\frac{1}{2(\mu-1)}}}\bigg(\sqrt{(1+x)^{\frac{1}{\mu-1}}-1} \ e^{-iw\frac{x}{1+x}}\\+i\int_0^{w\frac{x}{1+x}}\mathrm{d}w'\sqrt{\left(1-\frac{w'}{w}\right)^{-\frac{1}{\mu-1}}-1}\ e^{-iw'}\bigg)
\end{multline}
The forward spectrum (\ref{FS}), on the other hand, can be written with the change of variable $w'=\omega (t-t')$ as 
\begin{equation}\label{re_FS}
    \Gf^*(\omega,t)=1-\frac{c}{t^{\frac{1}{2(\mu-1)}}}i\int_{0}^{\infty}\mathrm{d}w'\sqrt{1-\left(1+\frac{w'}{w}\right)^{-\frac{1}{\mu-1}}} \ e^{-iw'}
\end{equation}
We now take the limits $w\equiv \omega t \gg 1$ and $w\frac{x}{1+x}=\omega (t-t_{\rm w})\gg1$ in (\ref{re_Gstar}), following \cite{sollich_aging_2017}. As shown there, the first term in brackets of (\ref{re_Gstar}) can be included into the integral over $w'$, with a constant integrand for $w'>wx/(1+x)$. As we take the limit $wx/(1+x)\gg 1$ we are left only with the integral up to infinity of the second term in brackets, which in addition for $w\gg1$ converges to the forward spectrum (\ref{re_FS}).

To find the asymptotic form (\ref{FS_17}) given in the main text one can exploit the large $w$-limit imposed above to simplify further. 
In~(\ref{re_FS}), one can then expand the argument in the square root as 
\begin{multline}
    f\left(\frac{w'}{w}\right)=\sqrt{1-\left(1+\frac{w'}{w}\right)^{-\frac{1}{\mu-1}}}\\=\sqrt{\frac{1}{\mu-1}}\sqrt{\frac{w'}{w}}+\mathcal{O}\left(\frac{w'}{w}\right)
\end{multline}
which leads to the form (\ref{FS_17}) in the main text.

One can proceed similarly for the case $\mu=1$, and show that the aging moduli~(\ref{aging_Gstar}) approach the forward spectrum~(\ref{FS}), where now the required limits are $\omega (t-t_{\rm w})\gg 1$ and $w\equiv \omega \sqrt{t}\gg 1$. To compute this forward spectrum, we consider $t_{\rm w}$ large enough for (\ref{lin_1}) to hold, that is
\begin{equation}
    G(t,t_{\rm w})=(1-G_{\infty}(t_{\rm w}))\sqrt{1-e^{-B\frac{x}{2}}}
\end{equation}
with $x=(t-t_{\rm w})/\sqrt{t_{\rm w}}$. We insert this into~(\ref{FS}) and obtain 
\begin{equation}
    \Gf(\omega,t)=1-(1-G_{\infty}(t))i\int_0^{\infty} \mathrm{d}w' \sqrt{1-e^{-\frac{Bw'}{2w}}} \ e^{-i w'}
\end{equation}
where we performed the change of variables $w'=\omega (t-t')$, and the rescaled frequency is $w\equiv\omega \sqrt{t}$. As was done for $1<\mu<2$ above, we now expand the square root as 
\begin{equation}
   f\left(\frac{w'}{w}\right)=\sqrt{1-e^{-\frac{Bw'}{2w}}}=\sqrt{\frac{B}{2}}\sqrt{\frac{w'}{w}}+\mathcal{O}\left(\frac{w'}{w}\right)
\end{equation}
where we have considered again $w\gg 1$. From here it is straightforward to derive expression (\ref{FS_1}) in the main text.

\begin{figure}\center{\includegraphics[width=0.42\textwidth]{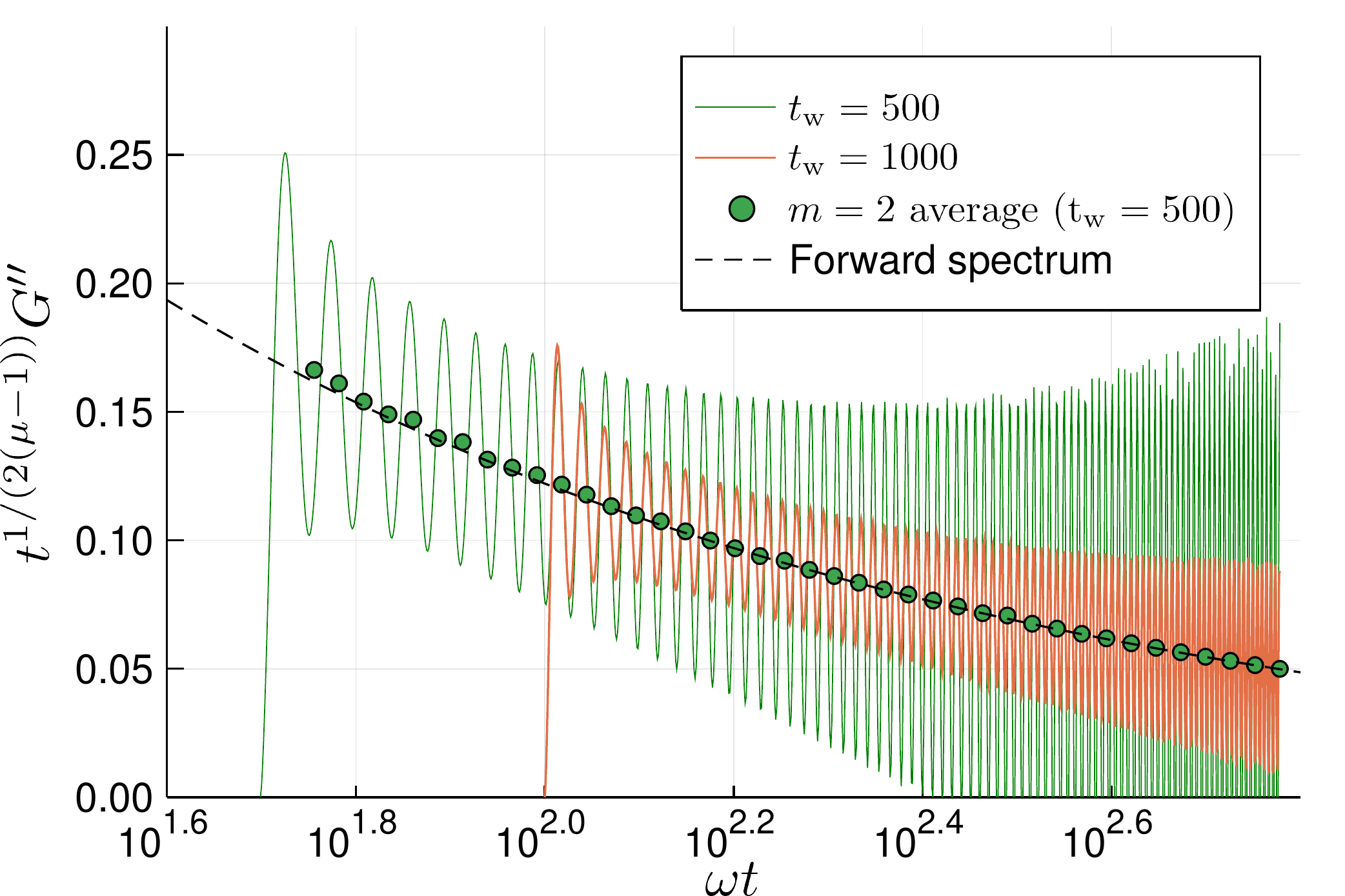}}
\caption{\label{fig:osci_FS}Comparison of the aging frequency response $G^{*}$ computed directly from (\ref{aging_Gstar}), with the averaged form $\bG$ (\ref{Gav_time}) (here averaged over $m=2$ periods), which cancels the oscillations (circles). Dashed line shows the forward spectrum (\ref{FS_17}). The growth of the oscillations for large $w$ is due to numerical instabilities in the oscillatory integral. Model parameter values as in Figs.~\ref{fig:G_17} and \ref{fig:frequency_17}.
}
\end{figure}

Finally, in Fig.~\ref{fig:osci_FS} (for the case $\mu=1.7$ and $A=0.15$ considered in the main text) we compare the averaged form $\bG(\omega,t,t_{\rm w})$ computed from (\ref{Gav_time}) with $G^*(\omega,t,t_{\rm w})$ calculated directly from (\ref{aging_Gstar}). Without the averaging, one sees that $G^*(\omega,t,t_{\rm w})$ does still approach the forward spectrum at long times, but presents oscillations around the asymptote with frequency $\omega$. As discussed in the main text and visible in the figure, the averaging cancels these oscillations and the asymptotic form is approached sooner. We note that the growing oscillations for small $t_{\rm w}$ are a numerical artifact due to the highly oscillating integrals.

\section{Details of MD simulations}\label{app:MD}
For comparison with the predictions derived from the mean field theory, we carry out numerical simulations using a model dense athermal solid (Sec.~\ref{sec:MD}). Here we summarize the model system and the simulation protocol.

We consider particles interacting via a pairwise repulsive harmonic potential $V_{ij}(r)=\frac{1}{2} k R^{3}\left(1-r/D_{ij}\right)^2 \theta \left(D_{ij}-r\right)$, where $r$ is the distance between particle $i$ and $j$. The system is bidisperse, with particles of radii $R$ and $1.4R$ in equal number, and $D_{ij}=R_i+R_j$. Such a bidisperse mixture helps to avoid crystallisation at high area fractions. Neglecting explicit hydrodynamic interactions, and in the absence of inertia, the \textit{unperturbed} dynamics of this system is simply a gradient descent in the energy landscape
\begin{equation}\label{grad}
    \frac{\mathrm{d}\mathbf{r}_i}{\mathrm{d}t}=-\frac{1}{\zeta}\sum_{j \neq i}\frac{\partial V\left(|\mathbf{r}_i-\mathbf{r}_j|\right)}{\partial \mathbf{r}_i}
\end{equation}
where $\mathbf{r}_i$ is the position vector of the $i$ th particle and $\zeta$ is the drag coefficient. By setting $k=R=\zeta=1$ we set the timescale $\zeta/(kR)=1$ in all the simulation results presented here. We implement the simulation in $2$d, using $N=40000$ particles compressed to area fraction $\phi=1$.

In the simulation we first quench the system from $T=\infty$ to $T=0$ and then allow it to relax athermally towards a force balanced inherent state. During this athermal aging process we collect samples that are aged up to time $t_{\rm w}$. We then implement a single step strain of amplitude $\gamma_0$ 
and measure the relaxation of the shear stress $\Sigma(t)$ for a time $\sim 10^6$. This time evolution happens in the presence of Lees-Edwards periodic boundary conditions \cite{lees_computer_1972} implementing the fixed strain, and using an adaptive Euler algorithm as deployed in \cite{chacko_slow_2019}.

\begin{figure}\center{\includegraphics[width=0.45\textwidth]{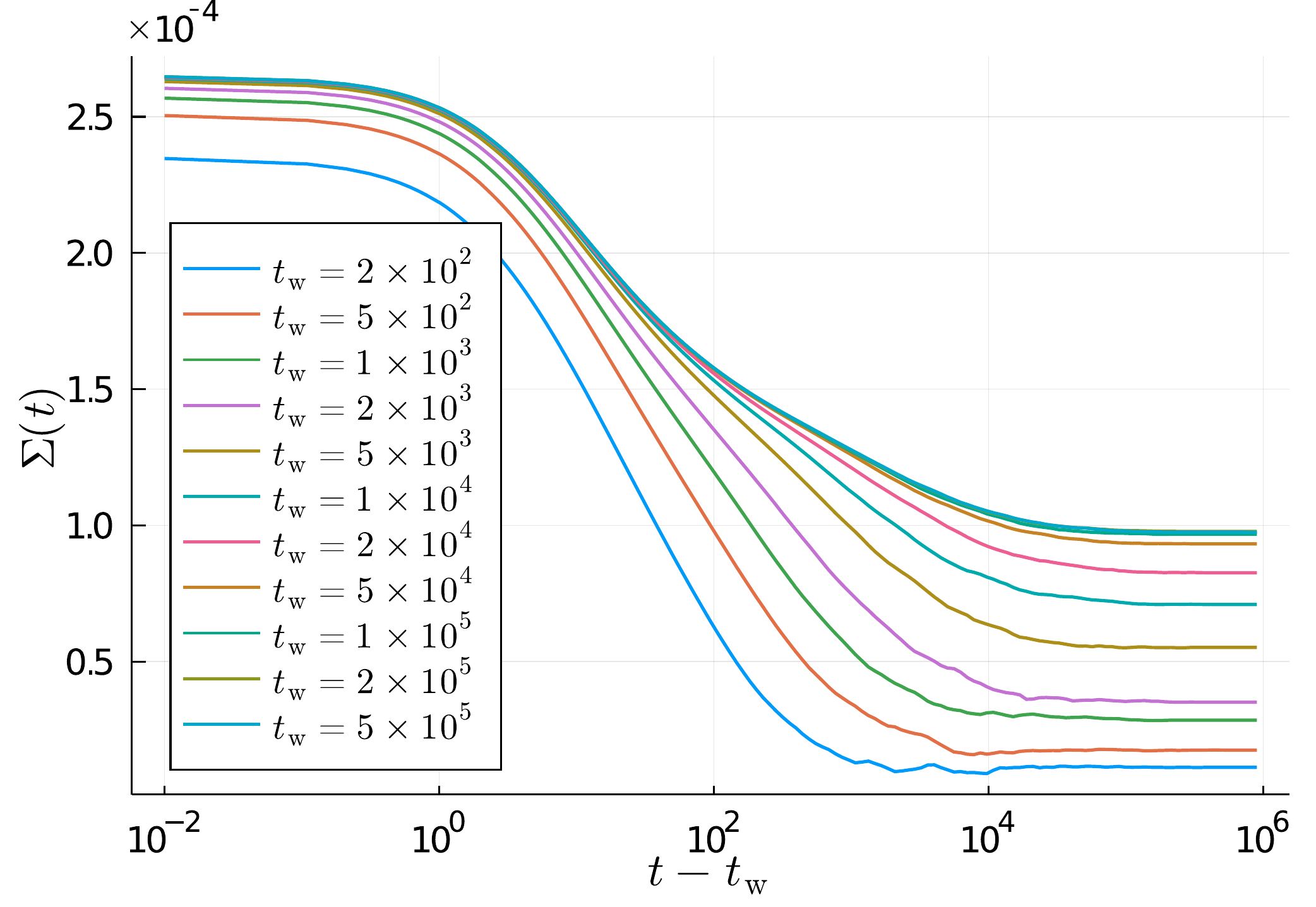}}
\caption{\label{fig:raw}Full stress relaxation measured in the MD simulations for step strain $\gamma_0=5 \times 10^{-3}$, at different waiting times $t_{\rm w}$ during the unperturbed relaxation.}
\end{figure}

Simulation results for $\gamma_0=5 \times 10^{-3}$ are shown in Fig.~\ref{fig:raw}. These are obtained by averaging over an ensemble of $N_{\rm rep}=128$ realizations of the random ($T=\infty$) initial condition; data for the smaller step strains shown in the paper are obtained with $N_{\rm rep}=1280$. For each realization we subtract the stress fluctuations of the unstrained $\gamma_0=0$ dynamics, which are due to the finite size. Note the non-affine stress relaxation present even for $t_{\rm w} \rightarrow \infty$, as detailed in the main text.

\section{Non-linear effects}\label{app:NL}
\begin{figure}\center{\includegraphics[width=0.42\textwidth]{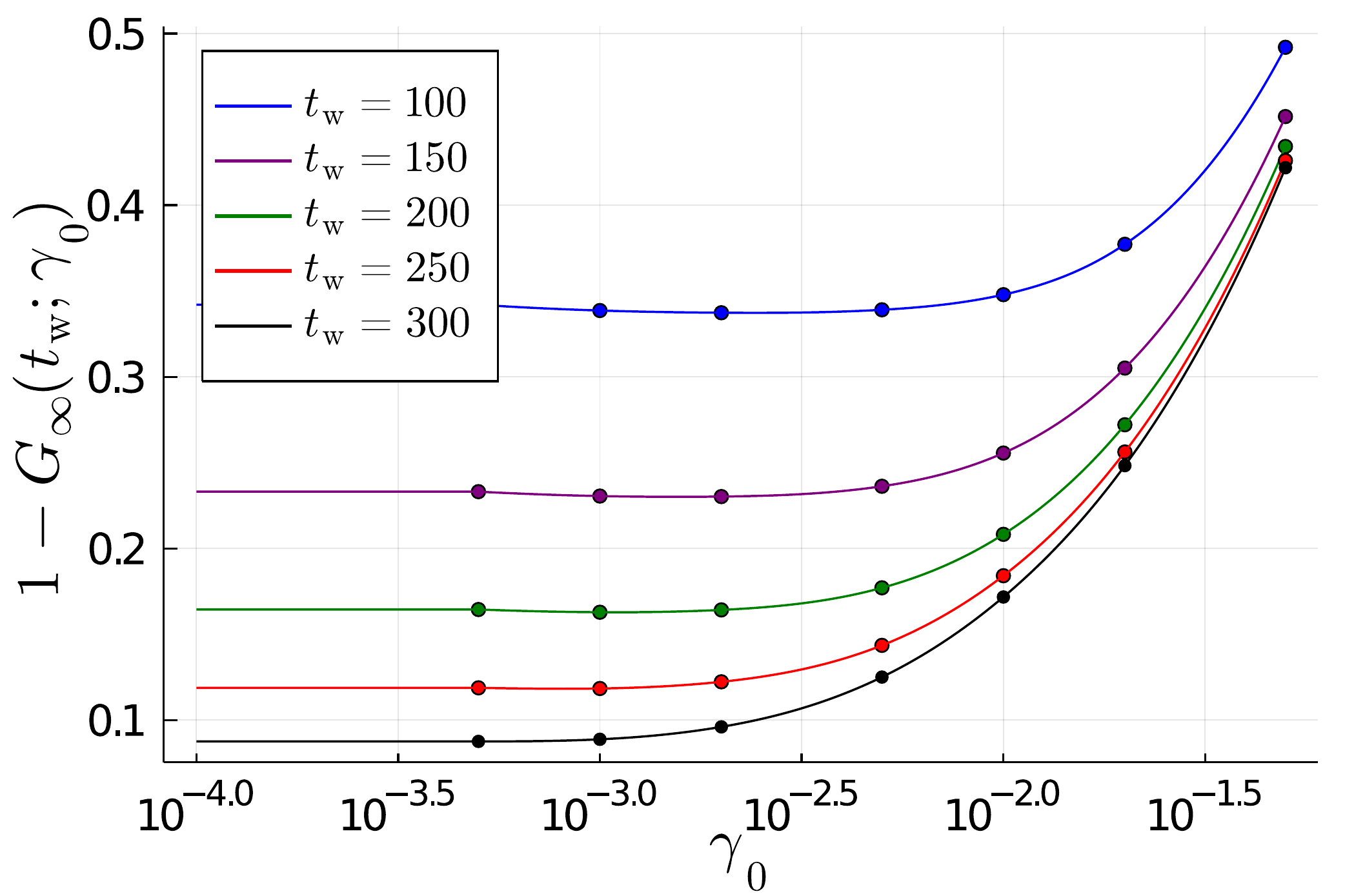}}
\caption{\label{fig:Ginf_spline}Total amount of stress relaxation for different step strain amplitudes and waiting times. The deviation from the linear response plateau values on the left occurs at smaller step strains as $t_{\rm w}$ increases. Lines show cubic spline interpolations as guides to the eye.}
\end{figure}

We show here three supplementary figures accompanying Sec.~\ref{sec:NL}. In Fig.~\ref{fig:Ginf_spline}, we exemplify how we interpolate the measured plateau values to obtain the full $1-G_{\infty}(t_{\rm w};\gamma_0)$ curve for each $t_{\rm w}$. This is then used to determine $\gamma_{\rm max}(t_{\rm w})$, which we recall was defined by setting a $10 \%$ threshold on the relative deviation of this curve with respect to the linear plateau for $\gamma_0\to 0$.

\begin{figure}\center{\includegraphics[width=0.42\textwidth]{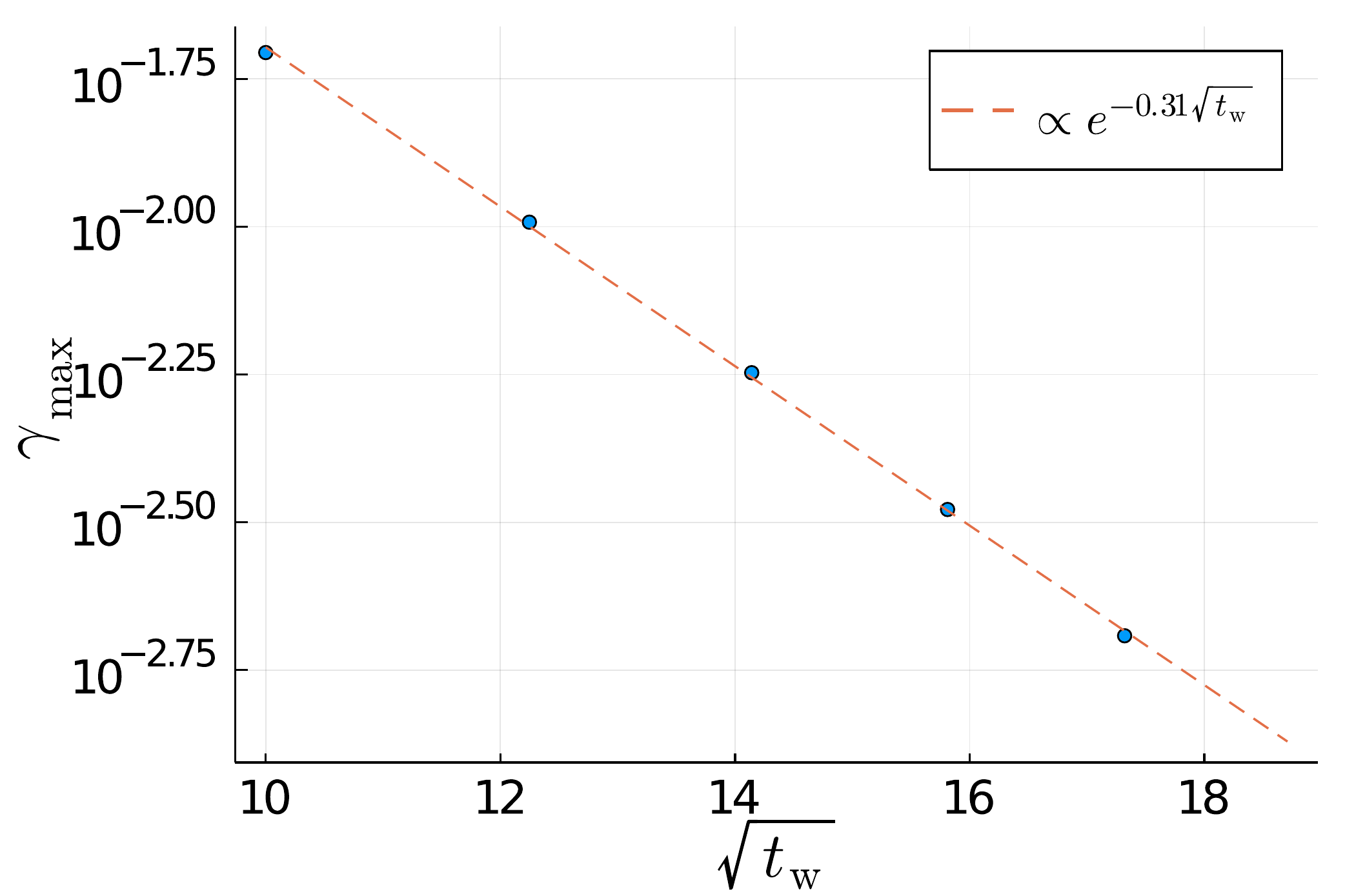}}
\caption{\label{fig:gmax}$\gamma_{\rm max}(t_{\rm w})$, obtained by fixing a 10 $\%$ threshold on the relative deviation of the amount of stress relaxed from the corresponding linear response value for each $t_{\rm w}$. The data agree well with a stretched exponential fit, where the fitted value of the decay constant $\sim 0.31$ is close to the theoretical prediction $(2/3)B_0$, 
which for $B_0=0.44$ would be $0.293$. }
\end{figure}

Fig.~\ref{fig:gmax} shows the $\gamma_{\rm max}(t_{\rm w})$ values determined in the aforementioned fashion. The decay for increasing $t_{\rm w}$, which leads to a narrowing of the linear response regime, roughly follows the prediction $\gamma_{\rm max}(t_{\rm w})\sim {\left(\Gamma(t_{\rm w})\right)}^{2/3}$.

\begin{figure}\center{\includegraphics[width=0.42\textwidth]{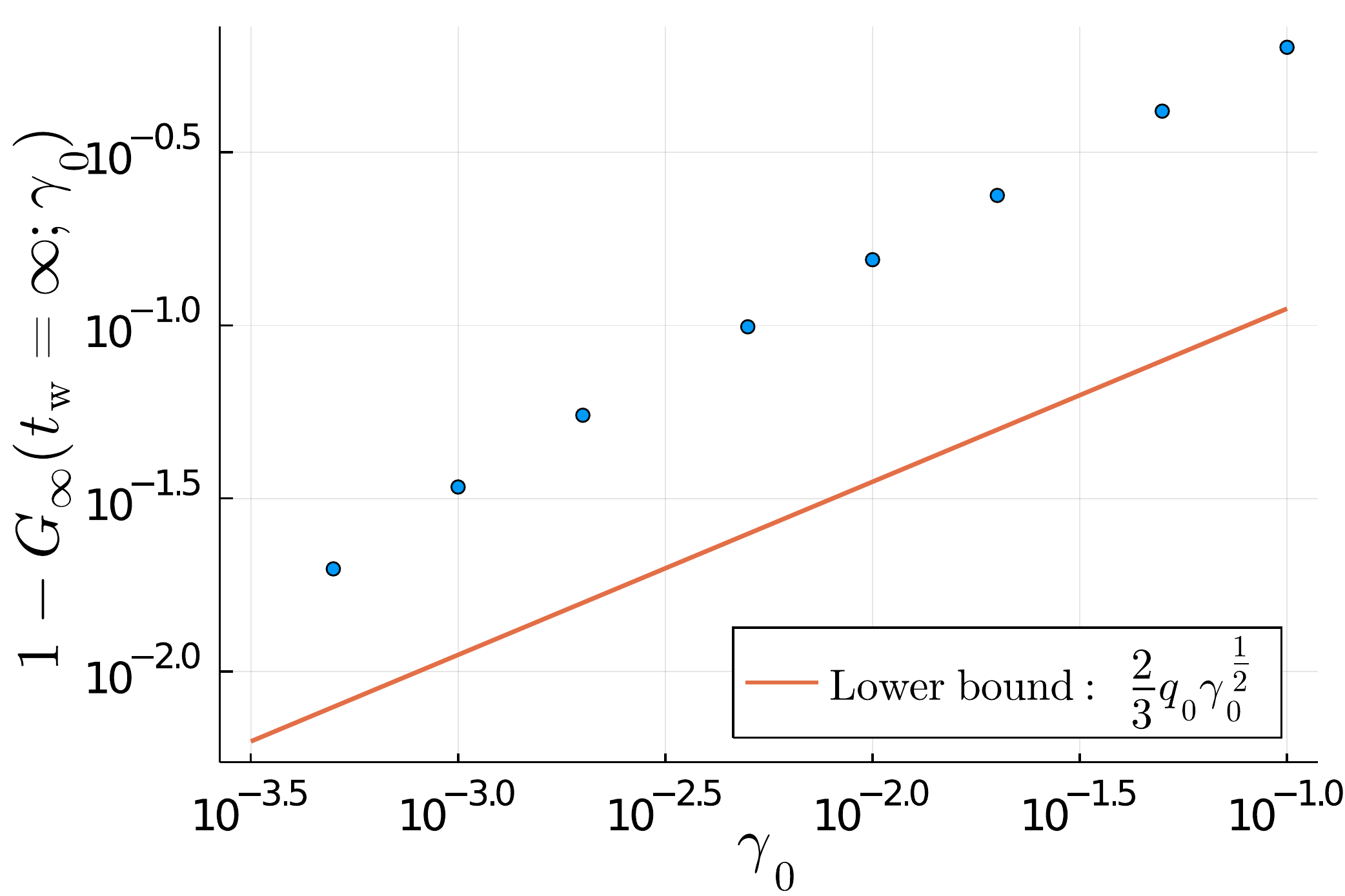}}
\caption{\label{fig:bound_modulus}Amount of stress relaxation in the frozen $t_{\rm w}\rightarrow \infty$ state, for the step strain values considered in Fig.~\ref{fig:tw_100_300}. Solid line shows the lower bound~(\ref{l_bound}) derived in the main text.}
\end{figure}

Finally, Fig.~\ref{fig:bound_modulus} concerns the relaxation for $t_{\rm w}\rightarrow \infty$, which we recall is a purely non-linear feature of the theory that disappears for $\gamma_0 \rightarrow 0$. In Fig.~\ref{fig:bound_modulus} we check that the amount of relaxation in the frozen state, $1-G_{\infty}(t_{\rm w}\rightarrow\infty;\gamma_0)$, indeed lies above the lower bound derived in the paper, and approaches it for decreasing $\gamma_0$.

\bibliography{ms}

\begin{thebibliography}{62}%
\makeatletter
\providecommand \@ifxundefined [1]{%
 \@ifx{#1\undefined}
}%
\providecommand \@ifnum [1]{%
 \ifnum #1\expandafter \@firstoftwo
 \else \expandafter \@secondoftwo
 \fi
}%
\providecommand \@ifx [1]{%
 \ifx #1\expandafter \@firstoftwo
 \else \expandafter \@secondoftwo
 \fi
}%
\providecommand \natexlab [1]{#1}%
\providecommand \enquote  [1]{``#1''}%
\providecommand \bibnamefont  [1]{#1}%
\providecommand \bibfnamefont [1]{#1}%
\providecommand \citenamefont [1]{#1}%
\providecommand \href@noop [0]{\@secondoftwo}%
\providecommand \href [0]{\begingroup \@sanitize@url \@href}%
\providecommand \@href[1]{\@@startlink{#1}\@@href}%
\providecommand \@@href[1]{\endgroup#1\@@endlink}%
\providecommand \@sanitize@url [0]{\catcode `\\12\catcode `\$12\catcode
  `\&12\catcode `\#12\catcode `\^12\catcode `\_12\catcode `\%12\relax}%
\providecommand \@@startlink[1]{}%
\providecommand \@@endlink[0]{}%
\providecommand \url  [0]{\begingroup\@sanitize@url \@url }%
\providecommand \@url [1]{\endgroup\@href {#1}{\urlprefix }}%
\providecommand \urlprefix  [0]{URL }%
\providecommand \Eprint [0]{\href }%
\providecommand \doibase [0]{https://doi.org/}%
\providecommand \selectlanguage [0]{\@gobble}%
\providecommand \bibinfo  [0]{\@secondoftwo}%
\providecommand \bibfield  [0]{\@secondoftwo}%
\providecommand \translation [1]{[#1]}%
\providecommand \BibitemOpen [0]{}%
\providecommand \bibitemStop [0]{}%
\providecommand \bibitemNoStop [0]{.\EOS\space}%
\providecommand \EOS [0]{\spacefactor3000\relax}%
\providecommand \BibitemShut  [1]{\csname bibitem#1\endcsname}%
\let\auto@bib@innerbib\@empty
\bibitem [{\citenamefont {Nicolas}\ \emph {et~al.}(2018)\citenamefont
  {Nicolas}, \citenamefont {Ferrero}, \citenamefont {Martens},\ and\
  \citenamefont {Barrat}}]{nicolas_deformation_2018}%
  \BibitemOpen
  \bibfield  {author} {\bibinfo {author} {\bibfnamefont {A.}~\bibnamefont
  {Nicolas}}, \bibinfo {author} {\bibfnamefont {E.~E.}\ \bibnamefont
  {Ferrero}}, \bibinfo {author} {\bibfnamefont {K.}~\bibnamefont {Martens}},\
  and\ \bibinfo {author} {\bibfnamefont {J.-L.}\ \bibnamefont {Barrat}},\
  }\href {https://doi.org/10.1103/RevModPhys.90.045006} {\bibfield  {journal}
  {\bibinfo  {journal} {Rev. Mod. Phys.}\ }\textbf {\bibinfo {volume} {90}},\
  \bibinfo {pages} {045006} (\bibinfo {year} {2018})}\BibitemShut {NoStop}%
\bibitem [{\citenamefont {Bonn}\ \emph {et~al.}(2017)\citenamefont {Bonn},
  \citenamefont {Denn}, \citenamefont {Berthier}, \citenamefont {Divoux},\ and\
  \citenamefont {Manneville}}]{bonn_yield_2017}%
  \BibitemOpen
  \bibfield  {author} {\bibinfo {author} {\bibfnamefont {D.}~\bibnamefont
  {Bonn}}, \bibinfo {author} {\bibfnamefont {M.~M.}\ \bibnamefont {Denn}},
  \bibinfo {author} {\bibfnamefont {L.}~\bibnamefont {Berthier}}, \bibinfo
  {author} {\bibfnamefont {T.}~\bibnamefont {Divoux}},\ and\ \bibinfo {author}
  {\bibfnamefont {S.}~\bibnamefont {Manneville}},\ }\href
  {https://doi.org/10.1103/RevModPhys.89.035005} {\bibfield  {journal}
  {\bibinfo  {journal} {Rev. Mod. Phys.}\ }\textbf {\bibinfo {volume} {89}},\
  \bibinfo {pages} {035005} (\bibinfo {year} {2017})}\BibitemShut {NoStop}%
\bibitem [{\citenamefont {Berthier}\ and\ \citenamefont
  {Biroli}(2011)}]{berthier_theoretical_2011}%
  \BibitemOpen
  \bibfield  {author} {\bibinfo {author} {\bibfnamefont {L.}~\bibnamefont
  {Berthier}}\ and\ \bibinfo {author} {\bibfnamefont {G.}~\bibnamefont
  {Biroli}},\ }\href {https://doi.org/10.1103/RevModPhys.83.587} {\bibfield
  {journal} {\bibinfo  {journal} {Rev. Mod. Phys.}\ }\textbf {\bibinfo {volume}
  {83}},\ \bibinfo {pages} {587} (\bibinfo {year} {2011})}\BibitemShut
  {NoStop}%
\bibitem [{\citenamefont {Argon}(1979)}]{argon_plastic_1979}%
  \BibitemOpen
  \bibfield  {author} {\bibinfo {author} {\bibfnamefont {A.~S.}\ \bibnamefont
  {Argon}},\ }\href {https://doi.org/10.1016/0001-6160(79)90055-5} {\bibfield
  {journal} {\bibinfo  {journal} {Acta Metallurgica}\ }\textbf {\bibinfo
  {volume} {27}},\ \bibinfo {pages} {47} (\bibinfo {year} {1979})}\BibitemShut
  {NoStop}%
\bibitem [{\citenamefont {Maloney}\ and\ \citenamefont
  {Lemaître}(2006)}]{maloney_amorphous_2006}%
  \BibitemOpen
  \bibfield  {author} {\bibinfo {author} {\bibfnamefont {C.~E.}\ \bibnamefont
  {Maloney}}\ and\ \bibinfo {author} {\bibfnamefont {A.}~\bibnamefont
  {Lemaître}},\ }\href {https://doi.org/10.1103/PhysRevE.74.016118} {\bibfield
   {journal} {\bibinfo  {journal} {Phys. Rev. E}\ }\textbf {\bibinfo {volume}
  {74}},\ \bibinfo {pages} {016118} (\bibinfo {year} {2006})}\BibitemShut
  {NoStop}%
\bibitem [{\citenamefont {Tanguy}\ \emph {et~al.}(2006)\citenamefont {Tanguy},
  \citenamefont {Leonforte},\ and\ \citenamefont
  {Barrat}}]{tanguy_plastic_2006}%
  \BibitemOpen
  \bibfield  {author} {\bibinfo {author} {\bibfnamefont {A.}~\bibnamefont
  {Tanguy}}, \bibinfo {author} {\bibfnamefont {F.}~\bibnamefont {Leonforte}},\
  and\ \bibinfo {author} {\bibfnamefont {J.~L.}\ \bibnamefont {Barrat}},\
  }\href {https://doi.org/10.1140/epje/i2006-10024-2} {\bibfield  {journal}
  {\bibinfo  {journal} {Eur. Phys. J. E}\ }\textbf {\bibinfo {volume} {20}},\
  \bibinfo {pages} {355} (\bibinfo {year} {2006})}\BibitemShut {NoStop}%
\bibitem [{\citenamefont {Puosi}\ \emph {et~al.}(2014)\citenamefont {Puosi},
  \citenamefont {Rottler},\ and\ \citenamefont
  {Barrat}}]{puosi_time-dependent_2014}%
  \BibitemOpen
  \bibfield  {author} {\bibinfo {author} {\bibfnamefont {F.}~\bibnamefont
  {Puosi}}, \bibinfo {author} {\bibfnamefont {J.}~\bibnamefont {Rottler}},\
  and\ \bibinfo {author} {\bibfnamefont {J.-L.}\ \bibnamefont {Barrat}},\
  }\href {https://doi.org/10.1103/PhysRevE.89.042302} {\bibfield  {journal}
  {\bibinfo  {journal} {Phys. Rev. E}\ }\textbf {\bibinfo {volume} {89}},\
  \bibinfo {pages} {042302} (\bibinfo {year} {2014})}\BibitemShut {NoStop}%
\bibitem [{\citenamefont {Lin}\ \emph {et~al.}(2014)\citenamefont {Lin},
  \citenamefont {Lerner}, \citenamefont {Rosso},\ and\ \citenamefont
  {Wyart}}]{lin_scaling_2014}%
  \BibitemOpen
  \bibfield  {author} {\bibinfo {author} {\bibfnamefont {J.}~\bibnamefont
  {Lin}}, \bibinfo {author} {\bibfnamefont {E.}~\bibnamefont {Lerner}},
  \bibinfo {author} {\bibfnamefont {A.}~\bibnamefont {Rosso}},\ and\ \bibinfo
  {author} {\bibfnamefont {M.}~\bibnamefont {Wyart}},\ }\href
  {https://doi.org/10.1073/pnas.1406391111} {\bibfield  {journal} {\bibinfo
  {journal} {Proceedings of the National Academy of Sciences}\ }\textbf
  {\bibinfo {volume} {111}},\ \bibinfo {pages} {14382} (\bibinfo {year}
  {2014})}\BibitemShut {NoStop}%
\bibitem [{\citenamefont {Lin}\ and\ \citenamefont
  {Wyart}(2016)}]{lin_mean-field_2016}%
  \BibitemOpen
  \bibfield  {author} {\bibinfo {author} {\bibfnamefont {J.}~\bibnamefont
  {Lin}}\ and\ \bibinfo {author} {\bibfnamefont {M.}~\bibnamefont {Wyart}},\
  }\href {https://doi.org/10.1103/PhysRevX.6.011005} {\bibfield  {journal}
  {\bibinfo  {journal} {Phys. Rev. X}\ }\textbf {\bibinfo {volume} {6}},\
  \bibinfo {pages} {011005} (\bibinfo {year} {2016})}\BibitemShut {NoStop}%
\bibitem [{\citenamefont {Liu}\ \emph {et~al.}(2016)\citenamefont {Liu},
  \citenamefont {Ferrero}, \citenamefont {Puosi}, \citenamefont {Barrat},\ and\
  \citenamefont {Martens}}]{liu_driving_2016}%
  \BibitemOpen
  \bibfield  {author} {\bibinfo {author} {\bibfnamefont {C.}~\bibnamefont
  {Liu}}, \bibinfo {author} {\bibfnamefont {E.~E.}\ \bibnamefont {Ferrero}},
  \bibinfo {author} {\bibfnamefont {F.}~\bibnamefont {Puosi}}, \bibinfo
  {author} {\bibfnamefont {J.-L.}\ \bibnamefont {Barrat}},\ and\ \bibinfo
  {author} {\bibfnamefont {K.}~\bibnamefont {Martens}},\ }\href
  {https://link.aps.org/doi/10.1103/PhysRevLett.116.065501} {\bibfield
  {journal} {\bibinfo  {journal} {Phys. Rev. Lett.}\ }\textbf {\bibinfo
  {volume} {116}},\ \bibinfo {pages} {065501} (\bibinfo {year}
  {2016})}\BibitemShut {NoStop}%
\bibitem [{\citenamefont {Fernández~Aguirre}\ and\ \citenamefont
  {Jagla}(2018)}]{fernandez_aguirre_critical_2018}%
  \BibitemOpen
  \bibfield  {author} {\bibinfo {author} {\bibfnamefont {I.}~\bibnamefont
  {Fernández~Aguirre}}\ and\ \bibinfo {author} {\bibfnamefont {E.~A.}\
  \bibnamefont {Jagla}},\ }\href {https://doi.org/10.1103/PhysRevE.98.013002}
  {\bibfield  {journal} {\bibinfo  {journal} {Phys. Rev. E}\ }\textbf {\bibinfo
  {volume} {98}},\ \bibinfo {pages} {013002} (\bibinfo {year}
  {2018})}\BibitemShut {NoStop}%
\bibitem [{\citenamefont {Ferrero}\ and\ \citenamefont
  {Jagla}(2019)}]{ferrero_criticality_2019}%
  \BibitemOpen
  \bibfield  {author} {\bibinfo {author} {\bibfnamefont {E.~E.}\ \bibnamefont
  {Ferrero}}\ and\ \bibinfo {author} {\bibfnamefont {E.~A.}\ \bibnamefont
  {Jagla}},\ }\href {https://doi.org/10.1039/C9SM01073D} {\bibfield  {journal}
  {\bibinfo  {journal} {Soft Matter}\ }\textbf {\bibinfo {volume} {15}},\
  \bibinfo {pages} {9041} (\bibinfo {year} {2019})}\BibitemShut {NoStop}%
\bibitem [{\citenamefont {Ferrero}\ and\ \citenamefont
  {Jagla}(2021)}]{ferrero_properties_2021}%
  \BibitemOpen
  \bibfield  {author} {\bibinfo {author} {\bibfnamefont {E.~E.}\ \bibnamefont
  {Ferrero}}\ and\ \bibinfo {author} {\bibfnamefont {E.~A.}\ \bibnamefont
  {Jagla}},\ }\href {https://doi.org/10.1088/1361-648X/abd73a} {\bibfield
  {journal} {\bibinfo  {journal} {J. Phys.: Condens. Matter}\ }\textbf
  {\bibinfo {volume} {33}},\ \bibinfo {pages} {124001} (\bibinfo {year}
  {2021})}\BibitemShut {NoStop}%
\bibitem [{\citenamefont {Ferrero}\ \emph {et~al.}(2021)\citenamefont
  {Ferrero}, \citenamefont {Kolton},\ and\ \citenamefont
  {Jagla}}]{ferrero_yielding_2021}%
  \BibitemOpen
  \bibfield  {author} {\bibinfo {author} {\bibfnamefont {E.~E.}\ \bibnamefont
  {Ferrero}}, \bibinfo {author} {\bibfnamefont {A.~B.}\ \bibnamefont
  {Kolton}},\ and\ \bibinfo {author} {\bibfnamefont {E.~A.}\ \bibnamefont
  {Jagla}},\ }\href {https://doi.org/10.1103/PhysRevMaterials.5.115602}
  {\bibfield  {journal} {\bibinfo  {journal} {Phys. Rev. Materials}\ }\textbf
  {\bibinfo {volume} {5}},\ \bibinfo {pages} {115602} (\bibinfo {year}
  {2021})}\BibitemShut {NoStop}%
\bibitem [{\citenamefont {Barlow}\ \emph {et~al.}(2020)\citenamefont {Barlow},
  \citenamefont {Cochran},\ and\ \citenamefont
  {Fielding}}]{barlow_ductile_2020}%
  \BibitemOpen
  \bibfield  {author} {\bibinfo {author} {\bibfnamefont {H.~J.}\ \bibnamefont
  {Barlow}}, \bibinfo {author} {\bibfnamefont {J.~O.}\ \bibnamefont
  {Cochran}},\ and\ \bibinfo {author} {\bibfnamefont {S.~M.}\ \bibnamefont
  {Fielding}},\ }\href {https://doi.org/10.1103/PhysRevLett.125.168003}
  {\bibfield  {journal} {\bibinfo  {journal} {Phys. Rev. Lett.}\ }\textbf
  {\bibinfo {volume} {125}},\ \bibinfo {pages} {168003} (\bibinfo {year}
  {2020})}\BibitemShut {NoStop}%
\bibitem [{\citenamefont {Parley}\ \emph {et~al.}(2022)\citenamefont {Parley},
  \citenamefont {Sastry},\ and\ \citenamefont {Sollich}}]{parley_mean_2022}%
  \BibitemOpen
  \bibfield  {author} {\bibinfo {author} {\bibfnamefont {J.~T.}\ \bibnamefont
  {Parley}}, \bibinfo {author} {\bibfnamefont {S.}~\bibnamefont {Sastry}},\
  and\ \bibinfo {author} {\bibfnamefont {P.}~\bibnamefont {Sollich}},\ }\href
  {http://arxiv.org/abs/2112.11578} {\bibfield  {journal} {\bibinfo  {journal}
  {arXiv:2112.11578 [cond-mat]}\ } (\bibinfo {year} {2022})}\BibitemShut
  {NoStop}%
\bibitem [{\citenamefont {Chacko}\ \emph {et~al.}(2019)\citenamefont {Chacko},
  \citenamefont {Sollich},\ and\ \citenamefont {Fielding}}]{chacko_slow_2019}%
  \BibitemOpen
  \bibfield  {author} {\bibinfo {author} {\bibfnamefont {R.}~\bibnamefont
  {Chacko}}, \bibinfo {author} {\bibfnamefont {P.}~\bibnamefont {Sollich}},\
  and\ \bibinfo {author} {\bibfnamefont {S.}~\bibnamefont {Fielding}},\ }\href
  {https://doi.org/10.1103/PhysRevLett.123.108001} {\bibfield  {journal}
  {\bibinfo  {journal} {Phys. Rev. Lett.}\ }\textbf {\bibinfo {volume} {123}},\
  \bibinfo {pages} {108001} (\bibinfo {year} {2019})}\BibitemShut {NoStop}%
\bibitem [{\citenamefont {Nishikawa}\ \emph {et~al.}(2022)\citenamefont
  {Nishikawa}, \citenamefont {Ozawa}, \citenamefont {Ikeda}, \citenamefont
  {Chaudhuri},\ and\ \citenamefont {Berthier}}]{nishikawa_relaxation_2022}%
  \BibitemOpen
  \bibfield  {author} {\bibinfo {author} {\bibfnamefont {Y.}~\bibnamefont
  {Nishikawa}}, \bibinfo {author} {\bibfnamefont {M.}~\bibnamefont {Ozawa}},
  \bibinfo {author} {\bibfnamefont {A.}~\bibnamefont {Ikeda}}, \bibinfo
  {author} {\bibfnamefont {P.}~\bibnamefont {Chaudhuri}},\ and\ \bibinfo
  {author} {\bibfnamefont {L.}~\bibnamefont {Berthier}},\ }\href
  {https://doi.org/10.1103/PhysRevX.12.021001} {\bibfield  {journal} {\bibinfo
  {journal} {Phys. Rev. X}\ }\textbf {\bibinfo {volume} {12}},\ \bibinfo
  {pages} {021001} (\bibinfo {year} {2022})}\BibitemShut {NoStop}%
\bibitem [{\citenamefont {Mandal}\ and\ \citenamefont
  {Sollich}(2020)}]{mandal_multiple_2020}%
  \BibitemOpen
  \bibfield  {author} {\bibinfo {author} {\bibfnamefont {R.}~\bibnamefont
  {Mandal}}\ and\ \bibinfo {author} {\bibfnamefont {P.}~\bibnamefont
  {Sollich}},\ }\href {https://doi.org/10.1103/PhysRevLett.125.218001}
  {\bibfield  {journal} {\bibinfo  {journal} {Phys. Rev. Lett.}\ }\textbf
  {\bibinfo {volume} {125}},\ \bibinfo {pages} {218001} (\bibinfo {year}
  {2020})}\BibitemShut {NoStop}%
\bibitem [{Note1()}]{Note1}%
  \BibitemOpen
  \bibinfo {note} {Athermal gradient descent dynamics has also been studied
  recently below and close to jamming, both in particle simulations~\cite
  {nishikawa_relaxation_2021,olsson_relaxation_2022} and from the perspective
  of dynamical mean field theory~\cite {manacorda_gradient_2022}}\BibitemShut
  {NoStop}%
\bibitem [{\citenamefont {Hunter}\ and\ \citenamefont
  {Weeks}(2012)}]{hunter_physics_2012}%
  \BibitemOpen
  \bibfield  {author} {\bibinfo {author} {\bibfnamefont {G.~L.}\ \bibnamefont
  {Hunter}}\ and\ \bibinfo {author} {\bibfnamefont {E.~R.}\ \bibnamefont
  {Weeks}},\ }\href {https://doi.org/10.1088/0034-4885/75/6/066501} {\bibfield
  {journal} {\bibinfo  {journal} {Rep. Prog. Phys.}\ }\textbf {\bibinfo
  {volume} {75}},\ \bibinfo {pages} {066501} (\bibinfo {year}
  {2012})}\BibitemShut {NoStop}%
\bibitem [{\citenamefont {Cloitre}\ \emph {et~al.}(2000)\citenamefont
  {Cloitre}, \citenamefont {Borrega},\ and\ \citenamefont
  {Leibler}}]{cloitre_rheological_2000}%
  \BibitemOpen
  \bibfield  {author} {\bibinfo {author} {\bibfnamefont {M.}~\bibnamefont
  {Cloitre}}, \bibinfo {author} {\bibfnamefont {R.}~\bibnamefont {Borrega}},\
  and\ \bibinfo {author} {\bibfnamefont {L.}~\bibnamefont {Leibler}},\ }\href
  {https://doi.org/10.1103/PhysRevLett.85.4819} {\bibfield  {journal} {\bibinfo
   {journal} {Physical Review Letters}\ }\textbf {\bibinfo {volume} {85}},\
  \bibinfo {pages} {4819} (\bibinfo {year} {2000})}\BibitemShut {NoStop}%
\bibitem [{\citenamefont {Cugliandolo}\ \emph {et~al.}(1994)\citenamefont
  {Cugliandolo}, \citenamefont {Kurchan},\ and\ \citenamefont
  {Ritort}}]{cugliandolo_evidence_1994}%
  \BibitemOpen
  \bibfield  {author} {\bibinfo {author} {\bibfnamefont {L.~F.}\ \bibnamefont
  {Cugliandolo}}, \bibinfo {author} {\bibfnamefont {J.}~\bibnamefont
  {Kurchan}},\ and\ \bibinfo {author} {\bibfnamefont {F.}~\bibnamefont
  {Ritort}},\ }\href {https://doi.org/10.1103/PhysRevB.49.6331} {\bibfield
  {journal} {\bibinfo  {journal} {Phys. Rev. B}\ }\textbf {\bibinfo {volume}
  {49}},\ \bibinfo {pages} {6331} (\bibinfo {year} {1994})}\BibitemShut
  {NoStop}%
\bibitem [{\citenamefont {Bouchaud}(1992)}]{bouchaud_weak_1992}%
  \BibitemOpen
  \bibfield  {author} {\bibinfo {author} {\bibfnamefont {J.~P.}\ \bibnamefont
  {Bouchaud}},\ }\href {https://doi.org/10.1051/jp1:1992238} {\bibfield
  {journal} {\bibinfo  {journal} {J. Phys. I France}\ }\textbf {\bibinfo
  {volume} {2}},\ \bibinfo {pages} {1705} (\bibinfo {year} {1992})}\BibitemShut
  {NoStop}%
\bibitem [{\citenamefont {Boettcher}\ \emph {et~al.}(2018)\citenamefont
  {Boettcher}, \citenamefont {Robe},\ and\ \citenamefont
  {Sibani}}]{boettcher_aging_2018}%
  \BibitemOpen
  \bibfield  {author} {\bibinfo {author} {\bibfnamefont {S.}~\bibnamefont
  {Boettcher}}, \bibinfo {author} {\bibfnamefont {D.~M.}\ \bibnamefont
  {Robe}},\ and\ \bibinfo {author} {\bibfnamefont {P.}~\bibnamefont {Sibani}},\
  }\href {https://doi.org/10.1103/PhysRevE.98.020602} {\bibfield  {journal}
  {\bibinfo  {journal} {Phys. Rev. E}\ }\textbf {\bibinfo {volume} {98}},\
  \bibinfo {pages} {020602} (\bibinfo {year} {2018})}\BibitemShut {NoStop}%
\bibitem [{\citenamefont {Parley}\ \emph {et~al.}(2020)\citenamefont {Parley},
  \citenamefont {Fielding},\ and\ \citenamefont {Sollich}}]{parley_aging_2020}%
  \BibitemOpen
  \bibfield  {author} {\bibinfo {author} {\bibfnamefont {J.~T.}\ \bibnamefont
  {Parley}}, \bibinfo {author} {\bibfnamefont {S.~M.}\ \bibnamefont
  {Fielding}},\ and\ \bibinfo {author} {\bibfnamefont {P.}~\bibnamefont
  {Sollich}},\ }\href {https://doi.org/10.1063/5.0033196} {\bibfield  {journal}
  {\bibinfo  {journal} {Physics of Fluids}\ }\textbf {\bibinfo {volume} {32}},\
  \bibinfo {pages} {127104} (\bibinfo {year} {2020})}\BibitemShut {NoStop}%
\bibitem [{\citenamefont {Picard}\ \emph {et~al.}(2004)\citenamefont {Picard},
  \citenamefont {Ajdari}, \citenamefont {Lequeux},\ and\ \citenamefont
  {Bocquet}}]{picard_elastic_2004}%
  \BibitemOpen
  \bibfield  {author} {\bibinfo {author} {\bibfnamefont {G.}~\bibnamefont
  {Picard}}, \bibinfo {author} {\bibfnamefont {A.}~\bibnamefont {Ajdari}},
  \bibinfo {author} {\bibfnamefont {F.}~\bibnamefont {Lequeux}},\ and\ \bibinfo
  {author} {\bibfnamefont {L.}~\bibnamefont {Bocquet}},\ }\href
  {https://doi.org/10.1140/epje/i2004-10054-8} {\bibfield  {journal} {\bibinfo
  {journal} {The European Physical Journal E}\ }\textbf {\bibinfo {volume}
  {15}},\ \bibinfo {pages} {371} (\bibinfo {year} {2004})}\BibitemShut
  {NoStop}%
\bibitem [{Note2()}]{Note2}%
  \BibitemOpen
  \bibinfo {note} {In taking the limit $\mu \rightarrow 2$, one scales $A$ to
  zero as $A\sim 2-\mu $ so that the second moment of the jump distribution
  $\alpha _{\protect \rm eff}=A/(2-\mu )\left (2A/\mu \right )^{2/\mu -1}$ goes
  to a finite limiting value $\alpha _{\protect \rm {HL}}$ corresponding to the
  coupling parameter of the HL model.}\BibitemShut {Stop}%
\bibitem [{\citenamefont {Agoritsas}\ \emph {et~al.}(2015)\citenamefont
  {Agoritsas}, \citenamefont {Bertin}, \citenamefont {Martens},\ and\
  \citenamefont {Barrat}}]{agoritsas_relevance_2015}%
  \BibitemOpen
  \bibfield  {author} {\bibinfo {author} {\bibfnamefont {E.}~\bibnamefont
  {Agoritsas}}, \bibinfo {author} {\bibfnamefont {E.}~\bibnamefont {Bertin}},
  \bibinfo {author} {\bibfnamefont {K.}~\bibnamefont {Martens}},\ and\ \bibinfo
  {author} {\bibfnamefont {J.-L.}\ \bibnamefont {Barrat}},\ }\href
  {https://doi.org/10.1140/epje/i2015-15071-x} {\bibfield  {journal} {\bibinfo
  {journal} {Eur. Phys. J. E}\ }\textbf {\bibinfo {volume} {38}},\ \bibinfo
  {pages} {71} (\bibinfo {year} {2015})}\BibitemShut {NoStop}%
\bibitem [{\citenamefont {Chacko}\ \emph {et~al.}(2021)\citenamefont {Chacko},
  \citenamefont {Landes}, \citenamefont {Biroli}, \citenamefont {Dauchot},
  \citenamefont {Liu},\ and\ \citenamefont
  {Reichman}}]{chacko_elastoplasticity_2021}%
  \BibitemOpen
  \bibfield  {author} {\bibinfo {author} {\bibfnamefont {R.~N.}\ \bibnamefont
  {Chacko}}, \bibinfo {author} {\bibfnamefont {F.~P.}\ \bibnamefont {Landes}},
  \bibinfo {author} {\bibfnamefont {G.}~\bibnamefont {Biroli}}, \bibinfo
  {author} {\bibfnamefont {O.}~\bibnamefont {Dauchot}}, \bibinfo {author}
  {\bibfnamefont {A.~J.}\ \bibnamefont {Liu}},\ and\ \bibinfo {author}
  {\bibfnamefont {D.~R.}\ \bibnamefont {Reichman}},\ }\href
  {https://doi.org/10.1103/PhysRevLett.127.048002} {\bibfield  {journal}
  {\bibinfo  {journal} {Phys. Rev. Lett.}\ }\textbf {\bibinfo {volume} {127}},\
  \bibinfo {pages} {048002} (\bibinfo {year} {2021})}\BibitemShut {NoStop}%
\bibitem [{\citenamefont {Popović}\ \emph {et~al.}(2021)\citenamefont
  {Popović}, \citenamefont {de~Geus}, \citenamefont {Ji},\ and\ \citenamefont
  {Wyart}}]{popovic_thermally_2021}%
  \BibitemOpen
  \bibfield  {author} {\bibinfo {author} {\bibfnamefont {M.}~\bibnamefont
  {Popović}}, \bibinfo {author} {\bibfnamefont {T.~W.~J.}\ \bibnamefont
  {de~Geus}}, \bibinfo {author} {\bibfnamefont {W.}~\bibnamefont {Ji}},\ and\
  \bibinfo {author} {\bibfnamefont {M.}~\bibnamefont {Wyart}},\ }\href
  {https://doi.org/10.1103/PhysRevE.104.025010} {\bibfield  {journal} {\bibinfo
   {journal} {Phys. Rev. E}\ }\textbf {\bibinfo {volume} {104}},\ \bibinfo
  {pages} {025010} (\bibinfo {year} {2021})}\BibitemShut {NoStop}%
\bibitem [{\citenamefont {Shang}\ \emph {et~al.}(2020)\citenamefont {Shang},
  \citenamefont {Guan},\ and\ \citenamefont {Barrat}}]{shang_elastic_2020}%
  \BibitemOpen
  \bibfield  {author} {\bibinfo {author} {\bibfnamefont {B.}~\bibnamefont
  {Shang}}, \bibinfo {author} {\bibfnamefont {P.}~\bibnamefont {Guan}},\ and\
  \bibinfo {author} {\bibfnamefont {J.-L.}\ \bibnamefont {Barrat}},\ }\href
  {https://doi.org/10.1073/pnas.1915070117} {\bibfield  {journal} {\bibinfo
  {journal} {PNAS}\ }\textbf {\bibinfo {volume} {117}},\ \bibinfo {pages} {86}
  (\bibinfo {year} {2020})}\BibitemShut {NoStop}%
\bibitem [{\citenamefont {Sollich}\ \emph {et~al.}(2017)\citenamefont
  {Sollich}, \citenamefont {Olivier},\ and\ \citenamefont
  {Bresch}}]{sollich_aging_2017}%
  \BibitemOpen
  \bibfield  {author} {\bibinfo {author} {\bibfnamefont {P.}~\bibnamefont
  {Sollich}}, \bibinfo {author} {\bibfnamefont {J.}~\bibnamefont {Olivier}},\
  and\ \bibinfo {author} {\bibfnamefont {D.}~\bibnamefont {Bresch}},\ }\href
  {https://doi.org/10.1088/1751-8121/aa6261} {\bibfield  {journal} {\bibinfo
  {journal} {J. Phys. A: Math. Theor.}\ }\textbf {\bibinfo {volume} {50}},\
  \bibinfo {pages} {165002} (\bibinfo {year} {2017})}\BibitemShut {NoStop}%
\bibitem [{\citenamefont {Fielding}\ \emph {et~al.}(2000)\citenamefont
  {Fielding}, \citenamefont {Sollich},\ and\ \citenamefont
  {Cates}}]{fielding_ageing_2000}%
  \BibitemOpen
  \bibfield  {author} {\bibinfo {author} {\bibfnamefont {S.~M.}\ \bibnamefont
  {Fielding}}, \bibinfo {author} {\bibfnamefont {P.}~\bibnamefont {Sollich}},\
  and\ \bibinfo {author} {\bibfnamefont {M.~E.}\ \bibnamefont {Cates}},\ }\href
  {https://doi.org/10.1122/1.551088} {\bibfield  {journal} {\bibinfo  {journal}
  {Journal of Rheology}\ }\textbf {\bibinfo {volume} {44}},\ \bibinfo {pages}
  {323} (\bibinfo {year} {2000})}\BibitemShut {NoStop}%
\bibitem [{\citenamefont {Purnomo}\ \emph {et~al.}(2008)\citenamefont
  {Purnomo}, \citenamefont {van~den Ende}, \citenamefont {Vanapalli},\ and\
  \citenamefont {Mugele}}]{purnomo_glass_2008}%
  \BibitemOpen
  \bibfield  {author} {\bibinfo {author} {\bibfnamefont {E.~H.}\ \bibnamefont
  {Purnomo}}, \bibinfo {author} {\bibfnamefont {D.}~\bibnamefont {van~den
  Ende}}, \bibinfo {author} {\bibfnamefont {S.~A.}\ \bibnamefont {Vanapalli}},\
  and\ \bibinfo {author} {\bibfnamefont {F.}~\bibnamefont {Mugele}},\ }\href
  {https://doi.org/10.1103/PhysRevLett.101.238301} {\bibfield  {journal}
  {\bibinfo  {journal} {Phys. Rev. Lett.}\ }\textbf {\bibinfo {volume} {101}},\
  \bibinfo {pages} {238301} (\bibinfo {year} {2008})}\BibitemShut {NoStop}%
\bibitem [{\citenamefont {Purnomo}\ \emph {et~al.}(2006)\citenamefont
  {Purnomo}, \citenamefont {Ende}, \citenamefont {Mellema},\ and\ \citenamefont
  {Mugele}}]{purnomo_linear_2006}%
  \BibitemOpen
  \bibfield  {author} {\bibinfo {author} {\bibfnamefont {E.~H.}\ \bibnamefont
  {Purnomo}}, \bibinfo {author} {\bibfnamefont {D.~v.~d.}\ \bibnamefont
  {Ende}}, \bibinfo {author} {\bibfnamefont {J.}~\bibnamefont {Mellema}},\ and\
  \bibinfo {author} {\bibfnamefont {F.}~\bibnamefont {Mugele}},\ }\href
  {https://doi.org/10.1209/epl/i2006-10234-2} {\bibfield  {journal} {\bibinfo
  {journal} {Europhys. Lett.}\ }\textbf {\bibinfo {volume} {76}},\ \bibinfo
  {pages} {74} (\bibinfo {year} {2006})}\BibitemShut {NoStop}%
\bibitem [{\citenamefont {Purnomo}\ \emph {et~al.}(2007)\citenamefont
  {Purnomo}, \citenamefont {van~den Ende}, \citenamefont {Mellema},\ and\
  \citenamefont {Mugele}}]{purnomo_rheological_2007}%
  \BibitemOpen
  \bibfield  {author} {\bibinfo {author} {\bibfnamefont {E.~H.}\ \bibnamefont
  {Purnomo}}, \bibinfo {author} {\bibfnamefont {D.}~\bibnamefont {van~den
  Ende}}, \bibinfo {author} {\bibfnamefont {J.}~\bibnamefont {Mellema}},\ and\
  \bibinfo {author} {\bibfnamefont {F.}~\bibnamefont {Mugele}},\ }\href
  {https://doi.org/10.1103/PhysRevE.76.021404} {\bibfield  {journal} {\bibinfo
  {journal} {Phys. Rev. E}\ }\textbf {\bibinfo {volume} {76}},\ \bibinfo
  {pages} {021404} (\bibinfo {year} {2007})}\BibitemShut {NoStop}%
\bibitem [{Note3()}]{Note3}%
  \BibitemOpen
  \bibinfo {note} {We note for clarity that this special choice of phase is
  made solely to simplify the expression (\ref {Gav_time}), and does not in
  itself contribute to reducing the oscillations in $G ^{*}(\omega
  ,t,t_{\protect \rm w})$. The reduction of oscillations is accomplished by the
  averaging, and is independent of the choice of phase $\phi $.}\BibitemShut
  {Stop}%
\bibitem [{\citenamefont {Zoia}\ \emph {et~al.}(2009)\citenamefont {Zoia},
  \citenamefont {Rosso},\ and\ \citenamefont
  {Majumdar}}]{zoia_asymptotic_2009}%
  \BibitemOpen
  \bibfield  {author} {\bibinfo {author} {\bibfnamefont {A.}~\bibnamefont
  {Zoia}}, \bibinfo {author} {\bibfnamefont {A.}~\bibnamefont {Rosso}},\ and\
  \bibinfo {author} {\bibfnamefont {S.~N.}\ \bibnamefont {Majumdar}},\ }\href
  {https://doi.org/10.1103/PhysRevLett.102.120602} {\bibfield  {journal}
  {\bibinfo  {journal} {Phys. Rev. Lett.}\ }\textbf {\bibinfo {volume} {102}},\
  \bibinfo {pages} {120602} (\bibinfo {year} {2009})}\BibitemShut {NoStop}%
\bibitem [{\citenamefont {Andersen}(1954)}]{andersen_fluctuations_1954}%
  \BibitemOpen
  \bibfield  {author} {\bibinfo {author} {\bibfnamefont {E.~S.}\ \bibnamefont
  {Andersen}},\ }\href {https://doi.org/10.7146/math.scand.a-10407} {\bibfield
  {journal} {\bibinfo  {journal} {MATHEMATICA SCANDINAVICA}\ }\textbf {\bibinfo
  {volume} {2}},\ \bibinfo {pages} {194} (\bibinfo {year} {1954})}\BibitemShut
  {NoStop}%
\bibitem [{\citenamefont {Bray}\ \emph {et~al.}(2013)\citenamefont {Bray},
  \citenamefont {Majumdar},\ and\ \citenamefont
  {Schehr}}]{bray_persistence_2013}%
  \BibitemOpen
  \bibfield  {author} {\bibinfo {author} {\bibfnamefont {A.~J.}\ \bibnamefont
  {Bray}}, \bibinfo {author} {\bibfnamefont {S.~N.}\ \bibnamefont {Majumdar}},\
  and\ \bibinfo {author} {\bibfnamefont {G.}~\bibnamefont {Schehr}},\ }\href
  {https://doi.org/10.1080/00018732.2013.803819} {\bibfield  {journal}
  {\bibinfo  {journal} {Advances in Physics}\ }\textbf {\bibinfo {volume}
  {62}},\ \bibinfo {pages} {225} (\bibinfo {year} {2013})}\BibitemShut
  {NoStop}%
\bibitem [{Note4()}]{Note4}%
  \BibitemOpen
  \bibinfo {note} {Here and in what follows we use, as in \cite
  {parley_aging_2020}, the steady state with $\Gamma =0.134$ as initial
  distribution for the unperturbed aging dynamics.}\BibitemShut {Stop}%
\bibitem [{Note5()}]{Note5}%
  \BibitemOpen
  \bibinfo {note} {We note that, as discussed in~\cite {parley_aging_2020}, in
  the unperturbed numerics the power law asymptote of $\Gamma (t)$ is
  eventually cut off exponentially by the fact that the required discretization
  of the $\sigma $-axis can no longer resolve the boundary layer.}\BibitemShut
  {Stop}%
\bibitem [{Note6()}]{Note6}%
  \BibitemOpen
  \bibinfo {note} {As done above for $\mu =1.7$, we extrapolate the asymptote
  of $\Gamma (t)$ to later times than we had access to in the unperturbed
  numerics, due to the same discretisation limit described there (the boundary
  layer becoming even harder to resolve for $\mu =1$).}\BibitemShut {Stop}%
\bibitem [{\citenamefont {Durian}(1997)}]{durian_bubble-scale_1997}%
  \BibitemOpen
  \bibfield  {author} {\bibinfo {author} {\bibfnamefont {D.~J.}\ \bibnamefont
  {Durian}},\ }\href {https://doi.org/10.1103/PhysRevE.55.1739} {\bibfield
  {journal} {\bibinfo  {journal} {Phys. Rev. E}\ }\textbf {\bibinfo {volume}
  {55}},\ \bibinfo {pages} {1739} (\bibinfo {year} {1997})}\BibitemShut
  {NoStop}%
\bibitem [{\citenamefont {Cugliandolo}\ and\ \citenamefont
  {Kurchan}(1994)}]{cugliandolo_out--equilibrium_1994}%
  \BibitemOpen
  \bibfield  {author} {\bibinfo {author} {\bibfnamefont {L.~F.}\ \bibnamefont
  {Cugliandolo}}\ and\ \bibinfo {author} {\bibfnamefont {J.}~\bibnamefont
  {Kurchan}},\ }\href {https://doi.org/10.1088/0305-4470/27/17/011} {\bibfield
  {journal} {\bibinfo  {journal} {J. Phys. A: Math. Gen.}\ }\textbf {\bibinfo
  {volume} {27}},\ \bibinfo {pages} {5749} (\bibinfo {year}
  {1994})}\BibitemShut {NoStop}%
\bibitem [{\citenamefont {Sollich}\ \emph {et~al.}(1997)\citenamefont
  {Sollich}, \citenamefont {Lequeux}, \citenamefont {Hébraud},\ and\
  \citenamefont {Cates}}]{sollich_rheology_1997}%
  \BibitemOpen
  \bibfield  {author} {\bibinfo {author} {\bibfnamefont {P.}~\bibnamefont
  {Sollich}}, \bibinfo {author} {\bibfnamefont {F.}~\bibnamefont {Lequeux}},
  \bibinfo {author} {\bibfnamefont {P.}~\bibnamefont {Hébraud}},\ and\
  \bibinfo {author} {\bibfnamefont {M.~E.}\ \bibnamefont {Cates}},\ }\href
  {https://doi.org/10.1103/PhysRevLett.78.2020} {\bibfield  {journal} {\bibinfo
   {journal} {Phys. Rev. Lett.}\ }\textbf {\bibinfo {volume} {78}},\ \bibinfo
  {pages} {2020} (\bibinfo {year} {1997})}\BibitemShut {NoStop}%
\bibitem [{\citenamefont {Sollich}(1998)}]{sollich_rheological_1998}%
  \BibitemOpen
  \bibfield  {author} {\bibinfo {author} {\bibfnamefont {P.}~\bibnamefont
  {Sollich}},\ }\href {https://doi.org/10.1103/PhysRevE.58.738} {\bibfield
  {journal} {\bibinfo  {journal} {Phys. Rev. E}\ }\textbf {\bibinfo {volume}
  {58}},\ \bibinfo {pages} {738} (\bibinfo {year} {1998})}\BibitemShut
  {NoStop}%
\bibitem [{noa()}]{noauthor_notitle_nodate}%
  \BibitemOpen
  \href@noop {} {\bibinfo  {journal} {P. Sollich, in CECAM Workshop (ACAM,
  Dublin, Ireland, 2011)}\ }\BibitemShut {NoStop}%
\bibitem [{\citenamefont {Puosi}\ \emph {et~al.}(2015)\citenamefont {Puosi},
  \citenamefont {Olivier},\ and\ \citenamefont {Martens}}]{puosi_probing_2015}%
  \BibitemOpen
\bibfield  {journal} {  }\bibfield  {author} {\bibinfo {author} {\bibfnamefont
  {F.}~\bibnamefont {Puosi}}, \bibinfo {author} {\bibfnamefont
  {J.}~\bibnamefont {Olivier}},\ and\ \bibinfo {author} {\bibfnamefont
  {K.}~\bibnamefont {Martens}},\ }\href {https://doi.org/10.1039/C5SM01694K}
  {\bibfield  {journal} {\bibinfo  {journal} {Soft Matter}\ }\textbf {\bibinfo
  {volume} {11}},\ \bibinfo {pages} {7639} (\bibinfo {year}
  {2015})}\BibitemShut {NoStop}%
\bibitem [{\citenamefont {Ruscher}\ and\ \citenamefont
  {Rottler}(2020)}]{ruscher_residual_2020}%
  \BibitemOpen
  \bibfield  {author} {\bibinfo {author} {\bibfnamefont {C.}~\bibnamefont
  {Ruscher}}\ and\ \bibinfo {author} {\bibfnamefont {J.}~\bibnamefont
  {Rottler}},\ }\href {https://doi.org/10.1039/D0SM01155J} {\bibfield
  {journal} {\bibinfo  {journal} {Soft Matter}\ }\textbf {\bibinfo {volume}
  {16}},\ \bibinfo {pages} {8940} (\bibinfo {year} {2020})}\BibitemShut
  {NoStop}%
\bibitem [{\citenamefont {Agarwal}\ and\ \citenamefont
  {Joshi}(2019)}]{agarwal_signatures_2019}%
  \BibitemOpen
  \bibfield  {author} {\bibinfo {author} {\bibfnamefont {M.}~\bibnamefont
  {Agarwal}}\ and\ \bibinfo {author} {\bibfnamefont {Y.~M.}\ \bibnamefont
  {Joshi}},\ }\href {https://doi.org/10.1063/1.5097779} {\bibfield  {journal}
  {\bibinfo  {journal} {Physics of Fluids}\ }\textbf {\bibinfo {volume} {31}},\
  \bibinfo {pages} {063107} (\bibinfo {year} {2019})}\BibitemShut {NoStop}%
\bibitem [{\citenamefont {Lidon}\ \emph {et~al.}(2017)\citenamefont {Lidon},
  \citenamefont {Villa},\ and\ \citenamefont
  {Manneville}}]{lidon_power-law_2017}%
  \BibitemOpen
  \bibfield  {author} {\bibinfo {author} {\bibfnamefont {P.}~\bibnamefont
  {Lidon}}, \bibinfo {author} {\bibfnamefont {L.}~\bibnamefont {Villa}},\ and\
  \bibinfo {author} {\bibfnamefont {S.}~\bibnamefont {Manneville}},\ }\href
  {https://doi.org/10.1007/s00397-016-0961-4} {\bibfield  {journal} {\bibinfo
  {journal} {Rheol Acta}\ }\textbf {\bibinfo {volume} {56}},\ \bibinfo {pages}
  {307} (\bibinfo {year} {2017})}\BibitemShut {NoStop}%
\bibitem [{\citenamefont {Hébraud}\ and\ \citenamefont
  {Lequeux}(1998)}]{hebraud_mode-coupling_1998}%
  \BibitemOpen
  \bibfield  {author} {\bibinfo {author} {\bibfnamefont {P.}~\bibnamefont
  {Hébraud}}\ and\ \bibinfo {author} {\bibfnamefont {F.}~\bibnamefont
  {Lequeux}},\ }\href {https://doi.org/10.1103/PhysRevLett.81.2934} {\bibfield
  {journal} {\bibinfo  {journal} {Physical Review Letters}\ }\textbf {\bibinfo
  {volume} {81}},\ \bibinfo {pages} {2934} (\bibinfo {year}
  {1998})}\BibitemShut {NoStop}%
\bibitem [{\citenamefont {Buldyrev}\ \emph {et~al.}(2001)\citenamefont
  {Buldyrev}, \citenamefont {Havlin}, \citenamefont {Kazakov}, \citenamefont
  {da~Luz}, \citenamefont {Raposo}, \citenamefont {Stanley},\ and\
  \citenamefont {Viswanathan}}]{buldyrev_average_2001}%
  \BibitemOpen
  \bibfield  {author} {\bibinfo {author} {\bibfnamefont {S.~V.}\ \bibnamefont
  {Buldyrev}}, \bibinfo {author} {\bibfnamefont {S.}~\bibnamefont {Havlin}},
  \bibinfo {author} {\bibfnamefont {A.~Y.}\ \bibnamefont {Kazakov}}, \bibinfo
  {author} {\bibfnamefont {M.~G.~E.}\ \bibnamefont {da~Luz}}, \bibinfo {author}
  {\bibfnamefont {E.~P.}\ \bibnamefont {Raposo}}, \bibinfo {author}
  {\bibfnamefont {H.~E.}\ \bibnamefont {Stanley}},\ and\ \bibinfo {author}
  {\bibfnamefont {G.~M.}\ \bibnamefont {Viswanathan}},\ }\href
  {https://doi.org/10.1103/PhysRevE.64.041108} {\bibfield  {journal} {\bibinfo
  {journal} {Phys. Rev. E}\ }\textbf {\bibinfo {volume} {64}},\ \bibinfo
  {pages} {041108} (\bibinfo {year} {2001})}\BibitemShut {NoStop}%
\bibitem [{\citenamefont {Zoia}\ \emph {et~al.}(2007)\citenamefont {Zoia},
  \citenamefont {Rosso},\ and\ \citenamefont {Kardar}}]{zoia_fractional_2007}%
  \BibitemOpen
  \bibfield  {author} {\bibinfo {author} {\bibfnamefont {A.}~\bibnamefont
  {Zoia}}, \bibinfo {author} {\bibfnamefont {A.}~\bibnamefont {Rosso}},\ and\
  \bibinfo {author} {\bibfnamefont {M.}~\bibnamefont {Kardar}},\ }\href
  {https://doi.org/10.1103/PhysRevE.76.021116} {\bibfield  {journal} {\bibinfo
  {journal} {Phys. Rev. E}\ }\textbf {\bibinfo {volume} {76}},\ \bibinfo
  {pages} {021116} (\bibinfo {year} {2007})}\BibitemShut {NoStop}%
\bibitem [{\citenamefont {Saitoh}\ \emph {et~al.}(2020)\citenamefont {Saitoh},
  \citenamefont {Hatano}, \citenamefont {Ikeda},\ and\ \citenamefont
  {Tighe}}]{saitoh_stress_2020}%
  \BibitemOpen
  \bibfield  {author} {\bibinfo {author} {\bibfnamefont {K.}~\bibnamefont
  {Saitoh}}, \bibinfo {author} {\bibfnamefont {T.}~\bibnamefont {Hatano}},
  \bibinfo {author} {\bibfnamefont {A.}~\bibnamefont {Ikeda}},\ and\ \bibinfo
  {author} {\bibfnamefont {B.~P.}\ \bibnamefont {Tighe}},\ }\href
  {https://doi.org/10.1103/PhysRevLett.124.118001} {\bibfield  {journal}
  {\bibinfo  {journal} {Phys. Rev. Lett.}\ }\textbf {\bibinfo {volume} {124}},\
  \bibinfo {pages} {118001} (\bibinfo {year} {2020})}\BibitemShut {NoStop}%
\bibitem [{Note7()}]{Note7}%
  \BibitemOpen
  \bibinfo {note} {There is, potentially, a contribution from relaxation around
  the origin $\sigma =0$, but we have checked numerically that this gives a
  sub-leading contribution.}\BibitemShut {Stop}%
\bibitem [{\citenamefont {Lees}\ and\ \citenamefont
  {Edwards}(1972)}]{lees_computer_1972}%
  \BibitemOpen
  \bibfield  {author} {\bibinfo {author} {\bibfnamefont {A.~W.}\ \bibnamefont
  {Lees}}\ and\ \bibinfo {author} {\bibfnamefont {S.~F.}\ \bibnamefont
  {Edwards}},\ }\href {https://doi.org/10.1088/0022-3719/5/15/006} {\bibfield
  {journal} {\bibinfo  {journal} {J. Phys. C: Solid State Phys.}\ }\textbf
  {\bibinfo {volume} {5}},\ \bibinfo {pages} {1921} (\bibinfo {year}
  {1972})}\BibitemShut {NoStop}%
\bibitem [{\citenamefont {Nishikawa}\ \emph {et~al.}(2021)\citenamefont
  {Nishikawa}, \citenamefont {Ikeda},\ and\ \citenamefont
  {Berthier}}]{nishikawa_relaxation_2021}%
  \BibitemOpen
  \bibfield  {author} {\bibinfo {author} {\bibfnamefont {Y.}~\bibnamefont
  {Nishikawa}}, \bibinfo {author} {\bibfnamefont {A.}~\bibnamefont {Ikeda}},\
  and\ \bibinfo {author} {\bibfnamefont {L.}~\bibnamefont {Berthier}},\ }\href
  {https://doi.org/10.1007/s10955-021-02710-8} {\bibfield  {journal} {\bibinfo
  {journal} {J Stat Phys}\ }\textbf {\bibinfo {volume} {182}},\ \bibinfo
  {pages} {37} (\bibinfo {year} {2021})}\BibitemShut {NoStop}%
\bibitem [{\citenamefont {Olsson}(2022)}]{olsson_relaxation_2022}%
  \BibitemOpen
  \bibfield  {author} {\bibinfo {author} {\bibfnamefont {P.}~\bibnamefont
  {Olsson}},\ }\href {https://doi.org/10.1103/PhysRevE.105.034902} {\bibfield
  {journal} {\bibinfo  {journal} {Phys. Rev. E}\ }\textbf {\bibinfo {volume}
  {105}},\ \bibinfo {pages} {034902} (\bibinfo {year} {2022})}\BibitemShut
  {NoStop}%
\bibitem [{\citenamefont {Manacorda}\ and\ \citenamefont
  {Zamponi}(2022)}]{manacorda_gradient_2022}%
  \BibitemOpen
  \bibfield  {author} {\bibinfo {author} {\bibfnamefont {A.}~\bibnamefont
  {Manacorda}}\ and\ \bibinfo {author} {\bibfnamefont {F.}~\bibnamefont
  {Zamponi}},\ }\href {http://arxiv.org/abs/2201.01161} {\bibfield  {journal}
  {\bibinfo  {journal} {arXiv:2201.01161 [cond-mat]}\ } (\bibinfo {year}
  {2022})}\BibitemShut {NoStop}%
\end{thebibliography}%
\bibliographystyle{apsrev4-2.bst}

\end{document}